# Taming Scope Extrusion in Gradual Imperative Metaprogramming

TIANYU CHEN, Indiana University, USA
DARSHAL SHETTY, Indiana University, USA
JEREMY G. SIEK, Indiana University, USA
CHAO-HONG CHEN, Meta, USA
WEIXI MA, Meta, USA
ARNAUD VENET, Meta, USA
ROCKY LIU, Meta, USA

Metaprogramming enables the generation of performant code, while gradual typing facilitates the smooth migration from untyped scripts to robust statically typed programs. However, combining these features with imperative state—specifically mutable references—introduces the classic peril of scope extrusion, where code fragments containing free variables escape their defining lexical context. While static type systems have employed environment classifiers to successfully tame this interaction, enforcing these invariants in a gradual language remains an open challenge.

This paper presents $\lambda^{\alpha,\star}_{\text{Ref}}$, the first gradual metaprogramming language that supports mutable references while guaranteeing scope safety. To put $\lambda^{\alpha,\star}_{\text{Ref}}$ on a firm foundation, we also develop its statically typed sister language, $\lambda^{\alpha}_{\text{Ref}}$, which introduces unrestricted subtyping for environment classifiers. We mechanize the proof of scope safety for $\lambda^{\alpha}_{\text{Ref}}$, contributing the first mechanized proof of scope safety for a statically typed language. Our key innovation, however, is the dynamic enforcement of the environment classifier discipline in $\lambda^{\alpha,\star}_{\text{Ref}}$, enabling the language to mediate between statically verified scopes and dynamically verified scopes. The dynamic enforcement is carried out in a novel cast calculus $\text{CC}^{\alpha,\star}_{\text{Ref}}$ that uses an extension of Henglein's Coercion Calculus to handle code types, classifier polymorphism, and subtype constraints. We prove that $\lambda^{\alpha,\star}_{\text{Ref}}$ satisfies type safety and scope safety. Finally, we provide a space-efficient implementation strategy for the dynamic scope checks, ensuring that the runtime overhead remains practical. All of our results are fully mechanized in Agda.

**ACM Reference Format:**
Tianyu Chen, Darshal Shetty, Jeremy G. Siek, Chao-Hong Chen, Weixi Ma, Arnaud Venet, and Rocky Liu. 2018. Taming Scope Extrusion in Gradual Imperative Metaprogramming. In *Proceedings of Make sure to enter the correct conference title from your rights confirmation email (Conference acronym 'XX)*. ACM, New York, NY, USA, 52 pages. https://doi.org/XXXXXXX.XXXXXXX

Authors' Contact Information: Tianyu Chen, Indiana University, Bloomington, USA, chen512@iu.edu; Darshal Shetty, Indiana University, Bloomington, USA, dcshetty@iu.edu; Jeremy G. Siek, Indiana University, Bloomington, USA, jsiek@iu.edu; Chao-Hong Chen, Meta, Menlo Park, USA, chaohong@meta.com; Weixi Ma, Meta, Menlo Park, USA, mavc@meta.com; Arnaud Venet, Meta, Menlo Park, USA, ajv@meta.com; Rocky Liu, Meta, Menlo Park, USA, rockyliu4@meta.com.

## 1 Introduction

The impetus of our work is a production problem at Meta: helping machine-learning engineers build and debug feature-engineering pipelines. These engineers use F3 [Ma et al. 2024], a DSL embedded in Python that generates statically typed data-pipeline descriptions. Although Python is dynamically typed, engineers add type annotations and check them with gradual type checkers such as Pyrefly (https://pyrefly.org/). Thus F3 combines metaprogramming with gradual typing in a production setting. Of course, the two principles at play here, metaprogramming and gradual typing, are of broad interest. Metaprogramming lets programs generate programs, enabling runtime specialization without sacrificing high-level abstractions [Davies and Pfenning 2001; Taha and Sheard 2000] and gradual typing lets developers mix dynamic and static code while migrating toward stronger static checks [Gronski et al. 2006; Matthews and Findler 2007; Siek and Taha 2006, 2007; Tobin-Hochstadt and Felleisen 2006].

One of the classic problems in language design for metaprogramming languages is how to prevent the generation and execution of ill-scoped code, that is, code with undefined variables. In general, one can study languages with arbitrary nesting of code and stages of computation, but here we only require a two-stage language to model F3. Interestingly, a two-stage language based on the simply-typed lambda calculus (STLC) with quote, splice, and a straightforward code type, is trivially guaranteed to generate well-scoped code. However, if you naively add **`run`** (aka. **`eval`**) to the language, then one can apply **`run`** to a code fragment with free variables (*open code*) and therefore trigger a runtime scope error. If you naively add mutable references, more trouble arises because open code can be smuggled out of its defining scope by assignment to a reference that was created prior to its scope, a problem known as *scope extrusion*. In general, we use the term *scope safety* for the property that a metaprogram is guaranteed only to generate well-scoped code.

Many programming language techniques have been invented to address scope safety. In general, they can be classified into three categories: *static* type systems that analyze metaprograms, *runtime* systems that monitor meta evaluation to catch the generation of ill-scope code, and *post-hoc* systems that check the generated code immediately after meta evaluation. These techniques come with the usual pros and cons of static and dynamic checking. The static type systems impose a conservative regime on metaprograms and come with an up-front development cost, but guarantee the complete absence of scope errors. Thus, developers can distribute their metaprograms to clients without fear of scope errors occurring when those metaprograms are run. The runtime systems typically do not impose up-front constraints on a metaprogram, but instead come with runtime overhead and lack the strong no-scope-error guarantee of static checking. However, they do aid in the debugging of metaprograms because they catch scope errors closer to their source. The post-hoc approach is simple to implement and better than silently generating ill-scoped code, but lacks the above-mentioned benefits of static and runtime scope checking.

Table 1 categorizes the language techniques that have been invented for ensuring scope safety, organized by the above-mentioned categories (static, runtime, post-hoc) and by the programming language setting.

Within the static category, researchers have addressed scope safety in languages with **`run`** by leveraging contextual modal types [Parreaux 2020] and inventing closed types [Moggi et al. 1999] and environment classifiers (ECs) [Taha and Nielsen 2003]. Contextual modal types record the typing context of code directly in the quote term and code type [Murase et al. 2023; Nanevski et al. 2008; Zyuzin and Nanevski 2021]; environment classifiers give names to scopes and track them in the type system [Taha and Nielsen 2003]. Early versions of MetaOCaml used environment classifiers for static scope checking, but that was not enough to guarantee scope safety because MetaOCaml includes mutable references [Calcagno et al. 2003; Lee et al. 2026]. Since then researchers have invented several techniques to ensure scope safety in the presence of references including refined environment classifiers [Isoda et al. 2024; Kiselyov et al. 2016], closed types for references [Calcagno et al. 2000], and weak separability [Westbrook et al. 2010].

| Paradigms | $\lambda$ with quote, splice, and run | $\lambda$ with quote, splice, and references/effects |
|---|---|---|
| Static | closed types [Moggi et al. 1999];<br>environment classifiers [Taha and Nielsen 2003];<br>contextual modal types [Parreaux 2020]. | closed types [Calcagno et al. 2000];<br>weak separability [Westbrook et al. 2010];<br>refined environment classifiers [Isoda et al. 2024; Kiselyov et al. 2016]. |
| Runtime | closedness checking [Yaguchi and Kameyama 2025]. | eager check of free vars. [Kiselyov 2014, 2023, 2026];<br>cause-for-concern check [Lee et al. 2026]. |
| Post-hoc | — | Calcagno et al. [2003]; Ma et al. [2024]; Sheard and Peyton Jones [2002]. |
| Gradual | — | gradual environment classifiers (this paper) |

Table 1. Language techniques for ensuring scope safety

Within the runtime category of Table 1, the **run** setting can be handled by dynamic closedness checking, which blocks attempts to execute open code [Yaguchi and Kameyama 2025].[1] In languages with mutable references or effects, the monitor must also prevent open code from escaping its defining scope through the store. MetaOCaml uses eager free-variable checks for this purpose [Kiselyov 2014, 2023, 2026]; MacoCaml uses cause-for-concern checks [Lee et al. 2026].

Within the category of post-hoc scope checking, example systems include F3 [Ma et al. 2024] and early versions of MetaOCaml [Calcagno et al. 2003; Lee et al. 2026], which implemented post-hoc checking as a backup to static checking with environment classifiers.

The prior work on gradual metaprogramming by Chen et al. [2025], not shown in the table, was conducted in a language with neither **run** nor mutable references, so scope safety was trivial.

Several open questions and challenges remain. The main open question addressed in this paper is how to design a gradual system for scope safety in non-trivial settings. By a gradual system, we specifically mean one in which scope checking can happen statically or dynamically in different regions of a program at the discretion of the programmer. We choose the setting of a language with mutable references, rather than **run**, because that is a better model of F3. We note that mutable references are the more difficult setting and it is usually straightforward to adapt scope checking techniques that work for mutable references to also handle languages with **run**. We choose refined environment classifiers [Kiselyov et al. 2016] as our static approach to scope safety and we invent a notion of runtime checking of environment classifiers in this paper. We did not use any of the prior runtime systems because there would have been an impedance mismatch with the static environment classifiers.[2] Thus, this paper answers a second open question, which is whether there exists a runtime counterpart to environment classifiers. Finally, while there have been pen-and-paper proofs of scope safety for several of these systems, there have not been any mechanized proofs of scope safety. Li et al. [2026] mechanize the semantics of a multi-stage programming language and prove type soundness and phase distinction but leave "the challenging problem of scope extrusion" as future work. We present mechanized proofs of scope safety for a fully static metalanguage and for a gradual metalanguage, both with mutable references.

[1] The language of Yaguchi and Kameyama [2025] is broadly categorized as gradual, but the technique employed for scope safety is fully dynamic.

[2] We conjecture that runtime environment classifiers will yield lower runtime overhead than runtime scope checking based on sets of free variables.

***Our Solution.*** We present $\lambda^{\alpha,\star}_{\text{Ref}}$, a gradually typed metaprogramming language that supports mutable references while guaranteeing scope safety for generated code. Our central technical idea is the dynamic enforcement of the environment-classifier discipline. In $\lambda^{\alpha,\star}_{\text{Ref}}$, environment classifiers are also represented at runtime by unique identifiers. When code is cast from a specific scope to an unknown scope, the identifier for the source scope is attached to the code value as runtime type information (RTTI). When code is later cast from an unknown scope back to a specific scope, $\lambda^{\alpha,\star}_{\text{Ref}}$ checks at runtime that the scope recorded in the RTTI is well scoped with respect to the target scope of the cast. This check prevents code whose free variables belong to an inaccessible scope from being treated as well scoped in the target scope. We formalize this design with a cast calculus, $\text{CC}^{\alpha,\star}_{\text{Ref}}$, whose coercions extend the Coercion Calculus of Henglein [1994] to account for code types, environment classifiers, and subtype constraints.

Naively implementing gradual typing can lead to catastrophic slowdowns in the runtime of programs [Herman et al. 2010; Kuhlenschmidt et al. 2019]. In a naive implementation, values can be wrapped in long chains of proxies while flowing through the static/dynamic boundary at runtime, causing unbounded space consumption. Accessing those values involves traversing the proxy chains, which causes catastrophic slowdowns. We discuss how to prevent such slowdowns by compressing proxy chains at runtime using the hypercoercion cast representation in $\text{CC}^{\alpha,\star}_{\text{Ref}}$. We show that compressing proxy chains avoids catastrophic slowdowns by proving a space consumption theorem which states that the space consumed by proxy chains doesn't grow in an unbounded manner.

***Contributions.*** This paper makes the following technical contributions:

(1) We present a fully mechanized proof of scope safety for a static metaprogramming language, namely $\lambda^{\alpha}_{\text{Ref}}$ (Corollary 7.8). The most noteworthy feature of $\lambda^{\alpha}_{\text{Ref}}$ is that it is the first meta-language to provide first-class subtyping with respect to environment classifiers (§ 3). The scope safety of $\lambda^{\alpha}_{\text{Ref}}$ is a direct corollary of that of the gradual language.
(2) We introduce the first gradual metaprogramming language with general mutable references that prevents scope extrusion, namely $\lambda^{\alpha,\star}_{\text{Ref}}$ (§ 4).
(3) We introduce the first mechanism for the dynamic enforcement of the environment classifier discipline, a necessary ingredient for our gradual language $\lambda^{\alpha,\star}_{\text{Ref}}$ and its cast calculus $\text{CC}^{\alpha,\star}_{\text{Ref}}$ (§ 5). We present the first mechanized scope safety proof for a gradually typed metalanguage (Theorem 7.7).
(4) We extend the Coercion Calculus to handle code types, environment classifiers, and subtype constraints (§ 5.3). This novel EC coercion calculus serves as our on-demand runtime monitor for detecting scope extrusion.
(5) We provide a space-efficient implementation strategy for the dynamic enforcement of ECs, which ensures that the overhead of scope safety remains practical (§ 8). We mechanize the space consumption theorem for $\lambda^{\alpha,\star}_{\text{Ref}}$ (Theorem 8.1).

The remainder of this paper is organized as follows. Section 2 provides an overview of the gradual enforcement of scope safety. Section 3 describes the fully static sister language $\lambda^{\alpha}_{\text{Ref}}$. Section 4 presents the formal syntax and static semantics of $\lambda^{\alpha,\star}_{\text{Ref}}$. Section 5 details the cast calculus and Section 6 defines the elaboration process. Section 7 presents the scope safety results. Section 8 describes our space-efficient implementation strategy. Section 9 concludes.

The Appendix, the Agda proofs, the example programs of $\lambda^{\alpha,\star}_{\text{Ref}}$ and its sister $\lambda^{\alpha}_{\text{Ref}}$, and a prototype implementation of $\lambda^{\alpha,\star}_{\text{Ref}}$ are all in the supplementary material of this paper. Table 2 of the Appendix maps the lemmas and theorems of this paper to Agda code.

## 2 Taming Scope Extrusion

In the tradition of Lisp [Steele 1990] and MetaML [Sheard 1998], we study a language that can manipulate code as data. The *quote* feature turns a piece of code into a value, here written with literal quotes (`` . . . ''), and the *splice* feature inserts the result of an expression into a code value, here written as a tilde (~). For example, the following program produces the code ``4 + 3''.

```
let c = ``3'' in ``4 + ~c''        ⟶   ``4 + 3''
```

The code in a quote may include function definitions so questions about variable scoping arise. In the following, the x in ``x'' is bound to the λ (x:Int) despite the intervening splice.

```
``λ (x:Int). ~(let c = ``x'' in c)''        ⟶   ``λ (x:Int). x''
```

This program produces the well-scoped code ``λ (x:Int). x''.

Of course, one must worry about generating code with occurrences of undefined variables. Standard rules about lexical scoping reject problematic metaprograms such as ``λ (x:Int). y'', because y is not in scope. However, rules regarding lexical scoping are not enough to prevent variables from escaping their defining scope, a phenomenon known as *scope extrusion*. The following classic example [Kiselyov et al. 2016] assigns ``x'' to a mutable reference, creates and discards the code ``λ (x:Int). x'', and then reads from the reference to produce an ill-scoped ``x''.

```
1 let r = ref ``0'' in
2   ``λ (x:Int). ~(r := ``x''; !r)'';
3   !r
                                   ⟶   ``x''        scope extrusion
```

There are many other language features that enable the escaping of variables from their defining scope, such as exceptions, continuations, and effect handlers. Also, in metalanguages with an operation to **run** a code value (aka. **eval**), variables don't need to escape their defining scope to be problematic. One can prematurely **run** a piece of code with free variables, as in the following.

```
``λ (x:Int). ~(run ``x''; ``x'')''        scope extrusion with run
```

The rest of the section is organized as follows. In Section 2.1, we review how environment classifiers guarantee scope safety fully statically in the presence of mutable references. In Section 2.2, we present examples of gradual environment classifiers, a new design of this paper that extends environment classifiers to combine both static and runtime scope enforcement.

### 2.1 Background: Environment Classifiers Enforce Scope Safety Statically

Taha and Nielsen [2003] introduced *environment classifiers* (ECs) in a *static* type system for a pure functional language to prevent **run** from executing code with free variables. The intuition is that environment classifiers act as scope identifiers so the type system can keep track of which variables can appear in which scopes, and which scopes can appear inside other scopes. Kiselyov et al. [2016] refined this type system to prevent scope extrusion through mutable references as well.

We review the core ideas of environment classifiers with some examples. Consider again the classic scope extrusion program, but this time annotated with environment classifiers.

```
1 let r : Ref ``Int''ε = ref ``0''ε in        scope extrusion with EC (statically rejected)
2   ``λ (x:Int) α. ~(r := ``x''α; !r)''ε;
3   !r
```

The empty scope (the top of the program) is identified by the EC written ε. Each λ in a quote is a binder for a new EC, such as α in λ (x:Int) α. Every quote is annotated with an EC to indicate which scope that quote may be spliced into. The type system makes sure that the variable occurrences inside a quote are well-scoped with respect to its EC annotation. For example, ``0''ε (line 1) is well-scoped because it does not contain any variable occurrences. On the other hand, ``x''α (line 2) is well-scoped because α refers to the scope λ (x:Int) α and x is bound in that scope. The type system tracks ECs by including an environment classifier in the code type. For example, ``Int''ε is the type for code that would produce an integer and that can be spliced into the empty scope. The type system correctly rejects the above example of scope extrusion. The type error points to the assignment ( r := ``x''α on line 2) because the type of the right-hand side (``Int''α) is not a subtype of the reference's element type (``Int''ε).

Subtyping for environment classifiers, α <: β, means that β is inside the scope of α. The empty scope ε is the bottom of the subtyping lattice. The following metaprogram demonstrates subtyping:

```
``λ (x:Int) α. λ (y:Int) β. ~(``x''α)''ε   ⟶   ``λ (x:Int) α. λ (y:Int) β. x''ε
```

The subtyping relation for this program has ε <: α <: β: the type system establishes ε <: α because it sees that λ (x:Int) α is inside a quote annotated with ε, and the type system establishes α <: β because λ (y:Int) β is inside λ (x:Int) α. The splice ~(``x'' α) expects ``Int''β but the type of the expression inside the splice is ``Int''α; that is okay because α <: β.

## 2.2 New: Gradual Environment Classifiers Combine Static and Dynamic Enforcement

Consider again the scope-extrusion example, but this time the annotation on the reference r uses the unknown environment classifier ★ (highlighted on line 1).

```
1 let r : Ref ``Int''★ = ref ``0''ε in        [scope extrusion with gradual EC (dynamically rejected)]
2   ``λ (x:Int) α. ~(r := ``x''α ; !r)''ε;   /* type checks, errors at runtime */
3   !r
```

With that change, the static type system of gradual $\lambda^{\alpha,\star}_{\text{Ref}}$ calculus accepts the program. In particular, the initialization of r involves the implicit conversion from Ref ``Int''ε to Ref ``Int''★, which is allowed because any environment classifier may be converted to ★. On the other hand, the type of !r is ``Int''★, which must be implicitly converted to ``Int''α, the type expected at the splice. That is also allowed because ★ may be converted to any classifier.

Now it is up to the runtime system of $\lambda^{\alpha,\star}_{\text{Ref}}$ to prevent scope extrusion by enforcing the environment classifier discipline. We take inspiration from the literature on gradual typing and polymorphism, which enforces parametricity at runtime using runtime-generated symbols [Ahmed et al. 2011, 2017; Igarashi et al. 2024; Labrada et al. 2022] together with the usual runtime checking of every implicit conversion between types. Here we use runtime-generated symbols to represent environment classifiers (i.e., scope locations) and we introduce new machinery to handle conversions on code types and environment classifiers.

Let us sketch the execution of the above example. The expression **ref** ``0''ε allocates a cell in memory containing a piece of code, the literal 0, and then returns the address of the cell. Next the conversion from Ref ``Int''ε to Ref ``Int''★ for the initialization of r causes the address to be wrapped in a proxy that will apply the appropriate conversion when reading from or writing to the address [Herman et al. 2010]. The evaluation of the quoted λ (x:Int) α will create an AST node for the λ, generate $\alpha_1$ to serve as a pointer to it, and then substitute $\alpha_1$ for α. The evaluation of ``x''$\alpha_1$ creates a piece of code and performs the implicit conversion from ``Int''$\alpha_1$ to ``Int''★, which attaches the EC $\alpha_1$ to the piece of code. The assignment r := ``x''$\alpha_1$ then asks the proxied

$$
\begin{array}{llcl}
\text{EC variables} & \alpha, \beta, \gamma & & \\
\text{ECs} & e & ::= & \alpha \mid \varepsilon \\
\text{code types} & T & ::= & \iota \mid T \to T \\
\text{meta types} & A, B & ::= & \iota \mid A \to B \mid \mathtt{Ref}\ A \mid \text{``}T\text{''}e \mid \forall \alpha. A \mid e{<:}e \Rightarrow A \\
\text{contexts} & \Gamma & ::= & \emptyset \mid \Gamma, \alpha \mid \Gamma, e_1{<:}e_2 \mid \Gamma, x : A \mid \Gamma, (x : T)^{\alpha} \\
\text{subtype contexts} & \Theta & ::= & \emptyset \mid \Theta, e_1{<:}e_2
\end{array}
$$

$\boxed{\Theta \vdash e_1 <: e_2}$

$$
\frac{e_1 <: e_2 \in \Theta}{\Theta \vdash e_1 <: e_2} \qquad \frac{}{\Theta \vdash \varepsilon <: e} \qquad \frac{}{\Theta \vdash e <: e} \qquad \frac{\Theta \vdash e_1 <: e_2 \qquad \Theta \vdash e_2 <: e_3}{\Theta \vdash e_1 <: e_3}
$$

$\boxed{\Theta \vdash A <: B}$

$$
\frac{\Theta \vdash e_1 <: e_2}{\Theta \vdash \text{``}T\text{''}e_1 <: \text{``}T\text{''}e_2} \qquad \frac{\Theta, e_1 <: e_2 \vdash A <: B}{\Theta \vdash (e_1{<:}e_2 \Rightarrow A) <: (e_1{<:}e_2 \Rightarrow B)} \qquad \ldots
$$

Fig. 1. Syntax and selected subtyping rules for $\lambda^{\alpha}_{\mathrm{Ref}}$ (excerpt of Figure 9 of the Appendix)

address to try to write the code with $\alpha_1$ into the heap cell of type ``Int''$\varepsilon$, which raises a runtime error because $\alpha_1 \not<: \varepsilon$. So indeed, the runtime system prevented scope extrusion.

The runtime checking of implicit type conversions can occur in many different locations in a program, not just in operations on mutable references. In this next example, the location of the implicit conversion is at a splice. The meta function `f` splices the parameter `c`, with an unknown EC, into the body of `λ (x:Int) α`, so there is an implicit conversion from ``Int''★ to ``Int''$\alpha$ just before the splice.

```
1 let f = λ (c:``Int''★). ``λ (x:Int) α. ~c ''ε in
2 let r = ref ``λ (z:Int) γ. z''ε in
3   ``λ (y:Int) β. ~(r := (f ``y''β); ``y''β)''ε;
4   !r
```

scope extrusion that errors at splice

The example creates the code `λ (y:Int) β` and calls `f` with the argument ``y''$\beta$. Once this argument is substituted into the body of `f`, the conversion to ``Int''$\alpha$ raises an error because $\beta \not<: \alpha$. Indeed, if there were no runtime checking, this program would produce `(λ (x:Int) β. y)`, with an ill-scoped occurrence of `y`.

## 3 Identifying the Static Sister Language $\lambda^{\alpha}_{\mathrm{Ref}}$

Every gradually typed language has a sister language that is statically typed. One school of design for gradually typed languages is to first identify or create such a sister language before designing the gradual language. The language <NJ> of Kiselyov et al. [2016] seems like a good fit in that it supports mutable references, environment classifiers, and environment classifier subtyping and polymorphism, which increase the expressivity of the language. However, we noticed that subtyping is restricted in <NJ> (§ 3.1) and that environment classifier polymorphism was not formally specified (§ 3.2). These observations led us to design a new core calculus for statically typed metaprogramming, $\lambda^{\alpha}_{\mathrm{Ref}}$. We present the type system for $\lambda^{\alpha}_{\mathrm{Ref}}$ in Section 3.3 and the reduction semantics in Section 3.4.

### 3.1 Unrestricted Subtyping for Classifiers

While the type system of <NJ> allows subtype conversion in some places, it is not allowed in other places. For example, the following program is rejected by <NJ> even though the normal subtyping rule for function types would allow it. The initialization of variable g from f requires a conversion from (Unit → ``Int''$\alpha$) to (Unit → ``Int''$\beta$), and indeed $\alpha <: \beta$.

```
1 ``λ (x:Int) α. λ (y:Int) β.                          <NJ> being restrictive
2    ~(let f : Unit → ``Int''α = λ (_:Unit). ``x''α in
3      let g : Unit → ``Int''β = f in
4        g ())''ε
```

We conjecture that the type system of Kiselyov et al. [2016] is restricted with respect to subtyping because it uses Hindley-Milner type inference, and the integration of the two requires complex engineering [Pottier 1998].

The meta language of $\lambda^{\alpha}_{\text{Ref}}$ does not employ Hindley-Milner type inference. The code language of $\lambda^{\alpha}_{\text{Ref}}$ uses bidirectional type inference to better align with its gradual counterpart $\lambda^{\alpha,\star}_{\text{Ref}}$, which requires bidirectional type inference for a technical reason that we describe in Section 6.

Figure 1 presents the syntax of types, EC subtyping, and selected subtyping rules for code and subtype-constrained types in $\lambda^{\alpha}_{\text{Ref}}$. The full type well-formedness and subtyping rules are defined in Figure 9 of the Appendix. We take $e$ to range over environment classifier expressions, $T$ to range over code types, and $A$ and $B$ to range over meta types. We write $\Theta$ for the subtype context, which syntactically is a sequence of EC subtype constraints, and $\text{Sub}(\Gamma)$ for the ordered subsequence of EC subtype constraints in $\Gamma$. The subtyping judgments use $\Theta$. The meta types include base types $(\iota \in \{\texttt{Int}, \texttt{Bool}, \ldots\})$, function types, and quote types ``$T$''$e$. We discuss the universal type $\forall \alpha. A$ and constrained type $e{<:}e{\Rightarrow}A$ in the next subsection. As usual, subtyping is reflexive and transitive.

## 3.2 Subtype-Bounded Classifier Polymorphism

Kiselyov et al. [2016] discuss the need for classifier polymorphism and subtype constraints. They implemented these features in MetaOCaml but did not include them in their formalization. We adapt their examples to our formalism and include these features in $\lambda^{\alpha}_{\text{Ref}}$.

**Classifier Polymorphism.** The following example demonstrates the use of classifier polymorphism. It defines a polymorphic function p and then uses it in two places with two different scopes $\beta$ and $\gamma$. Because p is parameterized over the EC $\alpha$, it can be instantiated with $\beta$ in the first case and instantiated with $\gamma$ in the second.

```
1 λ (x:``Int''ε).                                       classifier polymorphism example
2   let p = Λ α. λ (y:``Int''α). ``(~x) + (~y)''α in
3   let f = ``λ (z:Int) β. ~(p[β] ``z''β) + z''ε in
4   let g = ``λ (w:Int) γ. ~(p[γ] ``w''γ) * w''ε in
5     ...
```

**Subtype Constraints.** The need for subtype constraints can be seen in the following function named wrap that generates a function whose body is constructed by calling body with quoted x.

```
1 wrap = Λ α. λ (body : ∀ β. α <: β ⇒ ``Int''β → ``Int''β).      subtype constraint example
2          ``λ (x:Int) γ. ~(body[γ] • ``x''γ)''α
```

Note that wrap is parameterized by $\alpha$ and returns a quote annotated with $\alpha$. The intent is that the result of wrap will be spliced into scope $\alpha$ and that body may return code that uses variables from scope $\alpha$. The code returned by body may also use the variable x from scope $\gamma$. However, $\gamma$ is not in scope at the point where we must declare body, so we parameterize it with the EC variable $\beta$ and use a subtype constraint $\alpha <: \beta$ to allow body to return code that uses variables from scope $\alpha$.

Inside `wrap`, we write `body[γ]` to instantiate `body` at γ and then write • to discharge the subtype constraint α <: γ, which is satisfied because `λ (x:Int) γ` is inside a quote annotated with α.

To see how `wrap` can be used, suppose we want to generate the following code, where the inner λ, which binds `x` and is associated with the classifier γ, is created by `wrap`. Let ξ be the classifier of the outer λ that binds `u`.

```
``λ (u:Int) ξ. λ (x:Int) γ. u + x''        code to generate by calling “wrap”
```

We instantiate `wrap`'s α to ξ and create a function `g` that generates the body `u + x`. Because we instantiate `wrap` with ξ, the type of `g` must be

```
∀ β. ξ <: β ⇒ ``Int''β → ``Int''β
```

Following the types, here is the definition of `g`.

```
g = Λ β. ξ <: β ⇒ λ (z:``Int''β). ``u + ~z''β
```

Here is the completed code that uses `wrap`.

```
1 ``λ (u:Int) ξ.
2     ~(let g = Λ β. ξ <: β ⇒ λ (z:``Int''β). ``u + ~z''β in
3       wrap[ξ] g)''ε
```

⟶

```
``λ (u:Int) ξ.
   λ (x:Int) γ.
     u + x''ε
```

### 3.3 Type System for $\lambda^{\alpha}_{\mathrm{Ref}}$

Figure 2 defines the terms and typing rules of $\lambda^{\alpha}_{\mathrm{Ref}}$, which includes the discipline of environment classifiers that we discussed in Section 2.1. This type system adds first-class subtyping and formalizes the treatment of EC polymorphism and subtype constraints. The type system makes use of the type well-formedness and subtyping defined in Figure 9 of the Appendix. The context Γ consists of EC, code, and meta variables, as well as subtype constraints. There are three typing judgments. The typing judgments for code terms take the forms $\Gamma; e \vdash M^{\circ} \Rightarrow T$ and $\Gamma; e \vdash M^{\circ} \Leftarrow T$; they are for synthesis and checking in bidirectional typing, respectively. The EC $e$ in these judgments points to the nearest enclosing generated-code scope. The typing judgment for meta terms takes the form $\Gamma \vdash M : A$, where $A$ is a meta type.

**Type System of the Code Language.** The typing rule for code variables (⊢*code-var*) enforces the discipline that a variable may only occur in locations where its environment classifier $\alpha$ is a subtype of the nearest enclosing scope $e$ (highlighted), thereby ensuring that $e$ is lexically inside the scope of $\alpha$. The typing rule for code $\lambda$ (⊢*code-abs*) chooses a fresh EC $\alpha$ to associate with the scope of its parameter $x$ and links $\alpha$ to the nearest enclosing scope with the subtyping assumption $e <: \alpha$ (highlighted). In the body of this $\lambda$, $\alpha$ is the nearest enclosing scope. In the rule for splicing the result of a meta term (⊢*splice*), the meta term must produce code whose environment classifier is a subtype of the enclosing EC (highlighted). We do not allow splice to appear in synthesis positions, only checking, to synchronize with $\lambda^{\alpha,\star}_{\mathrm{Ref}}$ (see Section 6).

**Type System of the Meta Language.** For a quote annotated with EC $e$ (⊢*quote*), we synthesize the type of the code with $e$ as the enclosing EC. The application and reference assignment rules use subtyping exactly at the points where a synthesized type must flow into an expected type. The typing rules for EC abstraction (⊢*ec-abs*) and EC application (⊢*ec-app*) are analogous to the standard rules for type abstraction and type application in System F [Girard 1972, 1986; Reynolds 1983]. The rule for subtype constraint introduction (⊢*sub-intro*) extends the context Γ with the assumption

$$
\begin{array}{llcl}
\text{constants} & k & & \\
\text{term variables} & x, y, z & & \\
\text{code terms} & L^\circ, M^\circ, N^\circ & ::= & k \mid x \mid \lambda(x{:}T)^{\alpha}.\, M^\circ \mid M^\circ\ N^\circ \mid {\sim}M \\
\text{meta terms} & L, M, N & ::= & k \mid x \mid \lambda x{:}A.\, M \mid M\ N \mid \Lambda\alpha.\, M \mid M[e] \mid e_1{<:}e_2{\Rightarrow}M \mid M\bullet \\
 & & \mid & ``M^\circ{''}e \mid \mathtt{ref}\ M \mid M := N \mid\ !M
\end{array}
$$

$$\boxed{\Gamma; e \vdash M^\circ \Rightarrow T}$$

$$\frac{k : \iota}{\Gamma; e \vdash k \Rightarrow \iota}\ (\vdash\textit{code-const}) \qquad \frac{(x : T)^\alpha \in \Gamma \qquad \mathsf{Sub}(\Gamma) \vdash \alpha <: e}{\Gamma; e \vdash x \Rightarrow T}\ (\vdash\textit{code-var})$$

$$\frac{\Gamma,\ \alpha, e{<:}\alpha, (x : T_1)^\alpha; \alpha \vdash N^\circ \Rightarrow T_2 \qquad \alpha \notin \Gamma}{\Gamma; e \vdash \lambda(x{:}T_1)^\alpha.\, N^\circ \Rightarrow T_1 \to T_2}\ (\vdash\textit{code-abs}) \qquad \frac{\Gamma; e \vdash L^\circ \Rightarrow T_1 \to T_2 \qquad \Gamma; e \vdash M^\circ \Leftarrow T_1}{\Gamma; e \vdash L^\circ\ M^\circ \Rightarrow T_2}\ (\vdash\textit{code-app})$$

$$\boxed{\Gamma; e \vdash M^\circ \Leftarrow T}$$

$$\frac{\Gamma \vdash M : A \qquad \mathsf{Sub}(\Gamma) \vdash A <: ``T{''}e}{\Gamma; e \vdash {\sim}M \Leftarrow T}\ (\vdash\textit{splice}) \qquad \frac{\Gamma; e \vdash M^\circ \Rightarrow T}{\Gamma; e \vdash M^\circ \Leftarrow T}\ (\vdash\textit{check})$$

$$\boxed{\Gamma \vdash M : A}$$

$$\frac{k : \iota}{\Gamma \vdash k : \iota}\ (\vdash\textit{meta-const}) \qquad \frac{x : A \in \Gamma}{\Gamma \vdash x : A}\ (\vdash\textit{meta-var}) \qquad \frac{\Gamma \vdash A \qquad \Gamma, x : A \vdash M : B}{\Gamma \vdash \lambda x{:}A.\, M : A \to B}\ (\vdash\textit{meta-abs})$$

$$\frac{\Gamma \vdash L : A_1 \to A_2 \qquad \Gamma \vdash M : B \qquad \mathsf{Sub}(\Gamma) \vdash B <: A_1}{\Gamma \vdash L\ M : A_2}\ (\vdash\textit{meta-app}) \qquad \frac{\Gamma, \alpha \vdash M : A \qquad \alpha \notin \Gamma}{\Gamma \vdash \Lambda\alpha.\, M : \forall\alpha.\, A}\ (\vdash\textit{ec-abs}) \qquad \frac{\Gamma \vdash M : \forall\alpha.\, A \qquad \Gamma \vdash e}{\Gamma \vdash M[e] : A[\alpha := e]}\ (\vdash\textit{ec-app})$$

$$\frac{\Gamma \vdash e_1 \qquad \Gamma \vdash e_2 \qquad \Gamma, e_1{<:}e_2 \vdash M : A}{\Gamma \vdash e_1{<:}e_2{\Rightarrow}M : e_1{<:}e_2{\Rightarrow}A}\ (\vdash\textit{sub-intro}) \qquad \frac{\mathsf{Sub}(\Gamma) \vdash e_1 <: e_2 \qquad \Gamma \vdash M : e_1{<:}e_2{\Rightarrow}A}{\Gamma \vdash M\bullet : A}\ (\vdash\textit{sub-elim}) \qquad \frac{\Gamma \vdash e \qquad \Gamma; e \vdash M^\circ \Rightarrow T}{\Gamma \vdash ``M^\circ{''}e : ``T{''}e}\ (\vdash\textit{quote})$$

$$\frac{\Gamma \vdash M : A}{\Gamma \vdash \mathtt{ref}\ M : \mathtt{Ref}\ A}\ (\vdash\textit{ref}) \qquad \frac{\Gamma \vdash L : \mathtt{Ref}\ A \qquad \Gamma \vdash M : B \qquad \mathsf{Sub}(\Gamma) \vdash B <: A}{\Gamma \vdash L := M : \mathtt{Unit}}\ (\vdash\textit{assign}) \qquad \frac{\Gamma \vdash M : \mathtt{Ref}\ A}{\Gamma \vdash\ !M : A}\ (\vdash\textit{deref})$$

Fig. 2. Terms and bidirectional typing rules for $\lambda^\alpha_{\mathsf{Ref}}$, the static sister language

$e_1 <: e_2$ (highlighted). On the other hand, the rule for subtype constraint elimination ($\vdash$*sub-elim*) discharges the assumption by checking whether $e_1 <: e_2$ follows from the context $\Gamma$ (highlighted).

### 3.4 Operational Semantics of $\lambda^\alpha_{\mathsf{Ref}}$

Figure 3 defines the reduction rules for $\lambda^\alpha_{\mathsf{Ref}}$. The reduction rules for the meta language and code language are mutually recursive thanks to quote and splice. The reduction rules for quote and splice reduce any splices within the code, either by reducing a meta term in a splice ($\xi$-*splice*), or once that is finished, by canceling juxtaposed splices and quotes and integrating the resulting code (*splice*). The reduction rule for EC application ($\beta$-*ec*) is analogous to type application in System F, substituting the given EC for the EC variable. The reduction rule for subtype constraint elimination discards the subtype constraint ($\beta$-*sub*). The reduction rules for meta-level function application and mutable references are standard.

$$
\begin{array}{llcl}
\text{code values} & V^\circ & ::= & k \mid x \mid \lambda(x{:}T)^{\boldsymbol{\alpha}}.\, V^\circ \mid V^\circ\; V^\circ \\
\text{addresses} & a & \in & \mathbb{N} \\
\text{meta values} & V & ::= & k \mid \lambda x{:}A.\, M \mid a \mid \text{``}V^\circ\text{''}e \mid \Lambda\boldsymbol{\alpha}.\, M \mid e{<:}e{\Rightarrow}M \\
\text{code frames} & F^\circ & ::= & \lambda(x{:}T)^{\boldsymbol{\alpha}}.\, \square \mid \square\; M^\circ \mid V^\circ\; \square \\
\text{meta frames} & F & ::= & \square\; M \mid V\; \square \mid \mathtt{ref}\; \square \mid \square := M \mid V := \square \mid\; !\,\square \mid \square[e] \mid \square\bullet
\end{array}
$$

$$\boxed{\mu \mid M^\circ \longrightarrow^\circ \mu' \mid N^\circ}$$

$$
\frac{\mu \mid M^\circ \longrightarrow^\circ \mu' \mid N^\circ}{\mu \mid F^\circ[M^\circ] \longrightarrow^\circ \mu' \mid F^\circ[N^\circ]}\,(\xi\text{-code})
\qquad
\frac{\mu \mid M \longrightarrow^m \mu' \mid N}{\mu \mid {\sim}M \longrightarrow^\circ \mu' \mid {\sim}N}\,(\xi\text{-splice})
\qquad
\frac{}{\mu \mid {\sim}\text{``}V^\circ\text{''}e \longrightarrow^\circ \mu \mid V^\circ}\,(\text{splice})
$$

$$\boxed{\mu \mid M \longrightarrow^m \mu' \mid N}$$

$$
\frac{\mu \mid M \longrightarrow^m \mu' \mid N}{\mu \mid F[M] \longrightarrow^m \mu' \mid F[N]}\,(\xi\text{-meta})
\qquad
\frac{\mu \mid M^\circ \longrightarrow^\circ \mu' \mid N^\circ}{\mu \mid \text{``}M^\circ\text{''}e \longrightarrow^m \mu' \mid \text{``}N^\circ\text{''}e}\,(\xi\text{-quote})
$$

$$
\frac{}{\mu \mid (\lambda x{:}A.\, M)\; V \longrightarrow^m \mu \mid M[x := V]}\,(\beta)
\qquad
\frac{}{\mu \mid (\Lambda\alpha.\, M)[e] \longrightarrow^m \mu \mid M[\alpha := e]}\,(\beta\text{-ec})
$$

$$
\frac{}{\mu \mid (e_1{<:}e_2{\Rightarrow}M)\bullet \longrightarrow^m \mu \mid M}\,(\beta\text{-sub})
\qquad
\frac{a \notin \mathsf{dom}(\mu)}{\mu \mid \mathtt{ref}\; V \longrightarrow^m \mu[a \mapsto V] \mid a}\,(\text{ref})
$$

$$
\frac{\mu(a) = V}{\mu \mid\; !\,a \longrightarrow^m \mu \mid V}\,(\text{deref})
\qquad
\frac{}{\mu \mid a := V \longrightarrow^m \mu[a \mapsto V] \mid ()}\,(\text{assign})
$$

Fig. 3. Reduction semantics for $\lambda^{\alpha}_{\text{Ref}}$, the static sister language

## 4 The Gradual Metalanguage $\lambda^{\alpha,\star}_{\text{Ref}}$

The language $\lambda^{\alpha,\star}_{\text{Ref}}$ is a gradually typed metalanguage on top of a statically typed code language (the simply typed lambda calculus). The types of $\lambda^{\alpha,\star}_{\text{Ref}}$ (Figure 4) are obtained by extending the types of the sister language $\lambda^{\alpha}_{\text{Ref}}$ with (1) the statically unknown environment classifier $\star$ to form gradual classifiers $\hat{e}$, (2) the code type $\star$ to form gradual code types $\hat{T}$, and (3) the meta type $\star$ to form gradual meta types $\hat{A}$. Even though the code language is fully static, the gradual code type $\hat{T}$ includes $\star$ because the meta language is gradual and $\hat{T}$ is a metalanguage type *about* the code. As in $\lambda^{\alpha}_{\text{Ref}}$, $\Theta$ is the subtype context and $\text{Sub}(\Gamma)$ is the subsequence of subtype constraints in $\Gamma$.

$\lambda^{\alpha,\star}_{\text{Ref}}$ includes both gradual typing and subtyping, so its type system relies on a relation called *consistent subtyping*, written $\Theta \vdash \hat{A} \lesssim \hat{B}$, that composes subtyping with consistency. (Recall that consistency is the relation that enables implicit conversions to and from type $\star$ [Siek and Taha 2006, 2007].) Consistent subtyping is used in $\lambda^{\alpha,\star}_{\text{Ref}}$ to specify which implicit conversions are allowed during type checking and therefore must be checked at runtime. We show selected consistent subtyping rules in Figure 4 (full version in Figure 10 of the Appendix).

We present the type system for $\lambda^{\alpha,\star}_{\text{Ref}}$ in Section 4.1 and the operational semantics in Section 4.2.

### 4.1 Type System for $\lambda^{\alpha,\star}_{\text{Ref}}$

The syntax and selected typing rules of $\lambda^{\alpha,\star}_{\text{Ref}}$ are also shown in Figure 4. There are two major changes compared with the static sister language $\lambda^{\alpha}_{\text{Ref}}$: (1) gradual types and (2) the addition of blame labels. The parameter of a meta language $\lambda$ is annotated with a gradual meta type (so it may contain $\star$). The EC on a quote remains a classifier and not a gradual classifier, and the same for the classifier in EC application, which is required to ensure that runtime values are fully precise [Chen and Siek 2024]. Elimination forms such as splice, function application, EC application, subtype

$$
\begin{array}{llcl}
\text{gradual ECs} & \hat{e} & ::= & \star \mid e \\
\text{gradual code types} & \hat{T} & ::= & \star \mid \iota \mid \hat{T} \to \hat{T} \\
\text{gradual meta types} & \hat{A}, \hat{B} & ::= & \star \mid \iota \mid \hat{A} \to \hat{B} \mid \mathsf{Ref}\ \hat{A} \mid ``\hat{T}''\hat{e} \mid \forall \alpha.\, \hat{A} \mid e{<:}e \Rightarrow \hat{A} \\
\text{gradual contexts} & \Gamma & ::= & \emptyset \mid \Gamma, \alpha \mid \Gamma, e_1{<:}e_2 \mid \Gamma, x : \hat{A} \mid \Gamma, (x : T)^{\alpha} \\
\text{code terms} & L^{\circ}, M^{\circ}, N^{\circ} & ::= & k \mid x \mid \lambda(x{:}T)^{\alpha}.\, M^{\circ} \mid (L^{\circ}\ M^{\circ})^{\ell} \mid \sim^{\ell} M \\
\text{meta terms} & L, M, N & ::= & k \mid x \mid \lambda x{:}\hat{A}.\, M \mid (L\ M)^{\ell} \mid \Lambda \alpha.\, M \mid M[e]^{\ell} \mid e_1{<:}e_2 \Rightarrow M \mid M\bullet^{\ell} \\
 & & \mid & ``M^{\circ}''e \mid \mathsf{ref}\ M \mid L :=^{\ell} M \mid !^{\ell} M
\end{array}
$$

$$\boxed{\Theta \vdash \hat{A} \lesssim \hat{B}}$$

$$
\frac{\Theta \vdash \hat{T}_1 \lesssim \hat{T}_2 \qquad \Theta \vdash \hat{e}_1 \lesssim \hat{e}_2}{\Theta \vdash ``\hat{T}_1''\hat{e}_1 \lesssim ``\hat{T}_2''\hat{e}_2}
\qquad
\frac{\Theta, e_1{<:}e_2 \vdash \hat{A} \lesssim \hat{B}}{\Theta \vdash (e_1{<:}e_2 \Rightarrow \hat{A}) \lesssim (e_1{<:}e_2 \Rightarrow \hat{B})}
\qquad \cdots
$$

$$\boxed{\Gamma; e \vdash M^{\circ} \Rightarrow T}$$

$$\cdots \quad (\text{same as those of } \lambda^{\alpha}_{\mathsf{Ref}})$$

$$\boxed{\Gamma; e \vdash M^{\circ} \Leftarrow T}$$

$$
\frac{\Gamma \vdash M : \hat{A} \qquad \mathsf{Sub}(\Gamma) \vdash \hat{A} \lesssim ``T''e}{\Gamma; e \vdash \sim^{\ell} M \Leftarrow T}\ (\vdash\textit{splice})
\qquad \cdots
$$

$$\boxed{\Gamma \vdash M : \hat{A}}$$

$$
\frac{\Gamma \vdash e \qquad \Gamma \vdash M : \star}{\Gamma \vdash M[e]^{\ell} : \star}\ (\vdash\textit{ec-app-}\star)
\qquad
\frac{\Gamma \vdash M : \star}{\Gamma \vdash M\bullet^{\ell} : \star}\ (\vdash\textit{sub-elim-}\star)
\qquad \cdots
$$

Fig. 4. Gradual types, syntax, and selected consistent-subtyping and typing rules for $\lambda^{\alpha,\star}_{\mathsf{Ref}}$ (excerpt of Figures 10 and 11 of the Appendix)

constraint elimination, reference assignment, and dereferencing are annotated with blame labels ($\ell$). Blame labels are automatically added by the parser and they point to locations in the source code of surface programs (so programmers do not need to add them by hand).

The typing rules of $\lambda^{\alpha,\star}_{\mathsf{Ref}}$ were systematically derived from those of $\lambda^{\alpha}_{\mathsf{Ref}}$ by the following steps. The full typing rules are given in Figure 11 of the Appendix.

(1) Replace static types $A, B$ with gradual types $\hat{A}, \hat{B}$.
(2) Replace uses of subtyping with consistent subtyping, as illustrated by the highlighted premise of the splice rule in Figure 4.
(3) For every elimination form, add a second typing rule that handles the case where the term being eliminated has type $\star$, as illustrated by the EC-application and subtype-constraint elimination rules in Figure 4.

### 4.2 Operational Semantics for $\lambda^{\alpha,\star}_{\mathsf{Ref}}$

The operational semantics of $\lambda^{\alpha,\star}_{\mathsf{Ref}}$ is defined by translation to an intermediate language, the cast calculus $\mathsf{CC}^{\alpha,\star}_{\mathsf{Ref}}$ defined in Section 5. We define the translation, a type-directed elaboration, in Section 6, and the operational semantics of $\mathsf{CC}^{\alpha,\star}_{\mathsf{Ref}}$ in Section 5.5.

## 5 Cast Calculus $\mathsf{CC}^{\alpha,\star}_{\mathsf{Ref}}$

A cast calculus is an intermediate language that makes explicit every conversion between types. It includes a term $M\langle c\rangle$ that applies cast $c$ to the value of $M$. The cast $c$ has type $\hat{A} \Rightarrow \hat{B}$ and captures the runtime conversion from $\hat{A}$ to $\hat{B}$; as a result, $M\langle c\rangle$ is typed at $\hat{B}$ if $M$ has type $\hat{A}$. A metaprogramming calculus keeps track of code and code types, so we first talk about casts between code types in Section 5.1. $\lambda^{\alpha,\star}_{\mathsf{Ref}}$ incorporates subtype bounded classifier polymorphism, so we discuss runtime checking of subtype constraints in Section 5.2. We then instantiate our cast representation in Section 5.3 by defining three coercion calculi. Finally, we are ready to formalize the cast calculus $\mathrm{CC}^{\alpha,\star}_{\mathrm{Ref}}$ itself, for which we define a type system in Section 5.4 and reduction semantics in Section 5.5.

## 5.1 Casting the Code Type

When the source and target types of a cast has the same head type constructor, there are two standard approaches to applying the cast to a value [Siek and Chen 2021]. (There are several ways to represent a cast $c$. We postpone that choice and proceed abstractly for a little way before making the cast representation concrete.) For example, consider function values and casts between function types. Suppose $V$ is a value of function type $A \to B$ and $c$ casts from $A \to B$ to $A' \to B'$. The *active* approach is to wrap $V$ inside another $\lambda$ that checks the argument and return value, using the domain ($A' \Rightarrow A$) and codomain ($B \Rightarrow B'$) parts [3] of the cast $c$.

$$V\langle c\rangle \longrightarrow \lambda x.\,(V\ (x\langle \mathsf{dom}(c)\rangle))\langle \mathsf{cod}(c)\rangle$$

The active formulation is essentially an $\eta$-expansion with casts attached.

Alternatively, one can treat casts between two function types as *inert* and declare that $V\langle c\rangle$ is a value. But then one needs to add the following reduction rule for function application.

$$V\langle c\rangle\ W \longrightarrow (V\ (W\langle \mathsf{dom}(c)\rangle))\langle \mathsf{cod}(c)\rangle$$

For functions, the active and inert alternatives yield the same observable behavior, so the choice does not matter.

In contrast, for the code type, the choice between active and inert does matter for the semantics of the meta language. Yaguchi and Kameyama [2025] propose to treat casts between code types as active, pushing the cast under the quote, as in the following reduction rule. Suppose $V$ has type $``T_1''$, $c$ is a cast to the type $``T_2''$, and $\mathsf{under}(c)$ is a cast from $T_1$ to $T_2$.

$$V\ \langle c\rangle \longrightarrow ``(\sim V)\langle \mathsf{under}(c)\rangle''$$

In our setting, the code language is statically typed and does not include casts, so this reduction rule would not make sense syntactically. Even if we changed the code language to include casts, this reduction rule would still be problematic because it would cause the checking that should have happened during meta evaluation to be *delayed* until the evaluation of the generated code.

Instead, we treat casts between code types as inert. The next step is to add a reduction rule to deal with splice, the elimination form of code types. The following is an attempt at such a rule, but it has the same problem as the above. It delays the checking of $\mathsf{under}(c)$ to the runtime of the generated code. (Moving the cast outside the splice puts it in code.)

$$\sim(V\langle c\rangle) \longrightarrow (\sim V)\langle \mathsf{under}(c)\rangle$$

Taking a step back, recall that our code language is fully static, so the type expected inside the splice is fully static, and the quoted code inside the value $V$ is also fully static. Furthermore, we know that a sequence of casts whose source and target are fully static must either produce an error, because two or more of the casts involve inconsistent types, or the sequence of casts acts like the identity function. So all we need is a way to reduce a sequence of casts to one of these outcomes. Thankfully, prior research into the space-efficient implementation of gradual typing has already developed machinery for reducing sequences of casts, using the Coercion Calculus of Henglein [1994], which is a combinator language for expressing casts that involve the unknown type.

Here we describe the intuition of the Coercion Calculus. We will formally define coercions and the reduction semantics later in Section 5.3. First, values are allowed to carry at most one coercion. Let $U$ range over uncoerced values (i.e., not a value of the form $V\langle c\rangle$). The Coercion Calculus comes with a reduction system, which takes the form $c \longrightarrow d$. When a coercion is applied to a value that already has another coercion on it, the two coercions are composed $(c; d)$, forming a single coercion. So far we have just pushed the problem into the coercion representation. If a coercion reduces to the identity coercion (id), it can be thrown away. On the other hand, if a coercion reduces to the blame coercion ($\bot^{\ell}$), signaling a cast failure, then the program halts by raising the runtime error blame $\ell$. We summarize the four reduction rules that we discussed below.

[3]The cast representation must provide the operations dom and cod.

$$U\langle c\rangle \longrightarrow U\langle d\rangle \text{ if } c \longrightarrow d \qquad U\langle c\rangle\langle d\rangle \longrightarrow U\langle c; d\rangle \qquad U\langle \mathtt{id}\rangle \longrightarrow U \qquad U\langle \bot^{\ell}\rangle \longrightarrow \mathtt{blame}\ \ell$$

However, there is one more wrinkle to deal with: we must handle classifier subtyping. We could choose to handle subtyping implicitly, with a subsumption rule similar to the cast calculus of Siek and Taha [2007], or we could handle classifier subtyping explicitly, with an explicit coercion, analogous to the treatment of subtyping between security labels by Chen and Siek [2024]. We choose to take the latter approach, with explicit coercions that handle subtyping. With that decision, an EC coercion can reduce to a singleton subtype coercion ($e_1 \uparrow e_2$). So we need the following rule, which throws away an EC subtype coercion between a quote and splice (highlighted).

$$\sim(\text{``}V^{\circ}\text{''}e_1\ \langle \text{``}\mathtt{id}\text{''}\ (e_1 \uparrow e_2)\rangle) \longrightarrow V^{\circ}$$

## 5.2 Runtime Checking of Subtype Constraints

For the most part, the elaboration from $\lambda^{\alpha,\star}_{\text{Ref}}$ to $\text{CC}^{\alpha,\star}_{\text{Ref}}$ requires the insertion of explicit casts. However, sometimes the necessary runtime checking cannot be expressed with a cast. Consider the typing rule of $\lambda^{\alpha,\star}_{\text{Ref}}$ (on the left) and how we might create a corresponding elaboration rule. Also keep in mind the typing rule for $M \bullet$ from $\lambda^{\alpha}_{\text{Ref}}$ (on the right), because the rule in $\text{CC}^{\alpha,\star}_{\text{Ref}}$ will be similar.

$$\frac{\Gamma \vdash M : \star}{\Gamma \vdash M\bullet^{\ell} : \star}\ (\vdash\textit{sub-elim-}\star) \qquad\qquad \frac{\Gamma \vdash M : e_1 {<:} e_2 {\Rightarrow} \hat{A} \qquad \mathsf{Sub}(\Gamma) \vdash e_1 <: e_2}{\Gamma \vdash M\bullet : \hat{A}}$$

Here is a first attempt at the elaboration rule.

$$\frac{\Gamma \vdash M : \star \rightsquigarrow M'}{\Gamma \vdash M\bullet^{\ell} : \star \rightsquigarrow (M'\langle c\rangle)\bullet} \qquad \text{(strawman)}$$

The problem is that we need to generate a cast $c$ whose target type is of the form $e_1 {<:} e_2 {\Rightarrow} \hat{A}$, but we do not know what the subtype constraint will be on the value that $M$ reduces to. So instead of using a cast, we need a way to say, "Please dynamically check the subtype constraint on the value that $M$ reduces to." For this purpose we extend $\text{CC}^{\alpha,\star}_{\text{Ref}}$ with a new term of the form $M\bullet\star^{\ell}$. Unlike the rule for $M\bullet$, the typing rule for $M\bullet\star^{\ell}$ does not check a subtype constraint. Instead, the subtype constraint on the value is checked at runtime. Here is the finished elaboration rule.

$$\frac{\Gamma \vdash M : \star \rightsquigarrow M'}{\Gamma \vdash M\bullet^{\ell} : \star \rightsquigarrow M'\bullet\star^{\ell}}$$

## 5.3 Coercions

We now make concrete the cast representation. We use the framework of the Coercion Calculus [Henglein 1994], which provides solutions for much of the design, only requiring a few key additions. While not originally designed for the purpose of gradual typing, the Coercion Calculus is now a standard tool for developing gradually typed languages [Chen and Siek 2024; Igarashi et al. 2024; Siek and Chen 2021; Siek et al. 2009, 2015].

One minor wrinkle in the Coercion Calculus is that composition is defined as a binary operator that is meant to be associative, so the traditional reduction semantics depends on a notion of equivalence up to associativity and the semantics allows uses of this equivalence between every reduction step [Henglein 1994; Siek et al. 2009], which complicates the metatheory considerably. Instead, we

$$
\begin{array}{llcl}
\text{subtype contexts} & \Theta & ::= & \emptyset \mid \Theta, e_1 <: e_2 \\
\text{single EC coercions} & c^e, d^e & ::= & e! \mid e?^{\ell} \mid \hat{e}_1 \uparrow \hat{e}_2 \mid \bot^{\ell} \\
\text{EC coercion sequences} & \bar{c}^e, \bar{d}^e & ::= & [\,] \mid c^e :: \bar{c}^e \\
\text{single normal forms} & c^e_{\mathrm{NF}}, d^e_{\mathrm{NF}} & ::= & e! \mid e?^{\ell} \mid \bot^{\ell} \mid \hat{e}_1 \uparrow \hat{e}_2 \text{ where } \hat{e}_1 \neq \hat{e}_2 \\
\text{normal forms} & \bar{c}^e_{\mathrm{NF}}, \bar{d}^e_{\mathrm{NF}} & ::= & [\,] \mid [c^e_{\mathrm{NF}}] \mid c^e_{\mathrm{NF}} :: d^e_{\mathrm{NF}} :: \bar{d}^e_{\mathrm{NF}} \text{ where } \mathsf{IrredPair}(c^e_{\mathrm{NF}}, d^e_{\mathrm{NF}})
\end{array}
$$

$$\mathsf{Blame}(c^e) \triangleq \exists \ell.\ c^e = \bot^{\ell}$$

$$\boxed{\Theta \vdash \hat{e}_1 \mathrel{\hat{<:}} \hat{e}_2}$$

$$
\frac{}{\Theta \vdash \star \mathrel{\hat{<:}} \star}\ \textit{(ec-}\star\textit{)}
\qquad
\frac{\Theta \vdash e_1 <: e_2}{\Theta \vdash e_1 \mathrel{\hat{<:}} e_2}\ \textit{(ec-static)}
$$

$$\boxed{\Gamma \vdash c^e : \hat{e}_1 \Rightarrow \hat{e}_2}$$

$$
\frac{\Gamma \vdash e}{\Gamma \vdash e! : e \Rightarrow \star}\ \textit{(}\vdash\textit{ec-inj)}
\qquad
\frac{\Gamma \vdash e}{\Gamma \vdash e?^{\ell} : \star \Rightarrow e}\ \textit{(}\vdash\textit{ec-proj)}
$$

$$
\frac{\Gamma \vdash \hat{e}_1 \qquad \Gamma \vdash \hat{e}_2 \qquad \mathsf{Sub}(\Gamma) \vdash \hat{e}_1 \mathrel{\hat{<:}} \hat{e}_2}{\Gamma \vdash \hat{e}_1 \uparrow \hat{e}_2 : \hat{e}_1 \Rightarrow \hat{e}_2}\ \textit{(}\vdash\textit{ec-sub)}
\qquad
\frac{\Gamma \vdash \hat{e}_1 \qquad \Gamma \vdash \hat{e}_2}{\Gamma \vdash \bot^{\ell} : \hat{e}_1 \Rightarrow \hat{e}_2}\ \textit{(}\vdash\textit{ec-blame)}
$$

$$\boxed{\Gamma \vdash \bar{c}^e : \hat{e}_1 \Rightarrow \hat{e}_2}$$

$$
\frac{\Gamma \vdash \hat{e}}{\Gamma \vdash [\,] : \hat{e} \Rightarrow \hat{e}}\ \textit{(}\vdash\textit{ec-seq-empty)}
\qquad
\frac{\Gamma \vdash c^e : \hat{e}_1 \Rightarrow \hat{e}_2 \qquad \Gamma \vdash \bar{c}^e : \hat{e}_2 \Rightarrow \hat{e}_3}{\Gamma \vdash c^e :: \bar{c}^e : \hat{e}_1 \Rightarrow \hat{e}_3}\ \textit{(}\vdash\textit{ec-seq-cons)}
$$

$$\boxed{\mathsf{IrredPair}(c^e, d^e)}$$

$$
\frac{}{\mathsf{IrredPair}(e?^{\ell}, e!)}
\qquad
\frac{e' \neq e}{\mathsf{IrredPair}(e' \uparrow e, e!)}
\qquad
\frac{}{\mathsf{IrredPair}(e!, \star \uparrow \star)}
\qquad
\frac{}{\mathsf{IrredPair}(e?^{\ell}, e \uparrow e')}
$$

$$\boxed{\Theta \vdash c^e \longrightarrow \bar{d}^e}$$

$$
\frac{}{\Theta \vdash \hat{e} \uparrow \hat{e} \longrightarrow [\,]}\ \textit{(}\uparrow\textit{-id)}
$$

$$\boxed{\Theta \vdash c^e_1 ; c^e_2 \longrightarrow \bar{d}^e}$$

$$
\frac{\Theta \vdash e_1 <: e_2}{\Theta \vdash e_1!; e_2?^{\ell} \longrightarrow [e_1 \uparrow e_2]}\ \textit{(collapse)}
\qquad
\frac{\Theta \nvdash e_1 <: e_2}{\Theta \vdash e_1!; e_2?^{\ell} \longrightarrow [\bot^{\ell}]}\ \textit{(collide)}
$$

$$
\frac{}{\Theta \vdash (\hat{e}_1 \uparrow \hat{e}_2); (\hat{e}_2 \uparrow \hat{e}_3) \longrightarrow [\hat{e}_1 \uparrow \hat{e}_3]}\ \textit{(}\uparrow\textit{-collapse)}
\qquad
\frac{}{\Theta \vdash \bot^{\ell}; c^e \longrightarrow [\bot^{\ell}]}\ \textit{(}\bot\textit{-left)}
\qquad
\frac{\neg \mathsf{Blame}(c^e)}{\Theta \vdash c^e; \bot^{\ell} \longrightarrow [\bot^{\ell}]}\ \textit{(}\bot\textit{-right)}
$$

$$\boxed{\Theta \vdash \bar{c}^e \longrightarrow \bar{d}^e}$$

$$
\frac{\Theta \vdash c^e \longrightarrow \bar{d}^e}{\Theta \vdash c^e :: \bar{c}^e \longrightarrow \bar{d}^e +\!\!\!+\, \bar{c}^e}\ \textit{(reduce-head)}
\qquad
\frac{\Theta \vdash c^e_{\mathrm{NF}}; c^e \longrightarrow \bar{d}^e}{\Theta \vdash c^e_{\mathrm{NF}} :: c^e :: \bar{c}^e \longrightarrow \bar{d}^e +\!\!\!+\, \bar{c}^e}\ \textit{(reduce-neighbors)}
$$

$$
\frac{\mathsf{IrredPair}(c^e_{\mathrm{NF}}, d^e_{\mathrm{NF}}) \qquad \Theta \vdash d^e_{\mathrm{NF}} :: \bar{c}^e \longrightarrow \bar{d}^e}{\Theta \vdash c^e_{\mathrm{NF}} :: d^e_{\mathrm{NF}} :: \bar{c}^e \longrightarrow c^e_{\mathrm{NF}} :: \bar{d}^e}\ \textit{(reduce-tail)}
$$

Fig. 5. EC coercion sequences: syntax, typing, and reduction

choose a canonical representation so that equivalent coercions are syntactically equal [Chen and Siek 2024]. We represent a coercion as a sequence (i.e. list) of "single" coercions, where the identity coercion is represented by the empty list [] and coercion composition is list concatenation ($+\!\!\!+$).

We use this canonical representation of coercions in three new coercion calculi for: (1) coercions between gradual classifiers, (2) coercions between gradual code types, and (3) coercions between gradual meta types.

**Coercions Between Classifiers.** The grammar and typing of the EC coercion calculus are in Figure 5. We use $c^e, d^e$ for single EC coercions and $\bar{c}^e, \bar{d}^e$ for sequences of single EC coercions. Subscript NF marks normal forms. Syntactically, an EC coercion sequence is just a list where cons $(c^e :: \bar{c}^e)$ adds a single coercion to the front of a sequence. We use $[c^e]$ as a shorthand for a singleton coercion sequence. The typing judgment for a single EC coercion takes the form $\Gamma \vdash c^e : \hat{e}_1 \Rightarrow \hat{e}_2$. An injection single coercion $e!$ injects EC $e$ into $\star$. A projection $e?^{\ell}$ converts to EC $e$ out of $\star$. Note that projection carries a blame label ($\ell$), which is reported if the projection fails. The $\bot^{\ell}$ coercion halts the program and assigns blame, so it goes from one well-formed gradual EC to any other. The coercion $\hat{e}_1 \uparrow \hat{e}_2$ converts between two gradual ECs in the subtype relation. As is usual in the literature of gradual typing, this subtyping relation is neutral with respect to $\star$. The typing for an EC coercion sequence has the form $\Gamma \vdash \bar{c}^e : \hat{e}_1 \Rightarrow \hat{e}_2$ and is defined straightforwardly.

The reduction semantics of EC coercions is also defined in Figure 5. The reduction relation of EC coercion sequences takes the form $\Theta \vdash \bar{c}^e \longrightarrow \bar{d}^e$, where $\Theta$ is for justifying subtyping at runtime. The one reduction rule for a single EC coercion ($\uparrow$-*id*) reduces an identity subtype coercion into the empty sequence. There are five rules that reduce neighboring single coercions: *collapse* reduces a pair of injection and projection to a subtype coercion if $\Theta$ justifies the subtyping between the source and the target; *collide* blames the projection if the subtyping check fails; $\uparrow$-*collapse* compresses two subtype coercions with the same middle EC (subtyping is transitive); $\bot$-*left* and $\bot$-*right* say that a blame coercion always annihilates its neighbor. There are three reduction rules for coercion sequences: *reduce-head*, *reduce-neighbors*, and *reduce-tail*. The *reduce-head* rule reduces the head using single coercion reduction ($\Theta \vdash c^e \longrightarrow \bar{d}^e$). The *reduce-neighbors* rule reduces the pair of neighboring single coercions using the neighbors reduction relation ($\Theta \vdash c^e_1; c^e_2 \longrightarrow \bar{d}^e$). Finally, the *reduce-tail* rule reduces a coercion sequence by recursively reducing the tail.

**Coercions on Code Types.** Next we define coercions that convert from one gradual code type to another. The coercions on code types are a standard calculus, the Eager UD Coercion Calculus of Siek et al. [2009], which treats errors in an *eager* fashion, immediately propagating failing coercions (rules $\xi$-$\bot$-*fun-dom* and $\xi$-$\bot$-*fun-cod*). This eagerness is necessary because we need to reduce a coercion that involves code types to either an error or a subtype cast, as we discussed in Section 5.1. The syntax, typing rules, and reduction rules for code coercion sequences are in Figure 12 of the Appendix. The typing takes the form $\vdash \bar{c}^\circ : \hat{T}_1 \Rightarrow \hat{T}_2$ and the reduction relation is $\bar{c}^\circ \longrightarrow \bar{d}^\circ$.

**Coercions on Meta Types.** The third and final piece is coercions that convert from one meta type to another. The syntax, typing rules, and normal forms for meta coercions are defined in Figure 13 and their reduction rules are in Figure 14, both in the Appendix. The typing takes the form $\Gamma \vdash \bar{c} : \hat{A} \Rightarrow \hat{B}$ and the reduction relation is $\Theta \vdash \bar{c} \longrightarrow \bar{d}$. The meta coercions are a lazy coercion calculus (there is no reason to make them eager) and we introduce one new coercion for each of the three new type constructors (compared to prior work on coercions): coercions for code ($``\,\bar{c}^\circ\,''\,\bar{c}^e$), EC polymorphism ($\forall\alpha.\bar{c}$), and subtype constraints ($e{<:}e \Rightarrow \bar{c}$). The $\xi$-*code* and $\xi$-*code-ec* reduction rules dispatch to the reduction of code coercion sequences and EC coercion sequences that we just defined, respectively.

## 5.4 Type System for $\mathrm{CC}^{\alpha,\star}_{\mathbf{Ref}}$

The cast calculus $\mathrm{CC}^{\alpha,\star}_{\mathrm{Ref}}$ is an intermediate language for $\lambda^{\alpha,\star}_{\mathrm{Ref}}$ in which type conversions are explicit, so there is no use of subtype consistency as in $\lambda^{\alpha,\star}_{\mathrm{Ref}}$. Instead, $\mathrm{CC}^{\alpha,\star}_{\mathrm{Ref}}$ has a cast term $M\langle\bar{c}\rangle$ where the coercion sequence $\bar{c}$ records the type conversion being applied to $M$. The syntax of $\mathrm{CC}^{\alpha,\star}_{\mathrm{Ref}}$ is in Figure 6. The cast calculus also contains forms that arise at runtime: memory addresses ($a$), blame, subtype constraint checks ($M{\bullet}\star^{\ell}$), and code lambda values ($\bar{\lambda}(x{:}T)^{\alpha}.\,M^\circ$). The later represents a runtime-allocated abstract syntax tree. A code lambda value does not bind a static environment classifier but instead each code lambda value is associated with a unique runtime EC that can be used (like a pointer) to refer to its scope.

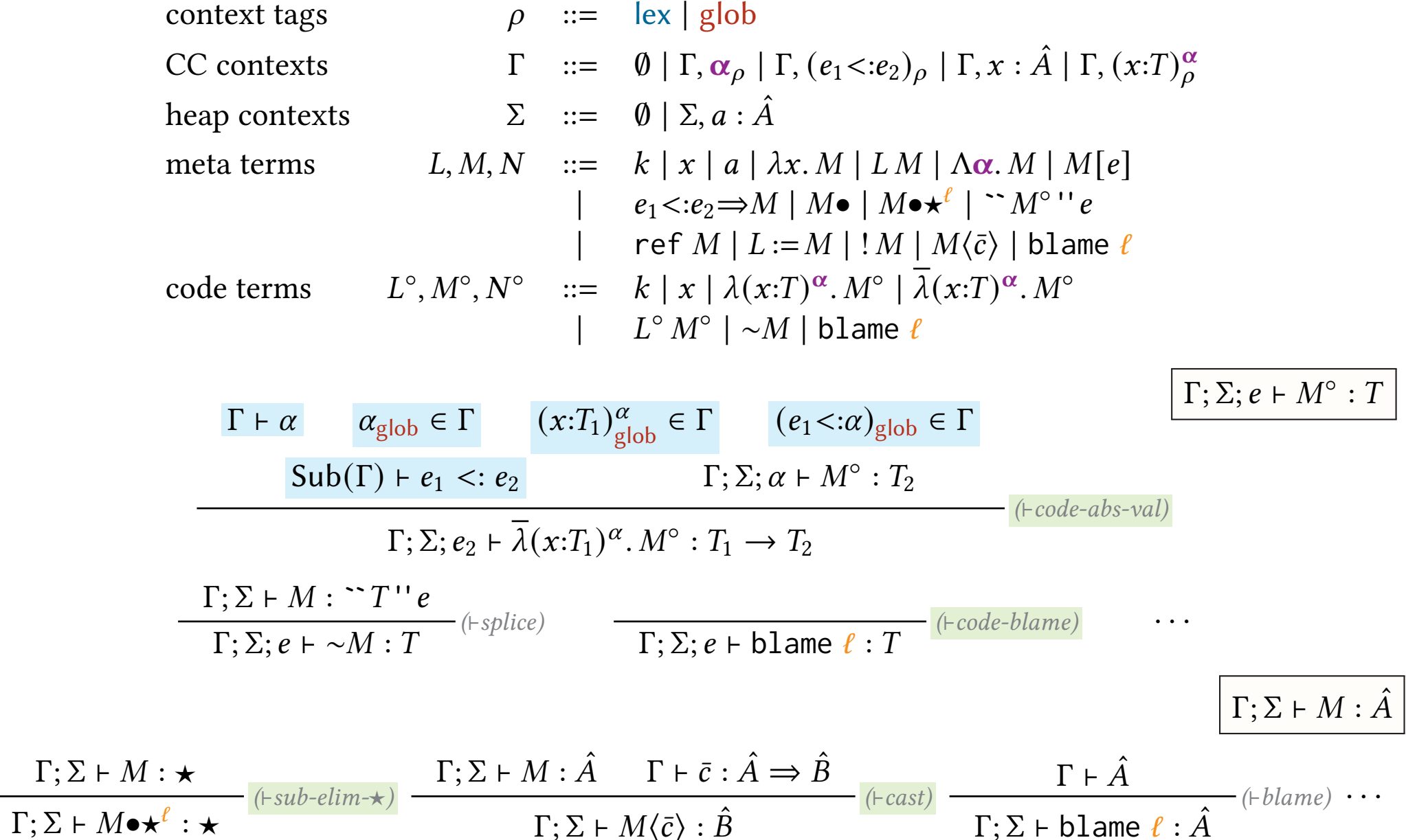

Fig. 6. Syntax and selected typing rules for cast calculus $CC^{\alpha,\star}_{Ref}$ (excerpt of Figure 16 of the Appendix)

The typing judgments for $CC^{\alpha,\star}_{Ref}$ take the forms $\Gamma; \Sigma \vdash M : \hat{A}$ for meta terms and $\Gamma; \Sigma; e \vdash M^{\circ} : T$ for code terms. The typing context $\Gamma$ contains EC variables, subtype constraints, meta variables, and code variables. The EC, subtype, and code-variable entries in $\Gamma$ are tagged as either lexical (lex) or global (glob). We use $\mathrm{Sub}(\Gamma)$ to denote the ordered subsequence of subtype constraints in $\Gamma$, ignoring their lexical/global tags. The heap typing context ($\Sigma$) maps addresses to types. Similar to the type system of $\lambda^{\alpha,\star}_{Ref}$, the classifier $e$ in the code judgment communicates the enclosing scope. Interesting typing rules are shown in Figure 6, and the complete rules are in Figure 16 of the Appendix. Well-formed typing contexts are defined in Figure 15 of the Appendix.

**Type System of the Code Language.** Many of the typing rules, including those for constants, variables, code lambdas, and function applications are similar to those of $\lambda^{\alpha}_{Ref}$ and $\lambda^{\alpha,\star}_{Ref}$. The main difference is the absence of both subtyping and subtype consistency. For example, in the splice rule the EC recorded on the code type must be equal to that of the enclosing scope. However, there are two new rules for terms that appear at runtime, ⊢*blame* and ⊢*code-abs-val* (highlighted). Blame is trivially well-typed. Typing a code lambda value, on the other hand, is complicated because the rule must capture the invariant established when the scope of the code lambda becomes global. There are three parts to that invariant: (1) $\Gamma \vdash \alpha$ and $\alpha_{glob} \in \Gamma$ say that the EC of that scope is well-formed and globally allocated; (2) $(x : T_1)^{\alpha}_{glob} \in \Gamma$ says the code variable has also become global; (3) finally, the subtyping premises $(e_1 <: \alpha)_{glob} \in \Gamma$ and $\mathrm{Sub}(\Gamma) \vdash e_1 <: e_2$ say that the code lambda may have been spliced deeper into the AST, so the current enclosing scope ($e_2$) might be larger than the original parent scope ($e_1$).

**Type System of the Meta Language.** Again, many of the typing rules are similar to those of $\lambda^{\alpha,\star}_{Ref}$, except that consistent subtyping is replaced by type equality. $CC^{\alpha,\star}_{Ref}$ includes memory

$$
\begin{array}{llcl}
\text{heaps} & \mu & ::= & \emptyset \mid \mu[a \mapsto V] \\
\text{meta configurations} & C, \mathcal{D} & ::= & \langle \Gamma; M; \mu \rangle \\
\text{code configurations} & C^\circ, \mathcal{D}^\circ & ::= & \langle \Gamma; M^\circ; \mu \rangle \\
\text{code frames} & F^\circ & ::= & \square\, M^\circ \mid V^\circ\, \square \\
\text{meta frames} & F & ::= & \square\, M \mid V\, \square \mid \mathsf{ref}\, \square \mid \square := M \mid V := \square \mid !\square \mid \square[e] \\
 & & \mid & \square\, \bullet \mid \square\, \bullet\star^{\ell} \mid \square\langle \bar{c} \rangle \\
\text{code values} & V^\circ & ::= & k \mid x \mid \overline{\lambda}(x{:}T)^{\boldsymbol{\alpha}}.\, V^\circ \mid V^\circ\, V^\circ \\
\text{uncoerced meta values} & U & ::= & k \mid a \mid \lambda x.\, M \mid \Lambda \boldsymbol{\alpha}.\, M \mid ``V^\circ{''}e \mid e_1{<:}e_2 {\Rightarrow} M \\
\text{meta values} & V, W & ::= & U \mid U\langle \bar{c}_{\mathrm{NF}} \rangle \text{ where } \mathsf{Inert}(\bar{c}_{\mathrm{NF}})
\end{array}
$$

$$\mathsf{Inert}(\bar{c}_{\mathrm{NF}}) \triangleq \neg \mathsf{Id}(\bar{c}_{\mathrm{NF}}) \wedge \neg \mathsf{Blame}(\bar{c}_{\mathrm{NF}})$$

Fig. 7. Configurations, evaluation contexts (frames), and values for $\mathrm{CC}^{\alpha,\star}_{\mathrm{Ref}}$

addresses, whose typing is standard. The $\vdash$*cast* rule (highlighted) is for explicit conversion: if $M$ is typed at $\hat{A}$ and $\bar{c}$ is a well-typed meta coercion sequence from $\hat{A}$ to $\hat{B}$, then $M\langle \bar{c} \rangle$ has type $\hat{B}$. Third, there is a runtime subtype check (highlighted), where both the term being checked and the entire $M\bullet\star^{\ell}$ are typed at $\star$. Finally, blame is trivially well-typed.

### 5.5 Operational Semantics for $\mathrm{CC}^{\alpha,\star}_{\mathrm{Ref}}$

The reduction judgments for $\mathrm{CC}^{\alpha,\star}_{\mathrm{Ref}}$ take two forms: code reduction $e \vdash C^\circ \longrightarrow^\circ \mathcal{D}^\circ$ reduces a code runtime configuration ($C^\circ$) to another under the enclosing classifier $e$. Similarly, meta reduction $C \longrightarrow^m \mathcal{D}$ reduces a meta configuration ($C$) to another. Configurations are defined in Figure 7. A configuration consists of a term, a heap, and a runtime context $\Gamma$, which contains global context entries generated by reducing code lambdas. Well-typed configurations are defined in Figure 17 of the Appendix: the runtime context ($\Gamma$), the heap context ($\Sigma$), the heap ($\mu$), and the term must all be well-typed. In addition, the runtime context $\Gamma$ must not contain meta-variable entries. Non-recursive evaluation contexts (“frames”) and values are also defined in Figure 7. A value can be either uncoerced or wrapped in a value-forming (“inert”) coercion sequence.

To simplify presentation, we group the reduction rules that change neither the heap nor the runtime context as pure reduction. Pure code reduction $M^\circ \longrightarrow^\circ N^\circ$ changes only a code term. Pure meta reduction $\Theta \vdash M \longrightarrow^m N$ changes only a meta term, but may read from the subtype context $\Theta$. Figure 8 presents the reduction rules that we are going to discuss below. The complete reduction rules are defined in Figure 18 (for code) and Figure 19 (for meta) of the Appendix.

**Reduction of Code Terms.** We focus on the new reduction rules of $\mathrm{CC}^{\alpha,\star}_{\mathrm{Ref}}$ compared to $\lambda^{\alpha}_{\mathrm{Ref}}$. The *splice-sub* rule handles the case where a quoted code value is wrapped in an inert code coercion. By typing, the coercion must be a subtype coercion, so it can be directly eliminated before the quote and splice cancel. The *code-λ* rule performs global allocation for the scope of a code lambda. It first allocates a fresh runtime EC $\beta$, which uniquely identifies the scope, as well as a fresh code variable $y$. It then extends $\Gamma$ with three things: (1) the allocated EC $\beta_{\mathrm{glob}}$, (2) the edge that connects $\beta$ to its parent scope $(e{<:}\beta)_{\mathrm{glob}}$, and (3) the global code-variable entry $(y{:}T)^{\beta}_{\mathrm{glob}}$ that registers the code variable $y$ to $\beta$. (The extension is highlighted in blue.) Finally, the reduction turns the code lambda into a code lambda value and renames the lexically bound variables ($\alpha$ and $x$) to the globally bound ones ($\beta$ and $y$). After a code lambda is turned into a code lambda value, reduction can happen in its body. The *ξ-code-$\overline{\lambda}$* rule reduces the body of a code lambda value.

**Reduction of Meta Terms.** We focus on the rules for explicit casts. The *cast-reduce* rule reduces the underlying coercion in a cast term by delegating to the reduction semantics of meta coercion

$$\boxed{M^\circ \longrightarrow^\circ N^\circ}$$

$$\frac{}{\sim((\text{``}V^\circ\text{''}e_1)\langle\text{``}[\,]\text{''}\ [e_1{\uparrow}e_2]\rangle) \longrightarrow^\circ V^\circ}\ (\textit{splice-sub}) \qquad \cdots$$

$$\boxed{e \vdash \mathcal{C}^\circ \longrightarrow^\circ \mathcal{D}^\circ}$$

$$\frac{\beta, y \notin \Gamma}{e \vdash \langle \Gamma; \lambda(x{:}T)^\alpha.\, N^\circ; \mu\rangle \longrightarrow^\circ \langle(\Gamma,\ \beta_{\text{glob}}, (e{<:}\beta)_{\text{glob}}, (y{:}T)^\beta_{\text{glob}});\overline{\lambda}(y{:}T)^\beta.\, N^\circ[\alpha := \beta][x := y]; \mu\rangle}\ (\textit{code-}\lambda)$$

$$\frac{\alpha \vdash \langle\Gamma; N^\circ; \mu\rangle \longrightarrow^\circ \langle\Gamma'; N^{\circ\prime}; \mu'\rangle}{e \vdash \langle\Gamma; \overline{\lambda}(x{:}T)^\alpha.\, N^\circ; \mu\rangle \longrightarrow^\circ \langle\Gamma'; \overline{\lambda}(x{:}T)^\alpha.\, N^{\circ\prime}; \mu'\rangle}\ (\xi\textit{-code-}\overline{\lambda}) \qquad \frac{M^\circ \longrightarrow^\circ N^\circ}{e \vdash \langle\Gamma; M^\circ; \mu\rangle \longrightarrow^\circ \langle\Gamma; N^\circ; \mu\rangle}\ (\textit{pure-code})$$

$$\cdots$$

$$\boxed{\Theta \vdash M \longrightarrow^m N}$$

$$\frac{}{\Theta \vdash U\langle[\,]\rangle \longrightarrow^m U}\ (\textit{cast-id}) \qquad \frac{\Theta \vdash \bar{c} \longrightarrow \bar{d}}{\Theta \vdash U\langle\bar{c}\rangle \longrightarrow^m U\langle\bar{d}\rangle}\ (\textit{cast-reduce})$$

$$\frac{}{\Theta \vdash U\langle\bar{c}\rangle\langle\bar{d}\rangle \longrightarrow^m U\langle\bar{c} \mathbin{+\!\!+} \bar{d}\rangle}\ (\textit{cast-compose}) \qquad \frac{\mathsf{Inert}([\bar{c} \to \bar{d}])}{\Theta \vdash (U\langle[\bar{c} \to \bar{d}]\rangle)\ W \longrightarrow^m (U\ (W\langle\bar{c}\rangle))\langle\bar{d}\rangle}\ (\beta\textit{-cast})$$

$$\frac{}{\Theta \vdash (U\langle[\forall\alpha.\, \bar{c}]\rangle)[e] \longrightarrow^m U[e]\langle\bar{c}[\alpha := e]\rangle}\ (\beta\textit{-ec-cast})$$

$$\frac{\mathsf{Inert}([\mathtt{Ref}\ \bar{c}\ \bar{d}])}{\Theta \vdash !(a\langle[\mathtt{Ref}\ \bar{c}\ \bar{d}]\rangle) \longrightarrow^m (!a)\langle\bar{d}\rangle}\ (\textit{deref-cast}) \qquad \frac{\mathsf{Inert}([\mathtt{Ref}\ \bar{c}\ \bar{d}])}{\Theta \vdash (a\langle[\mathtt{Ref}\ \bar{c}\ \bar{d}]\rangle) := V \longrightarrow^m a := (V\langle\bar{c}\rangle)}\ (\textit{assign-cast})$$

$$\frac{}{\Theta \vdash (U\langle[e_1{<:}e_2{\Rightarrow}\bar{c}]\rangle)\bullet \longrightarrow^m (U\bullet)\langle\bar{c}\rangle}\ (\beta\textit{-sub-cast})$$

$$\frac{\Theta \vdash e_1{<:}e_2 \qquad \mathsf{Inert}(\bar{c} \mathbin{+\!\!+} [(e_1{<:}e_2{\Rightarrow}\star)!])}{\Theta \vdash U\langle\bar{c} \mathbin{+\!\!+} [(e_1{<:}e_2{\Rightarrow}\star)!]\rangle\bullet\star^\ell \longrightarrow^m (U\langle\bar{c}\rangle)\bullet}\ (\textit{sub-cast-success})$$

$$\frac{\Theta \nvdash e_1{<:}e_2 \qquad \mathsf{Inert}(\bar{c} \mathbin{+\!\!+} [(e_1{<:}e_2{\Rightarrow}\star)!])}{\Theta \vdash U\langle\bar{c} \mathbin{+\!\!+} [(e_1{<:}e_2{\Rightarrow}\star)!]\rangle\bullet\star^\ell \longrightarrow^m \mathtt{blame}\ \ell}\ (\textit{sub-cast-sub-fail})$$

$$\frac{\mathsf{Inert}(\bar{c} \mathbin{+\!\!+} [G!]) \qquad \forall e_1\, e_2.\ G \neq (e_1{<:}e_2{\Rightarrow}\star)}{\Theta \vdash U\langle\bar{c} \mathbin{+\!\!+} [G!]\rangle\bullet\star^\ell \longrightarrow^m \mathtt{blame}\ \ell}\ (\textit{sub-cast-fail}) \qquad \cdots$$

$$\boxed{\mathcal{C} \longrightarrow^m \mathcal{D}}$$

$$\frac{\mathsf{Sub}(\Gamma) \vdash M \longrightarrow^m N}{\langle\Gamma; M; \mu\rangle \longrightarrow^m \langle\Gamma; N; \mu\rangle}\ (\textit{pure-meta}) \qquad \cdots$$

Fig. 8. Selected reduction rules for $\mathsf{CC}^{\alpha,\star}_{\mathsf{Ref}}$ (excerpt of Figures 18 and 19 of the Appendix)

sequences. The *cast-compose* rule concatenates two adjacent coercion sequences. The *cast-id* rule removes an identity cast. The rules for $\bullet\star^\ell$ perform runtime checking of subtype constraints. The check succeeds when the context justifies the subtype constraint stored in the injection coercion ($\Theta \vdash e_1 <: e_2$, rule *sub-cast-success*). Otherwise, the runtime subtype check fails with blame (rule *sub-cast-sub-fail*). If the injection is not from a subtype-constrained type at all, the check also fails (rule *sub-cast-fail*). The remaining rules deal with values wrapped in inert casts. For example, rule *$\beta$-cast* distributes a function coercion into domain and codomain parts. The other rules are analogous, and their shapes are dictated by type preservation.

## 6 Elaboration

This section defines the elaboration from the gradually typed surface language $\lambda^{\alpha,\star}_{\text{Ref}}$ into the cast calculus $\text{CC}^{\alpha,\star}_{\text{Ref}}$. The full elaboration is defined in Figure 21 of the Appendix.

**Explicating Type Conversions.** Recall that consistent subtyping is for implicit type conversion in $\lambda^{\alpha,\star}_{\text{Ref}}$, and coercions are for explicit type conversion in $\text{CC}^{\alpha,\star}_{\text{Ref}}$. Consequently, our primary job is to turn each use of consistent subtyping $\Theta \vdash \hat{A} \lesssim \hat{B}$ in the typing rules of $\lambda^{\alpha,\star}_{\text{Ref}}$ into a coercion sequence $\bar{c} : \hat{A} \Rightarrow \hat{B}$. The main coerce function takes two gradual meta types ($\hat{A}, \hat{B}$) that satisfy consistent subtyping as well as a blame label ($\ell$) and returns a meta coercion sequence ($\bar{c}$). The $\text{coerce}^{e}$ and $\text{coerce}^{\circ}$ functions are analogous, but for gradual ECs and gradual code types, respectively. We present the definitions of these three coerce functions in Figure 20 of the Appendix.

**Elaborating $\lambda^{\alpha,\star}_{\text{Ref}}$ into $\text{CC}^{\alpha,\star}_{\text{Ref}}$.** The elaboration is type directed, and the elaboration rules mimic the typing rules of $\lambda^{\alpha,\star}_{\text{Ref}}$. We discuss a few interesting rules here.

Consider the typing rule for splice. The consistent subtyping premise is highlighted.

$$\frac{\Gamma \vdash M : \hat{A} \qquad \text{Sub}(\Gamma) \vdash \hat{A} \lesssim \text{``}T\text{''}\, e}{\Gamma; e \vdash \sim^{\ell} M \Leftarrow T}$$

We replace consistent subtyping with the coercion generated by calling $\text{coerce}(\hat{A}, \text{``}T\text{''}\, e, \ell)$.

$$\frac{\Gamma \vdash M : \hat{A} \rightsquigarrow M'}{\Gamma; e \vdash \sim^{\ell} M \Leftarrow T \rightsquigarrow (\sim M' \langle\, \text{coerce}(\hat{A}, \text{``}T\text{''}\, e, \ell)\,\rangle)}$$

The above elaboration rule also explains why we need bidirectional typing for code. To generate a coercion, we need to know both the source type and the target type. In the case of splice, the target type $\text{``}T\text{''}\, e$ is formed from the expected code type $T$ and the expected classifier $e$.

Next we discuss an elimination form where the subject is of type $\star$. Recall the $\vdash$*ec-app*-$\star$ rule:

$$\frac{\Gamma \vdash M : \star \qquad \Gamma \vdash e}{\Gamma \vdash M[e]^{\ell} : \star}$$

The elaboration needs to preserve types. We observe that there is only one $\text{CC}^{\alpha,\star}_{\text{Ref}}$ typing rule $\vdash$*ec-app* for EC application, and it requires the subject to have a polymorphic type, not $\star$. So, the elaboration generates a coercion from $\star$ to $\forall\beta.\star$.

$$\frac{\Gamma \vdash M : \star \rightsquigarrow M' \qquad \Gamma \vdash e}{\Gamma \vdash M[e]^{\ell} : \star \rightsquigarrow (M' \langle\, \text{coerce}(\star, (\forall\beta.\star), \ell)\,\rangle)[e]}$$

Sometimes elaboration is not as simple as inserting a coercion. In the case of eliminating a subtype constraint, we have to introduce $M' \bullet\star^{\ell}$ that checks the constraint at runtime.

$$\frac{\Gamma \vdash M : \star \rightsquigarrow M'}{\Gamma \vdash M \bullet^{\ell} : \star \rightsquigarrow (\,M' \bullet\star^{\ell}\,)}$$

## 7 Type and Scope Safety

This section states the type and scope safety results for $\lambda^{\alpha,\star}_{\text{Ref}}$ and $\lambda^{\alpha}_{\text{Ref}}$. The proof factors through the cast calculus $\text{CC}^{\alpha,\star}_{\text{Ref}}$ and proceeds in four steps. First, we prove type safety for $\text{CC}^{\alpha,\star}_{\text{Ref}}$ by progress and preservation. Second, we connect $\lambda^{\alpha,\star}_{\text{Ref}}$ to $\text{CC}^{\alpha,\star}_{\text{Ref}}$ by proving that elaboration preserves types. These two steps establish type safety for evaluating $\lambda^{\alpha,\star}_{\text{Ref}}$, but type safety alone does not yet state that generated code is lexically well scoped: during metaevaluation, code lexical binders are converted into global runtime allocations. Third, we define reify, which restores lexical binding for code variables to produce a regular STLC term. Finally, we prove scope safety for $\lambda^{\alpha,\star}_{\text{Ref}}$ as a corollary of type safety for $\mathrm{CC}^{\alpha,\star}_{\mathrm{Ref}}$, elaboration preservation, and reification preservation. We also prove scope safety for the static sister $\lambda^{\alpha}_{\mathrm{Ref}}$ as a corollary of scope safety for $\lambda^{\alpha,\star}_{\mathrm{Ref}}$.

**Type Safety for $\mathrm{CC}^{\alpha,\star}_{\mathrm{Ref}}$.** We state preservation over the runtime configurations of Section 5.5, where the configuration typing judgments $\Sigma \vdash^{C} C : \hat{A}$ and $\Sigma; e \vdash^{C} C^{\circ} : T$ are defined in Figure 17. We write $\mathsf{Ctx}(\langle \Gamma; M; \mu \rangle) = \Gamma$ for projecting the context out of a configuration. Preservation also records that reduction allocates fresh runtime entries; the inclusion relations $\Gamma_1 \subseteq \Gamma_2$ and $\Sigma_1 \subseteq \Sigma_2$ are standard and defined in Figure 22 of the Appendix. We state preservation for $\mathrm{CC}^{\alpha,\star}_{\mathrm{Ref}}$:

Lemma 7.1 (Preservation for $\mathrm{CC}^{\alpha,\star}_{\mathrm{Ref}}$). *The following preservation properties hold.*

*(I) If $\Sigma_1 \vdash^{C} C_1 : \hat{A}$ and $C_1 \longrightarrow^{m} C_2$, then there exists $\Sigma_2$ such that $\Sigma_2 \vdash^{C} C_2 : \hat{A}$ and $\mathsf{Ctx}(C_1) \subseteq \mathsf{Ctx}(C_2)$ and $\Sigma_1 \subseteq \Sigma_2$.*

*(II) If $\Sigma_1; e \vdash^{C} C_1^{\circ} : T$, $\mathsf{Ctx}(C_1^{\circ}) \vdash e$, and $e \vdash C_1^{\circ} \longrightarrow^{\circ} C_2^{\circ}$, then there exists $\Sigma_2$ such that $\Sigma_2; e \vdash^{C} C_2^{\circ} : T$ and $\mathsf{Ctx}(C_1^{\circ}) \subseteq \mathsf{Ctx}(C_2^{\circ})$ and $\Sigma_1 \subseteq \Sigma_2$ and $\mathsf{Ctx}(C_2^{\circ}) \vdash e$.*

Proof. By induction on the reduction derivation, casing on the typing derivation of the configuration. Proof in `CCAlphaRefStarPreservation.agda`. □

Progress is stated over the same well-typed configurations. There are three cases: either the configuration takes a reduction step, or it contains a value, or it reports blame.

Lemma 7.2 (Progress for $\mathrm{CC}^{\alpha,\star}_{\mathrm{Ref}}$). *The following progress properties hold.*

*(I) If $\Sigma; e \vdash^{C} C^{\circ} : T$ and $C^{\circ} = \langle \Gamma; M^{\circ}; \mu \rangle$, then either (1) $\exists \mathcal{D}^{\circ}.\ e \vdash C^{\circ} \longrightarrow^{\circ} \mathcal{D}^{\circ}$, or (2) $M^{\circ}$ is a code value, or (3) $M^{\circ} = \mathtt{blame}\ \ell$ for some $\ell$.*

*(II) If $\Sigma \vdash^{C} C : \hat{A}$ and $C = \langle \Gamma; M; \mu \rangle$, then either (1) $\exists \mathcal{D}.\ C \longrightarrow^{m} \mathcal{D}$, or (2) $M$ is a meta value, or (3) $M = \mathtt{blame}\ \ell$ for some $\ell$.*

Proof. By inversion on configuration typing, followed by mutual induction on the code and meta term typing derivations. The proof uses canonical-form lemmas for values, decidability of EC subtyping in the runtime subtype elimination case (implemented as finite graph search), and progress of meta coercion sequences in the cast case. The proof is in `CCAlphaRefStarProgress.agda`. □

Having established type safety for $\mathrm{CC}^{\alpha,\star}_{\mathrm{Ref}}$, we reason about elaboration next.

**Elaboration Preserves Types.** Type-directed elaboration connects the surface language $\lambda^{\alpha,\star}_{\mathrm{Ref}}$ to $\mathrm{CC}^{\alpha,\star}_{\mathrm{Ref}}$ (Section 6) by inserting explicit coercion sequences at uses of consistent subtyping. We bridge $\lambda^{\alpha,\star}_{\mathrm{Ref}}$ and $\mathrm{CC}^{\alpha,\star}_{\mathrm{Ref}}$ by proving that the elaboration preserves types.

Lemma 7.3 (Elaboration Preserves Types). *If $\vdash \Gamma$, then the following properties hold:*

*(I) If $\Gamma \vdash M : \hat{A} \rightsquigarrow M'$, then $\Gamma; \emptyset \vdash M' : \hat{A}$.*

*(II) If $\Gamma \vdash e$ and $\Gamma; e \vdash M^{\circ} \Rightarrow T \rightsquigarrow M^{\circ\prime}$, then $\Gamma; \emptyset; e \vdash M^{\circ\prime} : T$.*

*(III) If $\Gamma \vdash e$ and $\Gamma; e \vdash M^{\circ} \Leftarrow T \rightsquigarrow M^{\circ\prime}$, then $\Gamma; \emptyset; e \vdash M^{\circ\prime} : T$.*

Proof. By induction on the elaboration derivation. Proof in `ElaborationPreserves.agda`. □

Together with progress and preservation for $\mathrm{CC}^{\alpha,\star}_{\mathrm{Ref}}$, Lemma 7.3 gives type safety for $\lambda^{\alpha,\star}_{\mathrm{Ref}}$. However, this does not yet say that generated code is scope safe. The remaining step is to reify code values back to ordinary STLC with lexical scopes.

**Reifying Generated Code.** The generated code value from metaevaluation contains globally allocated variables rather than lexical STLC variables. As a result, we must define a reification function that recovers lexical binders from the runtime context. The function has the shape $\mathsf{reify}(\Gamma, M^{\circ}) = M^{s}$.

It takes a context, a cast calculus code term, and returns an STLC term. It is defined through a helper reifyAt, which takes an environment $\rho$ that maps global variables to lexical variables.

$$
\begin{array}{llcl}
\text{STLC terms} & M^s, N^s & ::= & k \mid x \mid \lambda x{:}T.\, M^s \mid M^s\, N^s \\
\text{reification env} & \rho & ::= & \emptyset \mid \rho[x \mapsto y] \\
 & \mathsf{reify}(\Gamma, M^\circ) & = & \mathsf{reifyAt}(\Gamma, \emptyset, M^\circ) \\
 & \mathsf{reifyAt}(\Gamma, \rho, k) & = & k \\
 & \mathsf{reifyAt}(\Gamma, \rho, x) & = & \rho(x) \\
 & \mathsf{reifyAt}(\Gamma, \rho, \overline{\lambda}(x{:}T)^\alpha.\, M^\circ) & = & \lambda y{:}T.\, \mathsf{reifyAt}((\Gamma, \rho[x \mapsto y]), M^\circ) \\
 & & & \text{if } (x{:}T)^\alpha_{\mathsf{glob}} \in \Gamma \\
 & \mathsf{reifyAt}(\Gamma, \rho, L^\circ\, M^\circ) & = & (\mathsf{reifyAt}(\Gamma, \rho, L^\circ))\ (\mathsf{reifyAt}(\Gamma, \rho, M^\circ))
\end{array}
$$

We write $\Gamma^s \vdash^{\mathrm{STLC}} M^s : T$ for the typing of STLC, where $\Gamma^s$ is the standard STLC typing context.

Lemma 7.4 (Reification Preserves Types). *Suppose the runtime context is well-formed:* $\vdash \Gamma$ *and all code variable entries are global:* $\forall x, \alpha, T.\ (x{:}T)^\alpha_{\mathsf{lex}} \notin \Gamma$. *If* $\Gamma; \Sigma; \varepsilon \vdash M^\circ : T$ *and* $\mathsf{reify}(\Gamma, M^\circ) = M^s$ *then* $\emptyset \vdash^{\mathrm{STLC}} M^s : T$.

Proof. Successful reification first implies that $M^\circ$ has the shape of a code value. The proof then proceeds by induction on that value. The proof is in `ReifyPreserves.agda`. ☐

Now we can prove that reify is total when applied to a well-typed code value.

Lemma 7.5 (Reify is Total for Code Values). *Suppose the runtime context is well-formed:* $\vdash \Gamma$*, and all subtype-constraint entries are global:* $\forall e_1, e_2.\ (e_1 {<:} e_2)_{\mathsf{lex}} \notin \Gamma$*, and all code-variable entries are also global:* $\forall x, \alpha, T.\ (x{:}T)^\alpha_{\mathsf{lex}} \notin \Gamma$. *If* $V^\circ$ *is a well-typed code value:* $\Gamma; \Sigma; \varepsilon \vdash V^\circ : T$*, then there must exist an STLC term* $M^s$ *such that* $\mathsf{reify}(\Gamma, V^\circ) = M^s$ *and* $\emptyset \vdash^{\mathrm{STLC}} M^s : T$.

Proof. By induction on the code value. The proof is in `ScopeSafetyValue.agda`. ☐

**Scope Safety.** Now we are ready to present the main theoretical result of the paper: that $\lambda^{\alpha,\star}_{\mathrm{Ref}}$ is indeed guaranteed to generate well-scoped STLC code. We first define the obs function, which observes a value of code type, peels off any (inert) coercions and returns the underlying code value:

$$
\mathsf{obs}(\text{``}V^\circ\text{''}\,e) = V^\circ \qquad \mathsf{obs}((\text{``}V^\circ\text{''}\,e)\langle \bar{c} \rangle) = V^\circ.
$$

The following observation lemma shows that any meta value of code type has such a code value observation.

Lemma 7.6 (Observation for Code-Typed Values). *Suppose* $V$ *is a well-typed meta value:* $\Gamma; \Sigma \vdash V : \text{``}T\text{''}\,\varepsilon$. *Then there exists a code value* $V^\circ$ *such that* $\mathsf{obs}(V) = V^\circ$ *and* $\Gamma; \Sigma; \varepsilon \vdash V^\circ : T$.

Proof. The proof is `obs-code` in `CanonicalForms.agda`. ☐

We write $\longrightarrow^*$ for the reflexive transitive closure of single-step meta reduction. The scope safety theorem for $\lambda^{\alpha,\star}_{\mathrm{Ref}}$ says that if a metaprogram is well-typed at a code type with empty classifier, and metaevaluation finishes without error, then it must generate well-scoped code:

Theorem 7.7 (Scope Safety for $\lambda^{\alpha,\star}_{\mathrm{Ref}}$). *Let* $M$ *be a closed, well-typed* $\lambda^{\alpha,\star}_{\mathrm{Ref}}$ *term. Suppose* $\emptyset \vdash M : \text{``}T\text{''}\,\varepsilon \rightsquigarrow M'$ *and* $\langle \emptyset; M'; \emptyset \rangle \longrightarrow^* \langle \Gamma; N; \mu \rangle$. *If* $\langle \Gamma; N; \mu \rangle$ *is irreducible and* $N$ *is not blame, then there exist a code value* $V^\circ$ *and an STLC term* $M^s$ *such that*

$$
\mathsf{obs}(N) = V^\circ \qquad \mathsf{reify}(\Gamma, V^\circ) = M^s \qquad \emptyset \vdash^{\mathrm{STLC}} M^s : T.
$$

Proof. Elaboration preservation (Lemma 7.3) gives us a well-typed initial $\mathrm{CC}^{\alpha,\star}_{\mathrm{Ref}}$ configuration. Applying preservation (Lemma 7.1) at each step, we prove that the final configuration is well-typed. Progress (Lemma 7.2) proves that a non-blame, irreducible term is a value. Lemma 7.6 gives the observed code value $V^\circ$. Scope safety for code values (Lemma 7.5) gives the STLC typing derivation. Proof in `ScopeSafety.agda`. □

The non-blame premise is necessary for the gradual language: metaevaluation may fail a runtime cast or subtype check. The static sister $\lambda^{\alpha}_{\mathrm{Ref}}$, by contrast, rules out such failures statically. We model $\lambda^{\alpha}_{\mathrm{Ref}}$ as the fully static fragment of $\lambda^{\alpha,\star}_{\mathrm{Ref}}$: no annotation contains $\star$, either in types or ECs.

Corollary 7.8 (Scope Safety for the Static Sister $\lambda^{\alpha}_{\mathrm{Ref}}$). *Let $M$ be a closed, well-typed $\lambda^{\alpha}_{\mathrm{Ref}}$ term (so it is fully static). Suppose $\emptyset \vdash M :\ ``T''\varepsilon \rightsquigarrow M'$ and $\langle \emptyset; M'; \emptyset \rangle \longrightarrow^{*} \langle \Gamma; N; \mu \rangle$. If $\langle \Gamma; N; \mu \rangle$ is irreducible, then there exist a code value $V^\circ$ and an STLC term $M^s$ such that*

$$\mathsf{obs}(N) = V^\circ \qquad \mathsf{reify}(\Gamma, V^\circ) = M^s \qquad \emptyset \vdash^{\mathrm{STLC}} M^s : T.$$

Proof. A corollary of Theorem 7.7 plus: a fully static term elaborates to an error-free $\mathrm{CC}^{\alpha,\star}_{\mathrm{Ref}}$ term, and error-freeness is preserved by reduction. Proof in `StaticScopeSafety.agda`. □

## 8 Space-Efficiency

In Section 1 we mentioned that naively implementing gradual typing can lead to catastrophic slowdowns. An example of such a catastrophic slowdown is a gradually-typed quicksort program shown in Figure 23 of the Appendix. When the dynamically-typed array parameter of `sort` is passed as an argument to the statically-typed helper functions `swap` and `partition` at runtime, the array is wrapped in a proxy which applies the necessary cast when reading or writing elements of the array. In a naive implementation, these proxies stack on top of each other as `sort` is called recursively, thereby forming long proxy chains. These long proxy chains catastrophically slow down the quicksort algorithm by changing its worst-case time complexity from $O(n^2)$ to $O(n^3)$. This overhead is incurred because every read and write operation on the array traverses proxy chains of length $O(n)$. Thus, we need a way to compress these proxy chains during runtime to have a reasonable upper bound on how long the chains can grow.

Our space-efficient system is derived from a space-efficient GTLC that uses a variant of coercions called hypercoercions [Lu et al. 2020; Siek and Chen 2021]. Hypercoercions are a canonical form of coercion and they enable a compact bit-level representation of coercions for an implementation. In addition to the use of hypercoercions, to achieve space efficiency one must also adjust the order of evaluation for casts to compress two adjacent casts before diving further in. We will discuss these space-efficient implementation changes in detail.

The main theoretical result of this section is the space consumption theorem (Theorem 8.1), which shows that our implementation changes indeed make $\mathrm{CC}^{\alpha,\star}_{\mathrm{Ref}}$ space-efficient. This result is also mechanized in Agda.

**Space-efficient Coercions.** We discuss the syntax of hypercoercions and hypercoercion composition. There are three hypercoercion calculi for: meta (Figure 24), code (Figure 26), and EC (Figure 25). Their syntax shares a common form $h; m; t$ consisting of head, middle and tail coercions. The hypercoercion composition operator, written $c \mathbin{⨟} d$, returns the hypercoercion that represents applying $c$ first then $d$. We derived the equations for the composition operators by appending the two hypercoercion inputs $h_1; m_1; t_1$ and $h_2; m_2; t_2$, to get $h_1; m_1; t_1; h_2; m_2; t_2; []$ and then normalizing the sequence according to the regular coercion reduction rules. Hypercoercion composition is used in the space-efficient version of the *cast-compose* rule, which we discuss next.

**Space-efficient $\mathbf{CC}^{\alpha,\star}_{\mathbf{Ref}}$ Reductions.** To prevent long chains of casts from forming, we need to ensure that adjacent casts are compressed in an eager fashion, and that reduction is not allowed to proceed underneath two or more casts, as that could grow the chain. One way to specify the correct order of reduction is with specialized evaluation contexts [Herman et al. 2010]. However, to continue with the frame-based approach used in the rest of this paper, we follow the approach of Siek and Chen [2021]. They introduce a simple reduction context flag which can be either NonCast or Any: the NonCast flag indicates that the reduction can only be fired when the enclosing term is not a cast, whereas the Any flag indicates that the reduction can be fired under any enclosing term. Therefore we modify the meta reduction judgment of $\mathrm{CC}^{\alpha,\star}_{\mathrm{Ref}}$ (Figure 8) to have the form $\mathrm{ctx} \vdash C \longrightarrow^{m} \mathcal{D}$ where ctx is a reduction context. The following three reduction rules are important for demonstrating how the reduction rules use reduction contexts to compress adjacent casts.

$$\begin{aligned}
\mathrm{NonCast} \vdash \langle \Gamma; M\langle c\rangle\langle d\rangle; \mu\rangle &\longrightarrow^{m} \langle \Gamma; M\langle \mathrm{Sub}(\Gamma) \vdash c \mathbin{⨾} d\rangle; \mu\rangle \\
\mathrm{Any} \vdash \langle \Gamma; F[M]; \mu\rangle &\longrightarrow^{m} \langle \Gamma'; F[N]; \mu'\rangle \qquad \text{if } \mathrm{ctx} \vdash \langle \Gamma; M; \mu\rangle \longrightarrow^{m} \langle \Gamma'; N; \mu'\rangle \\
\mathrm{NonCast} \vdash \langle \Gamma; M\langle c\rangle; \mu\rangle &\longrightarrow^{m} \langle \Gamma'; N\langle c\rangle; \mu'\rangle \qquad \text{if } \mathrm{Any} \vdash \langle \Gamma; M; \mu\rangle \longrightarrow^{m} \langle \Gamma'; N; \mu'\rangle
\end{aligned}$$

The first reduction rule above is the space-efficient version of the *cast-compose* rule in Figure 8. It compresses adjacent casts. To prevent reduction under multiple casts, we restrict the grammar of frames $F$ by removing the cast frame. The second reduction rule, a congruence rule, goes under a non-cast frame $F$ and reduces the subterm in an Any flag. The third reduction rule handles reduction under a cast, but reduces the subterm in a NonCast flag. Figure 28 of the Appendix lists all of the reduction rules of our space-efficient variant of $\mathrm{CC}^{\alpha,\star}_{\mathrm{Ref}}$.

**Proving the Space Consumption Theorem.** The main result of this section, the space consumption theorem, states that the size of the configuration is always bounded above by the *ideal size* of the configuration, which is the size of the configuration after erasing all the cast terms (i.e. replace each $M\langle c\rangle$ with $M$).

Theorem 8.1 (Space Consumption). *If* $\emptyset \vdash M : \hat{A} \rightsquigarrow M'$, *then there exists* $k$ *such that for any configuration* $C$ *where* $\mathrm{ctx} \vdash \langle \emptyset; M'; \emptyset\rangle \longrightarrow^{*} C$, *we have* $\mathrm{size}(C) \leq k \cdot \mathrm{ideal_size}(C)$.

The proof of the space consumption theorem follows from the proof of the space consumption theorem for GTLC by Siek and Chen [2021]. The first major difference between the proofs for GTLC and $\lambda^{\alpha,\star}_{\mathrm{Ref}}$ is that the mutually recursive definition of meta terms and code terms leads to analogous versions of the same lemma for meta and code terms. From here on, we will not address the analogous code term statements for lemmas. The second major difference is that, due to mutable references, our lemmas will also refer to the size of the heap instead of just the size of the term in the reduction config. To understand the proof of Theorem 8.1, we will first discuss a few main lemmas which are modified versions of the main lemmas used in Siek and Chen [2021].

Lemma 8.2 helps us derive that the result of elaboration contains a bounded number of adjacent casts. This bound on the length of adjacent casts is denoted by the judgment $n \mid d \vdash^{m} M$ ok (which is also used by Siek and Chen [2021]), where $n$ denotes the upper bound on adjacent casts in the meta term $M$ and $d$ is a boolean which denotes whether $M$ is in a delayed context like under a $\lambda$ or $\Lambda$. The mutually recursive definition of meta terms forces us to define an analogous adjacent casts judgment for code terms $d \vdash^{\circ} M^{\circ}$ ok. The definitions of both the adjacent cast judgments can be found in Figure 29. The elaboration rules are the same as the ones discussed in Section 6, except we use a modified version of the coerce function for inserting hypercoercions. The definition of coerce for hypercoercions is in Figure 27 of the Appendix.

Lemma 8.2 (Elaboration Efficient).

*(1) If* $\Gamma \vdash M : \hat{A} \rightsquigarrow M'$, *then* $n \mid d \vdash^{m} M'$ ok *for some* $n \leq 1$.
*(2) If* $\Gamma; e \vdash M^{\circ} \Rightarrow T \rightsquigarrow M^{\circ\prime}$, *then* $d \vdash^{\circ} M^{\circ\prime}$ ok
*(3) If* $\Gamma; e \vdash M^{\circ} \Leftarrow T \rightsquigarrow M^{\circ\prime}$, *then* $d \vdash^{\circ} M^{\circ\prime}$ ok

We also define an adjacent casts judgment for heaps ($\mu$ ok) which states that all values in the heap $\mu$ satisfy the adjacent casts judgment for terms. Lemma 8.3 and Lemma 8.4 establish the fact that the adjacent casts judgments for both the term and heap in a reduction configuration are preserved under reduction. Lemma 8.4 is new because GTLC lacks mutable references.

Lemma 8.3 (Preserve Adjacent Casts).

*(1) If* $\mu_1$ ok, $n \mid \mathsf{false} \vdash^m M_1$ ok *and* $ctx \vdash \langle \Gamma_1; M_1; \mu_1 \rangle \longrightarrow^m \langle \Gamma_2; M_2; \mu_2 \rangle$, *then* $k \mid \mathsf{false} \vdash^m M_2$ ok *for some* $k \leq 2 + n$.

*(2) If* $\mu_1$ ok, $\mathsf{false} \vdash^\circ M_1^\circ$ ok *and* $e \vdash \langle \Gamma_1; M_1^\circ; \mu_1 \rangle \longrightarrow^\circ \langle \Gamma_2; M_2^\circ; \mu_2 \rangle$, *then* $\mathsf{false} \vdash^\circ M_2$ ok.

Lemma 8.4 (Preserve Adjacent Casts in Heap).

*(1) If* $\mu_1$ ok, $n \mid \mathsf{false} \vdash^m M_1$ ok *and* $ctx \vdash \langle \Gamma_1; M_1; \mu_1 \rangle \longrightarrow^m \langle \Gamma_2; M_2; \mu_2 \rangle$, *then* $\mu_2$ ok.

*(2) If* $\mu_1$ ok, $\mathsf{false} \vdash^\circ M_1^\circ$ ok *and* $e \vdash \langle \Gamma_1; M_1^\circ; \mu_1 \rangle \longrightarrow^\circ \langle \Gamma_2; M_2^\circ; \mu_2 \rangle$, *then* $\mu_2$ ok.

The following lemma is used to prove Lemma 8.6, especially in the cast composition reduction rule case where the hypercoercion cast composition operator is used to compress casts. It states that, unlike the $+\!\!+$ operator for coercion sequences, the hypercoercion composition operator does not allow its output's height to grow bigger than its input's heights. The height functions for terms are defined in Figure 31 of the Appendix.

Lemma 8.5 (Composition Height).

*(1) For any two code hypercoercions* $c_1^\circ$ *and* $c_2^\circ$, *we have:*
$\mathsf{height}(c_1^\circ \fatsemi c_2^\circ) \leq \max(\mathsf{height}(c_1^\circ), \mathsf{height}(c_2^\circ))$.

*(2) For any two code middle coercions* $m_1^\circ$ *and* $m_2^\circ$, *we have:*
$\mathsf{height}(m_1^\circ \fatsemi m_2^\circ) \leq \max(\mathsf{height}(m_1^\circ), \mathsf{height}(m_2^\circ))$.

*(3) For any two meta hypercoercions* $c_1$ *and* $c_2$, *we have:*
$\mathsf{height}(\Theta \vdash c_1 \fatsemi c_2) \leq \max(\mathsf{height}(c_1), \mathsf{height}(c_2))$.

*(4) For any two meta middle coercions* $m_1$ *and* $m_2$, *we have:*
$\mathsf{height}(\Theta \vdash m_1 \fatsemi m_2) \leq \max(\mathsf{height}(m_1), \mathsf{height}(m_2))$.

To denote the maximum height of any coercion in a term and in a configuration, we introduce the notation $\mathsf{c_height}(M)$ and $\mathsf{c_height}(\mathcal{D})$. The definitions of the c_height functions can be found in Figures 33 and 36 of the appendix. Lemma 8.6 states that the c_height of the configuration monotonically decreases under reduction.

Lemma 8.6 (Preserve Height).

*(1) If* $ctx \vdash C \longrightarrow^m \mathcal{D}$, *then* $\mathsf{c_height}(\mathcal{D}) \leq \mathsf{c_height}(C)$.

*(2) If* $e \vdash C^\circ \longrightarrow^\circ \mathcal{D}^\circ$, *then* $\mathsf{c_height}(\mathcal{D}^\circ) \leq \mathsf{c_height}(C^\circ)$.

Lemmas 8.7 and 8.8 are similar to their counterparts in the proof of Siek and Chen [2021]. The statements of these lemmas involve another notion of size: we write $\mathsf{num}(M)$ for the number of terms in the program, so $\mathsf{num}(M)$ ignores the size of the coercions but still counts one for each cast. Lemma 8.7 establishes that the size of a term is bounded by the c_height and the number of terms in the program. Lemma 8.8 says that the number of terms is bounded by the ideal size (ideal_size) of the term.

Lemma 8.7 (Size bounded by number of terms).

*(1)* $\mathsf{size}(M) \leq (9 \cdot 2^{\mathsf{c_height}(M)}) \cdot \mathsf{num}(M)$

*(2)* $\mathsf{size}(M^\circ) \leq (9 \cdot 2^{\mathsf{c_height}(M^\circ)}) \cdot \mathsf{num}(M^\circ)$

Lemma 8.8 (Number of terms bounded by ideal size).

(1) *If* $n \mid d \vdash^m M$ ok, *then* $\mathsf{num}(M) \leq n + 10 \cdot \mathsf{ideal_size}(M)$.
(2) *If* $d \vdash^\circ M^\circ$ ok, *then* $\mathsf{num}(M^\circ) \leq 10 \cdot \mathsf{ideal_size}(M^\circ)$.

The proof of Theorem 8.1 follows from these lemmas in a straightforward fashion, similar to the proof in Siek and Chen [2021] but with some extra reasoning about heaps. The mechanized proof for Theorem 8.1 is in `SpaceEfficiency/SpaceEfficient.agda`.

## 9 Conclusion

We presented $\lambda^{\alpha,\star}_{\mathrm{Ref}}$, the first *gradual* metaprogramming language to reconcile imperative state with robust scope safety. By dynamically enforcing environment classifiers with runtime-generated symbols, our system mediates between static and optimistic typing invariants. This mechanism is formalized in $\mathrm{CC}^{\alpha,\star}_{\mathrm{Ref}}$, a cast calculus extending Henglein's foundations with coercions on code types, subtype-constrained types, and classifier polymorphism. We fully mechanized the scope safety theorems for both $\lambda^{\alpha,\star}_{\mathrm{Ref}}$ and its static sister language $\lambda^{\alpha}_{\mathrm{Ref}}$. We further mechanized the space consumption theorem, thereby providing a space-efficient implementation strategy for $\lambda^{\alpha,\star}_{\mathrm{Ref}}$.

Ultimately, $\lambda^{\alpha,\star}_{\mathrm{Ref}}$ establishes a rigorous framework for building performant, gradual code generators that remain safe even in the presence of complex computational effects.

## Appendix

The Appendix is organized as follows:

(1) Table 2 maps the meta-theoretical results in this paper to their Agda mechanization.
(2) Figure 9 presents fully static types, type well-formedness, and subtyping for $\lambda^{\alpha}_{\text{Ref}}$.
(3) Figure 10 presents gradual types, type well-formedness, and consistent subtyping for $\lambda^{\alpha,\star}_{\text{Ref}}$.
(4) Figure 11 presents terms and bidirectional typing rules for $\lambda^{\alpha,\star}_{\text{Ref}}$.
(5) Figure 12 defines code coercion sequence syntax, typing, and reduction semantics.
(6) Figure 13 defines meta coercion sequence syntax, typing, and normal forms.
(7) Figure 14 defines meta coercion sequence reduction semantics.
(8) Figure 15 defines well-formed typing contexts of $\text{CC}^{\alpha,\star}_{\text{Ref}}$.
(9) Figure 16 gives the typing rules for $\text{CC}^{\alpha,\star}_{\text{Ref}}$.
(10) Figure 17 defines well-typed runtime configurations for $\text{CC}^{\alpha,\star}_{\text{Ref}}$.
(11) Figure 18 presents the code reduction rules for $\text{CC}^{\alpha,\star}_{\text{Ref}}$.
(12) Figure 19 presents the meta reduction rules for $\text{CC}^{\alpha,\star}_{\text{Ref}}$.
(13) Figure 20 defines the coerce functions that generate coercions.
(14) Figure 21 defines elaboration from $\lambda^{\alpha,\star}_{\text{Ref}}$ to $\text{CC}^{\alpha,\star}_{\text{Ref}}$.
(15) Figure 22 defines the projection functions and inclusion relations used by preservation.
(16) Figure 23 shows a gradual quicksort program with potential catastrophic slowdown.
(17) Figure 24 defines meta hypercoercion syntax, typing, and composition.
(18) Figure 25 defines EC hypercoercion syntax, typing, and composition.
(19) Figure 26 defines code hypercoercion syntax, typing, and composition.
(20) Figure 27 defines the coerce functions that generate hypercoercions.
(21) Figure 28 defines the space-efficient reduction rules for $\text{CC}^{\alpha,\star}_{\text{Ref}}$ that use hypercoercions.
(22) Figure 29 defines adjacent cast judgments for code and meta terms.
(23) Figure 30 defines the adjacent cast judgment for heaps.
(24) Figure 31 defines hypercoercion height.
(25) Figure 32 defines hypercoercion size.
(26) Figure 33 defines cast height for code and meta terms.
(27) Figure 34 defines size, num, and ideal_size functions for code and meta terms.
(28) Figure 35 defines size and height functions for heaps.
(29) Figure 36 defines size and height functions for configurations.

| Results | Agda identifiers | Agda files |
|---|---|---|
| **Section 7: Type and scope safety** | | |
| Preservation for $CC^{\alpha,\star}_{Ref}$ (Lemma 7.1) | preserve$^{o}$, preserve$^{m}$ | CCAlphaRefStarPreservation.agda |
| Progress for $CC^{\alpha,\star}_{Ref}$ (Lemma 7.2) | progress$^{o}$, progress$^{m}$ | CCAlphaRefStarProgress.agda |
| Elaboration preserves types (Lemma 7.3) | code-synth-preserves, code-check-preserves, meta-elab-preserves | ElaborationPreserves.agda |
| Reification preserves types (Lemma 7.4) | reify-preserves-types | ReifyPreserves.agda |
| Reification is total for code values (Lemma 7.5) | scope-safety-val | ScopeSafetyValue.agda |
| Observation for code-typed values (Lemma 7.6) | obs-code | CanonicalForms.agda |
| Scope safety for $\lambda^{\alpha,\star}_{Ref}$ (Theorem 7.7) | scope-safety | ScopeSafety.agda |
| Scope safety for $\lambda^{\alpha}_{Ref}$ (Corollary 7.8) | static-scope-safety | StaticScopeSafety.agda |
| **Section 8: Space efficiency** | | |
| Space consumption (Theorem 8.1) | space-consumption-elab | SpaceEfficiency/SpaceEfficient.agda |
| Elaboration efficient (Lemma 8.2) | elab-efficient, elab-efficient$^{o}$⇒, elab-efficient$^{o}$⇐ | SpaceEfficiency/Proofs/ElaborationEfficient.agda |
| Preserve adjacent casts (Lemma 8.3) | preserve-ok$^{m}$, preserve-ok$^{o}$ | SpaceEfficiency/Proofs/PreserveOK.agda |
| Preserve adjacent casts in heaps (Lemma 8.4) | preserve-heap-values-ok$^{m}$, preserve-heap-values-ok$^{o}$ | SpaceEfficiency/Proofs/PreserveOK.agda |
| Composition height (Lemma 8.5) | compose-height-code, compose-height-code-middle, compose-height-meta, compose-height-meta-middle | SpaceEfficiency/Proofs/HypercoercionCompositionBounded.agda |
| Preserve height (Lemma 8.6) | preserve-height$^{m}$, preserve-height$^{o}$, | SpaceEfficiency/Proofs/PreserveHeight.agda |
| Size bounded by number of terms (Lemma 8.7) | real-size-tsize$^{m}$, real-size-tsize$^{o}$ | SpaceEfficiency/Proofs/RealSizeTermAndHeapSize.agda |
| Number of terms bounded by ideal size (Lemma 8.8) | size-OK, size$^{o}$-OK | SpaceEfficiency/Proofs/SizeOK.agda |

Table 2. Mapping from meta-theoretical results to their Agda mechanization. File paths are relative to the root of the Agda development (the Theories/ directory)

$$
\begin{array}{llcl}
\text{EC variables} & \alpha, \beta, \gamma & & \\
\text{ECs} & e & ::= & \alpha \mid \varepsilon \\
\text{code types} & T & ::= & \iota \mid T \to T \\
\text{meta types} & A, B & ::= & \iota \mid A \to B \mid \mathsf{Ref}\ A \mid \text{``}T\text{''}\,e \mid \forall \alpha.\, A \mid e {<:} e \Rightarrow A \\
\text{contexts} & \Gamma & ::= & \emptyset \mid \Gamma, \alpha \mid \Gamma, e_1 {<:} e_2 \mid \Gamma, x : A \mid \Gamma, (x : T)^{\alpha} \\
\text{subtype contexts} & \Theta & ::= & \emptyset \mid \Theta, e_1 {<:} e_2
\end{array}
$$

$\boxed{\Gamma \vdash e}$

$$
\frac{}{\Gamma \vdash \varepsilon} \qquad \frac{\alpha \in \Gamma}{\Gamma \vdash \alpha}
$$

$\boxed{\Gamma \vdash A}$

$$
\frac{}{\Gamma \vdash \iota} \qquad \frac{\Gamma \vdash A \qquad \Gamma \vdash B}{\Gamma \vdash A \to B} \qquad \frac{\Gamma \vdash A}{\Gamma \vdash \mathsf{Ref}\ A}
$$

$$
\frac{\Gamma \vdash e}{\Gamma \vdash \text{``}T\text{''}\,e} \qquad \frac{\Gamma, \alpha \vdash A}{\Gamma \vdash \forall \alpha.\, A} \qquad \frac{\Gamma \vdash e_1 \qquad \Gamma \vdash e_2 \qquad \Gamma \vdash A}{\Gamma \vdash e_1 {<:} e_2 \Rightarrow A}
$$

$\boxed{\Theta \vdash e <: e}$

$$
\frac{e_1 <: e_2 \in \Theta}{\Theta \vdash e_1 <: e_2} \qquad \frac{}{\Theta \vdash \varepsilon <: e} \qquad \frac{}{\Theta \vdash e <: e} \qquad \frac{\Theta \vdash e_1 <: e_2 \qquad \Theta \vdash e_2 <: e_3}{\Theta \vdash e_1 <: e_3}
$$

$\boxed{\Theta \vdash A <: B}$

$$
\frac{}{\Theta \vdash \iota <: \iota} \qquad \frac{\Theta \vdash A' <: A \qquad \Theta \vdash B <: B'}{\Theta \vdash A \to B <: A' \to B'} \qquad \frac{\Theta \vdash A' <: A \qquad \Theta \vdash A <: A'}{\Theta \vdash \mathsf{Ref}\ A <: \mathsf{Ref}\ A'}
$$

$$
\frac{\Theta \vdash e_1 <: e_2}{\Theta \vdash \text{``}T\text{''}\,e_1 <: \text{``}T\text{''}\,e_2} \qquad \frac{\Theta \vdash A <: B}{\Theta \vdash \forall \alpha.\, A <: \forall \alpha.\, B} \qquad \frac{\Theta, e_1 <: e_2 \vdash A <: B}{\Theta \vdash (e_1 {<:} e_2 \Rightarrow A) <: (e_1 {<:} e_2 \Rightarrow B)}
$$

Fig. 9. Fully static types, type well-formedness, and subtyping in $\lambda^{\alpha}_{\mathsf{Ref}}$

| | | | |
|---|---|---|---|
| gradual ECs | $\hat{e}$ | $::=$ | $\star \mid e$ |
| gradual code types | $\hat{T}$ | $::=$ | $\star \mid \iota \mid \hat{T} \to \hat{T}$ |
| gradual meta types | $\hat{A}, \hat{B}$ | $::=$ | $\star \mid \iota \mid \hat{A} \to \hat{B} \mid \mathsf{Ref}\ \hat{A} \mid \text{``}\hat{T}\text{''}\,\hat{e} \mid \forall \alpha.\, \hat{A} \mid e{<:}e{\Rightarrow}\hat{A}$ |
| gradual contexts | $\Gamma$ | $::=$ | $\emptyset \mid \Gamma, \alpha \mid \Gamma, e_1{<:}e_2 \mid \Gamma, x : \hat{A} \mid \Gamma, (x : T)^{\alpha}$ |

$\boxed{\Gamma \vdash \hat{e}}$

$$\frac{}{\Gamma \vdash \star} \qquad \frac{}{\Gamma \vdash \varepsilon} \qquad \frac{\alpha \in \Gamma}{\Gamma \vdash \alpha}$$

$\boxed{\Gamma \vdash \hat{A}}$

$$\frac{}{\Gamma \vdash \star} \qquad \frac{}{\Gamma \vdash \iota} \qquad \frac{\Gamma \vdash \hat{A} \qquad \Gamma \vdash \hat{B}}{\Gamma \vdash \hat{A} \to \hat{B}} \qquad \frac{\Gamma \vdash \hat{A}}{\Gamma \vdash \mathsf{Ref}\ \hat{A}}$$

$$\frac{\Gamma \vdash \hat{e}}{\Gamma \vdash \text{``}\hat{T}\text{''}\,\hat{e}} \qquad \frac{\Gamma, \alpha \vdash \hat{A}}{\Gamma \vdash \forall \alpha.\, \hat{A}} \qquad \frac{\Gamma \vdash e_1 \qquad \Gamma \vdash e_2 \qquad \Gamma \vdash \hat{A}}{\Gamma \vdash e_1{<:}e_2{\Rightarrow}\hat{A}}$$

$\boxed{\Theta \vdash \hat{e} \lesssim \hat{e}}$

$$\frac{}{\Theta \vdash \hat{e} \lesssim \star} \qquad \frac{}{\Theta \vdash \star \lesssim \hat{e}} \qquad \frac{\Theta \vdash e_1 <: e_2}{\Theta \vdash e_1 \lesssim e_2}$$

$\boxed{\Theta \vdash \hat{T} \lesssim \hat{T}}$

$$\frac{}{\Theta \vdash \hat{T} \lesssim \star} \qquad \frac{}{\Theta \vdash \star \lesssim \hat{T}} \qquad \frac{}{\Theta \vdash \iota \lesssim \iota}$$

$$\frac{\Theta \vdash \hat{T}_1' \lesssim \hat{T}_1 \qquad \Theta \vdash \hat{T}_2 \lesssim \hat{T}_2'}{\Theta \vdash \hat{T}_1 \to \hat{T}_2 \lesssim \hat{T}_1' \to \hat{T}_2'}$$

$\boxed{\Theta \vdash \hat{A} \lesssim \hat{B}}$

$$\frac{}{\Theta \vdash \hat{A} \lesssim \star} \qquad \frac{}{\Theta \vdash \star \lesssim \hat{B}} \qquad \frac{}{\Theta \vdash \iota \lesssim \iota} \qquad \frac{\Theta \vdash \hat{A}' \lesssim \hat{A} \qquad \Theta \vdash \hat{B} \lesssim \hat{B}'}{\Theta \vdash \hat{A} \to \hat{B} \lesssim \hat{A}' \to \hat{B}'}$$

$$\frac{\Theta \vdash \hat{A}' \lesssim \hat{A} \qquad \Theta \vdash \hat{A} \lesssim \hat{A}'}{\Theta \vdash \mathsf{Ref}\ \hat{A} \lesssim \mathsf{Ref}\ \hat{A}'} \qquad \frac{\Theta \vdash \hat{T}_1 \lesssim \hat{T}_2 \qquad \Theta \vdash \hat{e}_1 \lesssim \hat{e}_2}{\Theta \vdash \text{``}\hat{T}_1\text{''}\,\hat{e}_1 \lesssim \text{``}\hat{T}_2\text{''}\,\hat{e}_2}$$

$$\frac{\Theta \vdash \hat{A} \lesssim \hat{B}}{\Theta \vdash \forall \alpha.\, \hat{A} \lesssim \forall \alpha.\, \hat{B}} \qquad \frac{\Theta, e_1{<:}e_2 \vdash \hat{A} \lesssim \hat{B}}{\Theta \vdash (e_1{<:}e_2{\Rightarrow}\hat{A}) \lesssim (e_1{<:}e_2{\Rightarrow}\hat{B})}$$

Fig. 10. Gradual types, type well-formedness, and consistent subtyping for $\lambda^{\alpha,\star}_{\mathsf{Ref}}$

$$
\begin{array}{llcl}
\text{code terms} & L^\circ, M^\circ, N^\circ & ::= & k \mid x \mid \lambda(x{:}T)^{\alpha}.\, M^\circ \mid (L^\circ\ M^\circ)^{\ell} \mid \sim^{\ell} M \\
\text{meta terms} & L, M, N & ::= & k \mid x \mid \lambda x{:}\hat{A}.\, M \mid (L\ M)^{\ell} \mid \Lambda\alpha.\, M \mid M[e]^{\ell} \mid e_1{<:}e_2{\Rightarrow}M \mid M{\bullet}^{\ell} \\
 & & \mid & \text{``}M^\circ\text{''}e \mid \mathtt{ref}\ M \mid L :=^{\ell} M \mid !^{\ell}\, M
\end{array}
$$

$$\boxed{\Gamma; e \vdash M^\circ \Rightarrow T}$$

$$\frac{k : \iota}{\Gamma; e \vdash k \Rightarrow \iota}\ (\vdash\textit{code-const}) \qquad \frac{(x : T)^{\alpha} \in \Gamma \qquad \mathrm{Sub}(\Gamma) \vdash \alpha <: e}{\Gamma; e \vdash x \Rightarrow T}\ (\vdash\textit{code-var})$$

$$\frac{\begin{array}{c}\Gamma, \alpha, e{<:}\alpha, (x : T_1)^{\alpha}; \alpha \vdash N^\circ \Rightarrow T_2 \\ \alpha \notin \Gamma\end{array}}{\Gamma; e \vdash \lambda(x{:}T_1)^{\alpha}.\, N^\circ \Rightarrow T_1 \to T_2}\ (\vdash\textit{code-abs}) \qquad \frac{\begin{array}{c}\Gamma; e \vdash L^\circ \Rightarrow T_1 \to T_2 \\ \Gamma; e \vdash M^\circ \Leftarrow T_1\end{array}}{\Gamma; e \vdash (L^\circ\ M^\circ)^{\ell} \Rightarrow T_2}\ (\vdash\textit{code-app})$$

$$\boxed{\Gamma; e \vdash M^\circ \Leftarrow T}$$

$$\frac{\Gamma \vdash M : \hat{A} \qquad \mathrm{Sub}(\Gamma) \vdash \hat{A} \lesssim \text{``}T\text{''}e}{\Gamma; e \vdash \sim^{\ell} M \Leftarrow T}\ (\vdash\textit{splice}) \qquad \frac{\Gamma; e \vdash M^\circ \Rightarrow T}{\Gamma; e \vdash M^\circ \Leftarrow T}\ (\vdash\textit{check})$$

$$\boxed{\Gamma \vdash M : \hat{A}}$$

$$\frac{k : \iota}{\Gamma \vdash k : \iota}\ (\vdash\textit{meta-const}) \qquad \frac{x : \hat{A} \in \Gamma}{\Gamma \vdash x : \hat{A}}\ (\vdash\textit{meta-var}) \qquad \frac{\Gamma \vdash \hat{A} \qquad \Gamma, x : \hat{A} \vdash M : \hat{B}}{\Gamma \vdash \lambda x{:}\hat{A}.\, M : \hat{A} \to \hat{B}}\ (\vdash\textit{meta-abs})$$

$$\frac{\begin{array}{c}\Gamma \vdash L : \hat{A}_1 \to \hat{A}_2 \\ \Gamma \vdash M : \hat{B} \qquad \mathrm{Sub}(\Gamma) \vdash \hat{B} \lesssim \hat{A}_1\end{array}}{\Gamma \vdash (L\ M)^{\ell} : \hat{A}_2}\ (\vdash\textit{meta-app}) \qquad \frac{\begin{array}{c}\Gamma \vdash L : \star \\ \Gamma \vdash M : \hat{A}\end{array}}{\Gamma \vdash (L\ M)^{\ell} : \star}\ (\vdash\textit{meta-app-}\star) \qquad \frac{\begin{array}{c}\Gamma, \alpha \vdash M : \hat{A} \\ \alpha \notin \Gamma\end{array}}{\Gamma \vdash \Lambda\alpha.\, M : \forall\alpha.\, \hat{A}}\ (\vdash\textit{ec-abs})$$

$$\frac{\begin{array}{c}\Gamma \vdash e \\ \Gamma \vdash M : \forall\alpha.\, \hat{A}\end{array}}{\Gamma \vdash M[e]^{\ell} : \hat{A}[\alpha := e]}\ (\vdash\textit{ec-app}) \qquad \frac{\begin{array}{c}\Gamma \vdash e \\ \Gamma \vdash M : \star\end{array}}{\Gamma \vdash M[e]^{\ell} : \star}\ (\vdash\textit{ec-app-}\star) \qquad \frac{\begin{array}{c}\Gamma \vdash e_1 \qquad \Gamma \vdash e_2 \\ \Gamma, e_1{<:}e_2 \vdash M : \hat{A}\end{array}}{\Gamma \vdash e_1{<:}e_2{\Rightarrow}M : e_1{<:}e_2{\Rightarrow}\hat{A}}\ (\vdash\textit{sub-intro})$$

$$\frac{\begin{array}{c}\mathrm{Sub}(\Gamma) \vdash e_1 <: e_2 \\ \Gamma \vdash M : e_1{<:}e_2{\Rightarrow}\hat{A}\end{array}}{\Gamma \vdash M{\bullet}^{\ell} : \hat{A}}\ (\vdash\textit{sub-elim}) \qquad \frac{\Gamma \vdash M : \star}{\Gamma \vdash M{\bullet}^{\ell} : \star}\ (\vdash\textit{sub-elim-}\star) \qquad \frac{\begin{array}{c}\Gamma \vdash e \\ \Gamma; e \vdash M^\circ \Rightarrow T\end{array}}{\Gamma \vdash \text{``}M^\circ\text{''}e : \text{``}T\text{''}e}\ (\vdash\textit{quote})$$

$$\frac{\Gamma \vdash M : \hat{A}}{\Gamma \vdash \mathtt{ref}\ M : \mathtt{Ref}\ \hat{A}}\ (\vdash\textit{ref}) \qquad \frac{\begin{array}{c}\Gamma \vdash L : \mathtt{Ref}\ \hat{A} \\ \Gamma \vdash M : \hat{B} \qquad \mathrm{Sub}(\Gamma) \vdash \hat{B} \lesssim \hat{A}\end{array}}{\Gamma \vdash L :=^{\ell} M : \mathtt{Unit}}\ (\vdash\textit{assign}) \qquad \frac{\begin{array}{c}\Gamma \vdash L : \star \\ \Gamma \vdash M : \hat{A}\end{array}}{\Gamma \vdash L :=^{\ell} M : \mathtt{Unit}}\ (\vdash\textit{assign-}\star)$$

$$\frac{\Gamma \vdash M : \mathtt{Ref}\ \hat{A}}{\Gamma \vdash !^{\ell}\, M : \hat{A}}\ (\vdash\textit{deref}) \qquad \frac{\Gamma \vdash M : \star}{\Gamma \vdash !^{\ell}\, M : \star}\ (\vdash\textit{deref-}\star)$$

Fig. 11. Terms and bidirectional typing rules for $\lambda^{\alpha,\star}_{\mathrm{Ref}}$

New rules (compared to $\lambda^{\alpha}_{\mathrm{Ref}}$ of Figure 2) are highlighted in green. The premises changes from subtyping to consistent subtyping are highlighted in blue. Orange text ($\ell$) indicates blame labels.

$$
\begin{array}{lrcl}
\text{ground code types} & G^\circ, H^\circ & ::= & \iota \mid \star \to \star \\
\text{single code coercions} & c^\circ, d^\circ & ::= & G^\circ! \mid G^\circ?^{\ell} \mid \bar{c}^\circ \to \bar{d}^\circ \mid \bot^{\ell} \\
\text{code coercion sequences} & \bar{c}^\circ, \bar{d}^\circ & ::= & [\,] \mid c^\circ :: \bar{c}^\circ \\
\text{single normal forms} & c^\circ_{\mathrm{NF}}, d^\circ_{\mathrm{NF}} & ::= & G^\circ! \mid G^\circ?^{\ell} \mid \bot^{\ell} \mid (\bar{c}^\circ_{\mathrm{NF}} \to \bar{d}^\circ_{\mathrm{NF}})^{\dagger} \\
\text{normal forms} & \bar{c}^\circ_{\mathrm{NF}}, \bar{d}^\circ_{\mathrm{NF}} & ::= & [\,] \mid [c^\circ_{\mathrm{NF}}] \mid (c^\circ_{\mathrm{NF}} :: d^\circ_{\mathrm{NF}} :: \bar{d}^\circ_{\mathrm{NF}})^{\ddagger}
\end{array}
$$

$^{\dagger}\neg(\mathsf{Id}(\bar{c}^\circ_{\mathrm{NF}}) \wedge \mathsf{Id}(\bar{d}^\circ_{\mathrm{NF}})), \neg\mathsf{Blame}(\bar{c}^\circ_{\mathrm{NF}}), \neg\mathsf{Blame}(\bar{d}^\circ_{\mathrm{NF}})$; $\quad ^{\ddagger}\mathsf{IrredPair}(c^\circ_{\mathrm{NF}}, d^\circ_{\mathrm{NF}})$.

$$
\mathsf{Id}(\bar{c}^\circ) \triangleq \bar{c}^\circ = [\,] \qquad \mathsf{Blame}(\bar{c}^\circ) \triangleq \exists \ell.\ \bar{c}^\circ = [\bot^{\ell}]
$$

$\boxed{\vdash c^\circ : \hat{T}_1 \Rightarrow \hat{T}_2}$

$$
\frac{}{\vdash G^\circ! : G^\circ \Rightarrow \star}\ (\vdash !) \qquad \frac{}{\vdash G^\circ?^{\ell} : \star \Rightarrow G^\circ}\ (\vdash ?)
$$

$$
\frac{\vdash \bar{c}^\circ : \hat{T}_3 \Rightarrow \hat{T}_1 \qquad \vdash \bar{d}^\circ : \hat{T}_2 \Rightarrow \hat{T}_4}{\vdash \bar{c}^\circ \to \bar{d}^\circ : (\hat{T}_1 \to \hat{T}_2) \Rightarrow (\hat{T}_3 \to \hat{T}_4)}\ (\vdash \mathit{fun}) \qquad \frac{}{\vdash \bot^{\ell} : \hat{T}_1 \Rightarrow \hat{T}_2}\ (\vdash \bot)
$$

$\boxed{\vdash \bar{c}^\circ : \hat{T}_1 \Rightarrow \hat{T}_2}$

$$
\frac{}{\vdash [\,] : \hat{T} \Rightarrow \hat{T}}\ (\vdash \mathit{empty}) \qquad \frac{\vdash c^\circ : \hat{T}_1 \Rightarrow \hat{T}_2 \qquad \vdash \bar{c}^\circ : \hat{T}_2 \Rightarrow \hat{T}_3}{\vdash c^\circ :: \bar{c}^\circ : \hat{T}_1 \Rightarrow \hat{T}_3}\ (\vdash \mathit{cons})
$$

$\boxed{c^\circ \longrightarrow \bar{d}^\circ}$

$$
\frac{}{[\,] \to [\,] \longrightarrow [\,]}\ (\mathit{id\text{-}fun})
$$

$$
\frac{\bar{c}^\circ_1 \longrightarrow \bar{d}^\circ_1}{\bar{c}^\circ_1 \to \bar{c}^\circ_2 \longrightarrow [\bar{d}^\circ_1 \to \bar{c}^\circ_2]}\ (\xi\text{-}\mathit{fun\text{-}dom}) \qquad \frac{\neg\mathsf{Blame}(\bar{c}^\circ_{\mathrm{NF}}) \qquad \bar{c}^\circ \longrightarrow \bar{d}^\circ}{\bar{c}^\circ_{\mathrm{NF}} \to \bar{c}^\circ \longrightarrow [\bar{c}^\circ_{\mathrm{NF}} \to \bar{d}^\circ]}\ (\xi\text{-}\mathit{fun\text{-}cod})
$$

$$
\frac{}{[\bot^{\ell}] \to \bar{c}^\circ \longrightarrow [\bot^{\ell}]}\ (\xi\text{-}\bot\text{-}\mathit{fun\text{-}dom}) \qquad \frac{\neg\mathsf{Blame}(\bar{c}^\circ_{\mathrm{NF}})}{\bar{c}^\circ_{\mathrm{NF}} \to [\bot^{\ell}] \longrightarrow [\bot^{\ell}]}\ (\xi\text{-}\bot\text{-}\mathit{fun\text{-}cod})
$$

$\boxed{c^\circ_1 ; c^\circ_2 \longrightarrow \bar{d}^\circ}$

$$
\frac{}{G^\circ!; G^\circ?^{\ell} \longrightarrow [\,]}\ (\mathit{collapse}) \qquad \frac{G^\circ \neq H^\circ}{G^\circ!; H^\circ?^{\ell} \longrightarrow [\bot^{\ell}]}\ (\mathit{collide})
$$

$$
\frac{}{(\bar{c}^\circ_1 \to \bar{c}^\circ_2); (\bar{d}^\circ_1 \to \bar{d}^\circ_2) \longrightarrow [(\bar{d}^\circ_1 \mathbin{+\!\!+} \bar{c}^\circ_1) \to (\bar{c}^\circ_2 \mathbin{+\!\!+} \bar{d}^\circ_2)]}\ (\mathit{seq\text{-}fun})
$$

$$
\frac{}{\bot^{\ell}; c^\circ \longrightarrow [\bot^{\ell}]}\ (\bot\text{-}\mathit{left}) \qquad \frac{}{G^\circ!; \bot^{\ell} \longrightarrow [\bot^{\ell}]}\ (\bot\text{-}\mathit{right\text{-}!}) \qquad \frac{}{(\bar{c}^\circ_1 \to \bar{c}^\circ_2); \bot^{\ell} \longrightarrow [\bot^{\ell}]}\ (\bot\text{-}\mathit{right\text{-}fun})
$$

$\boxed{\bar{c}^\circ \longrightarrow \bar{d}^\circ}$

$$
\frac{c^\circ \longrightarrow \bar{d}^\circ}{c^\circ :: \bar{c}^\circ \longrightarrow \bar{d}^\circ \mathbin{+\!\!+} \bar{c}^\circ}\ (\mathit{reduce\text{-}head}) \qquad \frac{c^\circ_{\mathrm{NF}}; c^\circ \longrightarrow \bar{d}^\circ}{c^\circ_{\mathrm{NF}} :: c^\circ :: \bar{c}^\circ \longrightarrow \bar{d}^\circ \mathbin{+\!\!+} \bar{c}^\circ}\ (\mathit{reduce\text{-}neighbors})
$$

$$
\frac{\mathsf{IrredPair}(c^\circ_{\mathrm{NF}}, d^\circ_{\mathrm{NF}}) \qquad d^\circ_{\mathrm{NF}} :: \bar{c}^\circ \longrightarrow \bar{d}^\circ}{c^\circ_{\mathrm{NF}} :: d^\circ_{\mathrm{NF}} :: \bar{c}^\circ \longrightarrow c^\circ_{\mathrm{NF}} :: \bar{d}^\circ}\ (\mathit{reduce\text{-}tail})
$$

$\boxed{\mathsf{IrredPair}(c^\circ, d^\circ)}$

$$
\frac{}{\mathsf{IrredPair}(G^\circ?^{\ell}, G^\circ!)} \qquad \frac{}{\mathsf{IrredPair}(G^\circ?^{\ell_1}, \bot^{\ell_2})}
$$

$$
\frac{}{\mathsf{IrredPair}((\star \to \star)?^{\ell}, \bar{c}^\circ_1 \to \bar{c}^\circ_2)} \qquad \frac{}{\mathsf{IrredPair}(\bar{c}^\circ_1 \to \bar{c}^\circ_2, (\star \to \star)!)}
$$

Fig. 12. Code coercion sequences: syntax, typing, and reduction semantics

$$
\begin{array}{llcl}
\text{ground meta types} & G, H & ::= & \text{``}\star\text{''}\star \mid \forall\alpha.\star \mid e{<:}e{\Rightarrow}\star \mid \iota \mid \star \to \star \mid \mathsf{Ref}\ \star \\
\text{single meta coercions} & c, d & ::= & G! \mid G?^{\ell} \mid \bot^{\ell} \mid \bar{c} \to \bar{d} \mid \text{``}\bar{c}^{\circ}\text{''}\bar{c}^{e} \mid \mathsf{Ref}\ \bar{c}\ \bar{d} \mid \forall\alpha.\bar{c} \mid e_1{<:}e_2{\Rightarrow}\bar{c} \\
\text{meta coercion sequences} & \bar{c}, \bar{d} & ::= & [] \mid c :: \bar{c} \\
\text{single normal forms} & c_{\mathrm{NF}}, d_{\mathrm{NF}} & ::= & G! \mid G?^{\ell} \mid \bot^{\ell} \mid (\bar{c}_{\mathrm{NF}} \to \bar{d}_{\mathrm{NF}})^{\dagger 1} \mid (\text{``}\bar{c}^{\circ}_{\mathrm{NF}}\text{''}\bar{c}^{e}_{\mathrm{NF}})^{\dagger 2} \mid (\mathsf{Ref}\ \bar{c}_{\mathrm{NF}}\ \bar{d}_{\mathrm{NF}})^{\dagger 3} \\
 & & \mid & (\forall\alpha.\bar{c}_{\mathrm{NF}})^{\dagger 4} \mid (e_1{<:}e_2{\Rightarrow}\bar{c}_{\mathrm{NF}})^{\dagger 5} \\
\text{normal forms} & \bar{c}_{\mathrm{NF}}, \bar{d}_{\mathrm{NF}} & ::= & [] \mid [c_{\mathrm{NF}}] \mid (c_{\mathrm{NF}} :: d_{\mathrm{NF}} :: \bar{d}_{\mathrm{NF}})^{\dagger 6}
\end{array}
$$

$^{\dagger 1}\neg(\mathsf{Id}(\bar{c}_{\mathrm{NF}}) \wedge \mathsf{Id}(\bar{d}_{\mathrm{NF}}))$; $^{\dagger 2}\neg(\mathsf{Id}(\bar{c}^{\circ}_{\mathrm{NF}}) \wedge \mathsf{Id}(\bar{c}^{e}_{\mathrm{NF}}))$, $\neg\mathsf{Blame}(\bar{c}^{\circ}_{\mathrm{NF}})$, $\neg\mathsf{Blame}(\bar{c}^{e}_{\mathrm{NF}})$; $^{\dagger 3}\neg(\mathsf{Id}(\bar{c}_{\mathrm{NF}}) \wedge \mathsf{Id}(\bar{d}_{\mathrm{NF}}))$;
$^{\dagger 4}\neg\mathsf{Id}(\bar{c}_{\mathrm{NF}})$; $^{\dagger 5}\neg\mathsf{Id}(\bar{c}_{\mathrm{NF}})$; $^{\dagger 6}\mathsf{IrredPair}(c_{\mathrm{NF}}, d_{\mathrm{NF}})$.

$$
\mathsf{Id}(\bar{c}) \triangleq \bar{c} = [] \qquad \mathsf{Blame}(c) \triangleq \exists \ell.\ c = \bot^{\ell} \qquad \mathsf{Proj}(c) \triangleq \exists G, \ell.\ c = G?^{\ell}
$$

$\boxed{\Gamma \vdash c : \hat{A} \Rightarrow \hat{B}}$

$$
\frac{\Gamma \vdash G}{\Gamma \vdash G! : G \Rightarrow \star}\ (\vdash !) \qquad \frac{\Gamma \vdash G}{\Gamma \vdash G?^{\ell} : \star \Rightarrow G}\ (\vdash ?) \qquad \frac{\Gamma \vdash \hat{A} \quad \Gamma \vdash \hat{B}}{\Gamma \vdash \bot^{\ell} : \hat{A} \Rightarrow \hat{B}}\ (\vdash \bot)
$$

$$
\frac{\Gamma \vdash \bar{c} : \hat{A}' \Rightarrow \hat{A} \quad \Gamma \vdash \bar{d} : \hat{B} \Rightarrow \hat{B}'}{\Gamma \vdash \bar{c} \to \bar{d} : (\hat{A} \to \hat{B}) \Rightarrow (\hat{A}' \to \hat{B}')}\ (\vdash \mathit{fun}) \qquad \frac{\vdash \bar{c}^{\circ} : \hat{T}_1 \Rightarrow \hat{T}_2 \quad \Gamma \vdash \bar{c}^{e} : \hat{e}_1 \Rightarrow \hat{e}_2}{\Gamma \vdash \text{``}\bar{c}^{\circ}\text{''}\bar{c}^{e} : \text{``}\hat{T}_1\text{''}\hat{e}_1 \Rightarrow \text{``}\hat{T}_2\text{''}\hat{e}_2}\ (\vdash \mathit{code})
$$

$$
\frac{\Gamma \vdash \bar{c} : \hat{B} \Rightarrow \hat{A} \quad \Gamma \vdash \bar{d} : \hat{A} \Rightarrow \hat{B}}{\Gamma \vdash \mathsf{Ref}\ \bar{c}\ \bar{d} : \mathsf{Ref}\ \hat{A} \Rightarrow \mathsf{Ref}\ \hat{B}}\ (\vdash \mathit{ref}) \qquad \frac{\Gamma, \alpha \vdash \bar{c} : \hat{A} \Rightarrow \hat{B}}{\Gamma \vdash \forall\alpha.\bar{c} : \forall\alpha.\hat{A} \Rightarrow \forall\alpha.\hat{B}}\ (\vdash \forall)
$$

$$
\frac{\Gamma \vdash e_1 \quad \Gamma \vdash e_2 \quad \Gamma, e_1{<:}e_2 \vdash \bar{c} : \hat{A} \Rightarrow \hat{B}}{\Gamma \vdash e_1{<:}e_2{\Rightarrow}\bar{c} : (e_1{<:}e_2{\Rightarrow}\hat{A}) \Rightarrow (e_1{<:}e_2{\Rightarrow}\hat{B})}\ (\vdash \mathit{sub})
$$

$\boxed{\Gamma \vdash \bar{c} : \hat{A} \Rightarrow \hat{B}}$

$$
\frac{\Gamma \vdash \hat{A}}{\Gamma \vdash [] : \hat{A} \Rightarrow \hat{A}}\ (\vdash \mathit{empty}) \qquad \frac{\Gamma \vdash c : \hat{A} \Rightarrow \hat{B} \quad \Gamma \vdash \bar{c} : \hat{B} \Rightarrow \hat{C}}{\Gamma \vdash c :: \bar{c} : \hat{A} \Rightarrow \hat{C}}\ (\vdash \mathit{cons})
$$

$\boxed{\mathsf{IrredPair}(c, d)}$

$$
\frac{}{\mathsf{IrredPair}(G?^{\ell}, G!)} \qquad \frac{}{\mathsf{IrredPair}(G?^{\ell}, \bot^{\ell'})}
$$

$$
\frac{}{\mathsf{IrredPair}((\star \to \star)?^{\ell}, \bar{c} \to \bar{d})} \qquad \frac{}{\mathsf{IrredPair}(\bar{c} \to \bar{d}, (\star \to \star)!)}
$$

$$
\frac{}{\mathsf{IrredPair}((\text{``}\star\text{''}\star)?^{\ell}, \text{``}\bar{c}^{\circ}\text{''}\bar{c}^{e})} \qquad \frac{}{\mathsf{IrredPair}(\text{``}\bar{c}^{\circ}\text{''}\bar{c}^{e}, (\text{``}\star\text{''}\star)!)}
$$

$$
\frac{}{\mathsf{IrredPair}((\mathsf{Ref}\ \star)?^{\ell}, \mathsf{Ref}\ \bar{c}\ \bar{d})} \qquad \frac{}{\mathsf{IrredPair}(\mathsf{Ref}\ \bar{c}\ \bar{d}, (\mathsf{Ref}\ \star)!)}
$$

$$
\frac{}{\mathsf{IrredPair}((\forall\alpha.\star)?^{\ell}, \forall\alpha.\bar{c})} \qquad \frac{}{\mathsf{IrredPair}(\forall\alpha.\bar{c}, (\forall\alpha.\star)!)}
$$

$$
\frac{}{\mathsf{IrredPair}((e_1{<:}e_2{\Rightarrow}\star)?^{\ell}, e_1{<:}e_2{\Rightarrow}\bar{c})} \qquad \frac{}{\mathsf{IrredPair}(e_1{<:}e_2{\Rightarrow}\bar{c}, (e_1{<:}e_2{\Rightarrow}\star)!)}
$$

Fig. 13. Meta coercion sequences: syntax and typing, and normal forms

$$\boxed{\Theta \vdash c \longrightarrow \bar{d}}$$

$$\frac{}{\Theta \vdash [] \to [] \longrightarrow []}\;(\textit{id-fun}) \qquad \frac{}{\Theta \vdash \text{``}[]\text{''}[] \longrightarrow []}\;(\textit{id-code}) \qquad \frac{}{\Theta \vdash \mathsf{Ref}\ []\ [] \longrightarrow []}\;(\textit{id-ref})$$

$$\frac{}{\Theta \vdash \forall\alpha.[] \longrightarrow []}\;(\textit{id-forall}) \qquad \frac{}{\Theta \vdash e_1{<:}e_2{\Rightarrow}[] \longrightarrow []}\;(\textit{id-sub})$$

$$\frac{\Theta \vdash \bar{c}_1 \longrightarrow \bar{d}_1}{\Theta \vdash \bar{c}_1 \to \bar{c}_2 \longrightarrow [\bar{d}_1 \to \bar{c}_2]}\;(\xi\textit{-fun-dom}) \qquad \frac{\Theta \vdash \bar{c} \longrightarrow \bar{d}}{\Theta \vdash \bar{c}_{\mathrm{NF}} \to \bar{c} \longrightarrow [\bar{c}_{\mathrm{NF}} \to \bar{d}]}\;(\xi\textit{-fun-cod})$$

$$\frac{\Theta \vdash \bar{c}_1 \longrightarrow \bar{d}_1}{\Theta \vdash \mathsf{Ref}\ \bar{c}_1\ \bar{c}_2 \longrightarrow [\mathsf{Ref}\ \bar{d}_1\ \bar{c}_2]}\;(\xi\textit{-ref-in}) \qquad \frac{\Theta \vdash \bar{c} \longrightarrow \bar{d}}{\Theta \vdash \mathsf{Ref}\ \bar{c}_{\mathrm{NF}}\ \bar{c} \longrightarrow [\mathsf{Ref}\ \bar{c}_{\mathrm{NF}}\ \bar{d}]}\;(\xi\textit{-ref-out})$$

$$\frac{\Theta \vdash \bar{c} \longrightarrow \bar{d}}{\Theta \vdash \forall\alpha.\bar{c} \longrightarrow [\forall\alpha.\bar{d}]}\;(\xi\textit{-forall}) \qquad \frac{\Theta, e_1{<:}e_2 \vdash \bar{c} \longrightarrow \bar{d}}{\Theta \vdash e_1{<:}e_2{\Rightarrow}\bar{c} \longrightarrow [e_1{<:}e_2{\Rightarrow}\bar{d}]}\;(\xi\textit{-sub})$$

$$\frac{\bar{c}^{\circ} \longrightarrow \bar{d}^{\circ}}{\Theta \vdash \text{``}\bar{c}^{\circ}\text{''}\bar{c}^{e} \longrightarrow [\text{``}\bar{d}^{\circ}\text{''}\bar{c}^{e}]}\;(\xi\textit{-code}) \qquad \frac{\neg\mathsf{Blame}(\bar{c}^{\circ}_{\mathrm{NF}}) \qquad \Theta \vdash \bar{c}^{e} \longrightarrow \bar{d}^{e}}{\Theta \vdash \text{``}\bar{c}^{\circ}_{\mathrm{NF}}\text{''}\bar{c}^{e} \longrightarrow [\text{``}\bar{c}^{\circ}_{\mathrm{NF}}\text{''}\bar{d}^{e}]}\;(\xi\textit{-code-ec})$$

$$\frac{}{\Theta \vdash \text{``}[\bot^{\ell}]\text{''}\bar{c}^{e} \longrightarrow [\bot^{\ell}]}\;(\xi\textit{-code-}\bot) \qquad \frac{\neg\mathsf{Blame}(\bar{c}^{\circ}_{\mathrm{NF}})}{\Theta \vdash \text{``}\bar{c}^{\circ}_{\mathrm{NF}}\text{''}[\bot^{\ell}] \longrightarrow [\bot^{\ell}]}\;(\xi\textit{-code-ec-}\bot)$$

$$\boxed{\Theta \vdash c_1; c_2 \longrightarrow \bar{d}}$$

$$\frac{}{\Theta \vdash G!; G?^{\ell} \longrightarrow []}\;(\textit{collapse}) \qquad \frac{G \neq H}{\Theta \vdash G!; H?^{\ell} \longrightarrow [\bot^{\ell}]}\;(\textit{collide})$$

$$\frac{}{\Theta \vdash (\bar{c}_1 \to \bar{c}_2); (\bar{d}_1 \to \bar{d}_2) \longrightarrow [(\bar{d}_1 \mathbin{+\!\!+} \bar{c}_1) \to (\bar{c}_2 \mathbin{+\!\!+} \bar{d}_2)]}\;(\textit{seq-fun})$$

$$\frac{}{\Theta \vdash \mathsf{Ref}\ \bar{c}_1\ \bar{c}_2; \mathsf{Ref}\ \bar{d}_1\ \bar{d}_2 \longrightarrow [\mathsf{Ref}\ (\bar{d}_1 \mathbin{+\!\!+} \bar{c}_1)\ (\bar{c}_2 \mathbin{+\!\!+} \bar{d}_2)]}\;(\textit{seq-ref})$$

$$\frac{}{\Theta \vdash \text{``}\bar{c}^{\circ}\text{''}\bar{c}^{e}; \text{``}\bar{d}^{\circ}\text{''}\bar{d}^{e} \longrightarrow [\text{``}\bar{c}^{\circ} \mathbin{+\!\!+} \bar{d}^{\circ}\text{''}(\bar{c}^{e} \mathbin{+\!\!+} \bar{d}^{e})]}\;(\textit{seq-code})$$

$$\frac{}{\Theta \vdash (\forall\alpha.\bar{c}); (\forall\alpha.\bar{d}) \longrightarrow [\forall\alpha.(\bar{c} \mathbin{+\!\!+} \bar{d})]}\;(\textit{seq-forall})$$

$$\frac{}{\Theta \vdash (e_1{<:}e_2{\Rightarrow}\bar{c}); (e_1{<:}e_2{\Rightarrow}\bar{d}) \longrightarrow [e_1{<:}e_2{\Rightarrow}(\bar{c} \mathbin{+\!\!+} \bar{d})]}\;(\textit{seq-sub})$$

$$\frac{}{\Theta \vdash \bot^{\ell}; c \longrightarrow [\bot^{\ell}]}\;(\bot\textit{-left}) \qquad \frac{\neg\mathsf{Blame}(c) \qquad \neg\mathsf{Proj}(c)}{\Theta \vdash c; \bot^{\ell} \longrightarrow [\bot^{\ell}]}\;(\bot\textit{-right})$$

$$\boxed{\Theta \vdash \bar{c} \longrightarrow \bar{d}}$$

$$\frac{\Theta \vdash c \longrightarrow \bar{d}}{\Theta \vdash c :: \bar{c} \longrightarrow \bar{d} \mathbin{+\!\!+} \bar{c}}\;(\textit{reduce-head}) \qquad \frac{\Theta \vdash c_{\mathrm{NF}}; c \longrightarrow \bar{d}}{\Theta \vdash c_{\mathrm{NF}} :: c :: \bar{c} \longrightarrow \bar{d} \mathbin{+\!\!+} \bar{c}}\;(\textit{reduce-neighbors})$$

$$\frac{\mathsf{IrredPair}(c_{\mathrm{NF}}, d_{\mathrm{NF}}) \qquad \Theta \vdash d_{\mathrm{NF}} :: \bar{c} \longrightarrow \bar{d}}{\Theta \vdash c_{\mathrm{NF}} :: d_{\mathrm{NF}} :: \bar{c} \longrightarrow c_{\mathrm{NF}}; \bar{d}}\;(\textit{reduce-tail})$$

Fig. 14. Meta coercion sequences: reduction semantics

$\boxed{\vdash \Gamma}$

$$\frac{}{\vdash \emptyset}\ \textit{(wf-empty)} \qquad \frac{\vdash \Gamma \qquad \Gamma \vdash \hat{A}}{\vdash \Gamma, x : \hat{A}}\ \textit{(wf-meta-var)} \qquad \frac{\vdash \Gamma \qquad \alpha \notin \Gamma}{\vdash \Gamma, \alpha_{\text{lex}}}\ \textit{(wf-ec-lex)} \qquad \frac{\vdash \Gamma \qquad \alpha \notin \Gamma}{\vdash \Gamma, \alpha_{\text{glob}}}\ \textit{(wf-ec-glob)}$$

$$\frac{\vdash \Gamma \qquad \Gamma \vdash e_1 \qquad \Gamma \vdash e_2}{\vdash \Gamma, (e_1 <: e_2)_{\text{lex}}}\ \textit{(wf-sub-lex)} \qquad \frac{\begin{array}{ccc}\vdash \Gamma & \Gamma \vdash e & \Gamma \vdash \alpha \\ \alpha_{\text{glob}} \in \Gamma & \multicolumn{2}{c}{\forall e'. (e' <: \alpha)_{\text{glob}} \notin \Gamma}\end{array}}{\vdash \Gamma, (e <: \alpha)_{\text{glob}}}\ \textit{(wf-sub-glob)}$$

$$\frac{\vdash \Gamma \qquad \Gamma \vdash \alpha}{\vdash \Gamma, (x{:}T)^{\alpha}_{\text{lex}}}\ \textit{(wf-code-var-lex)} \qquad \frac{\begin{array}{cc}\vdash \Gamma & \Gamma \vdash \alpha \\ \multicolumn{2}{c}{\forall y, S. (y{:}S)^{\alpha}_{\text{glob}} \notin \Gamma}\end{array}}{\vdash \Gamma, (x{:}T)^{\alpha}_{\text{glob}}}\ \textit{(wf-code-var-glob)}$$

Fig. 15. Well-formed typing contexts of $\mathsf{CC}^{\alpha,\star}_{\mathsf{Ref}}$

$$\boxed{\Gamma; \Sigma; e \vdash M^\circ : T}$$

$$\frac{k : \iota}{\Gamma; \Sigma; e \vdash k : \iota}\ (\vdash\textit{code-const}) \qquad \frac{(x{:}T)^{\alpha}_{\rho} \in \Gamma \qquad \mathsf{Sub}(\Gamma) \vdash \alpha <: e}{\Gamma; \Sigma; e \vdash x : T}\ (\vdash\textit{code-var})$$

$$\frac{\alpha \notin \Gamma \qquad \Gamma, \alpha_{\mathsf{lex}}, (e{<:}\alpha)_{\mathsf{lex}}, (x{:}T_1)^{\alpha}_{\mathsf{lex}}; \Sigma; \alpha \vdash M^\circ : T_2}{\Gamma; \Sigma; e \vdash \lambda(x{:}T_1)^{\alpha}.\, M^\circ : T_1 \rightarrow T_2}\ (\vdash\textit{code-abs})$$

$$\frac{\Gamma \vdash \alpha \qquad \alpha_{\mathsf{glob}} \in \Gamma \qquad (x{:}T_1)^{\alpha}_{\mathsf{glob}} \in \Gamma \qquad (e_1{<:}\alpha)_{\mathsf{glob}} \in \Gamma \qquad \mathsf{Sub}(\Gamma) \vdash e_1 <: e_2 \qquad \Gamma; \Sigma; \alpha \vdash M^\circ : T_2}{\Gamma; \Sigma; e_2 \vdash \overline{\lambda}(x{:}T_1)^{\alpha}.\, M^\circ : T_1 \rightarrow T_2}\ (\vdash\textit{code-abs-val})$$

$$\frac{\Gamma; \Sigma; e \vdash L^\circ : T_1 \rightarrow T_2 \qquad \Gamma; \Sigma; e \vdash M^\circ : T_1}{\Gamma; \Sigma; e \vdash L^\circ\, M^\circ : T_2}\ (\vdash\textit{code-app})$$

$$\frac{\Gamma; \Sigma \vdash M : ``T''e}{\Gamma; \Sigma; e \vdash {\sim}M : T}\ (\vdash\textit{splice}) \qquad \frac{}{\Gamma; \Sigma; e \vdash \mathtt{blame}\ \ell : T}\ (\vdash\textit{code-blame})$$

$$\boxed{\Gamma; \Sigma \vdash M : \hat{A}}$$

$$\frac{k : \iota}{\Gamma; \Sigma \vdash k : \iota}\ (\vdash\textit{meta-const}) \qquad \frac{x : \hat{A} \in \Gamma}{\Gamma; \Sigma \vdash x : \hat{A}}\ (\vdash\textit{meta-var}) \qquad \frac{a : \hat{A} \in \Sigma}{\Gamma; \Sigma \vdash a : \mathtt{Ref}\ \hat{A}}\ (\vdash\textit{addr})$$

$$\frac{\Gamma \vdash \hat{A} \qquad \Gamma, x : \hat{A}; \Sigma \vdash M : \hat{B}}{\Gamma; \Sigma \vdash \lambda x.\, M : \hat{A} \rightarrow \hat{B}}\ (\vdash\textit{meta-abs}) \qquad \frac{\Gamma; \Sigma \vdash L : \hat{A} \rightarrow \hat{B} \qquad \Gamma; \Sigma \vdash M : \hat{A}}{\Gamma; \Sigma \vdash L\, M : \hat{B}}\ (\vdash\textit{meta-app})$$

$$\frac{\Gamma, \alpha_{\mathsf{lex}}; \Sigma \vdash M : \hat{A} \qquad \alpha \notin \Gamma}{\Gamma; \Sigma \vdash \Lambda\alpha.\, M : \forall\alpha.\, \hat{A}}\ (\vdash\textit{ec-abs}) \qquad \frac{\Gamma; \Sigma \vdash M : \forall\alpha.\, \hat{A} \qquad \Gamma \vdash e}{\Gamma; \Sigma \vdash M[e] : \hat{A}[\alpha := e]}\ (\vdash\textit{ec-app})$$

$$\frac{\Gamma \vdash e_1 \qquad \Gamma \vdash e_2 \qquad \Gamma, (e_1{<:}e_2)_{\mathsf{lex}}; \Sigma \vdash M : \hat{A}}{\Gamma; \Sigma \vdash e_1{<:}e_2{\Rightarrow}M : e_1{<:}e_2{\Rightarrow}\hat{A}}\ (\vdash\textit{sub-intro})$$

$$\frac{\Gamma; \Sigma \vdash M : e_1{<:}e_2{\Rightarrow}\hat{A} \qquad \mathsf{Sub}(\Gamma) \vdash e_1 <: e_2}{\Gamma; \Sigma \vdash M\bullet : \hat{A}}\ (\vdash\textit{sub-elim})$$

$$\frac{\Gamma; \Sigma \vdash M : \star}{\Gamma; \Sigma \vdash M{\bullet}\star^{\ell} : \star}\ (\vdash\textit{sub-elim-}\star) \qquad \frac{\Gamma; \Sigma; e \vdash M^\circ : T \qquad \Gamma \vdash e}{\Gamma; \Sigma \vdash ``M^\circ''e : ``T''e}\ (\vdash\textit{quote})$$

$$\frac{\Gamma; \Sigma \vdash M : \hat{A}}{\Gamma; \Sigma \vdash \mathtt{ref}\ M : \mathtt{Ref}\ \hat{A}}\ (\vdash\textit{ref}) \qquad \frac{\Gamma; \Sigma \vdash L : \mathtt{Ref}\ \hat{A} \qquad \Gamma; \Sigma \vdash M : \hat{A}}{\Gamma; \Sigma \vdash L := M : \mathtt{Unit}}\ (\vdash\textit{assign})$$

$$\frac{\Gamma; \Sigma \vdash M : \mathtt{Ref}\ \hat{A}}{\Gamma; \Sigma \vdash\ !M : \hat{A}}\ (\vdash\textit{deref}) \qquad \frac{\Gamma; \Sigma \vdash M : \hat{A} \qquad \Gamma \vdash \bar{c} : \hat{A} \Rightarrow \hat{B}}{\Gamma; \Sigma \vdash M\langle\bar{c}\rangle : \hat{B}}\ (\vdash\textit{cast}) \qquad \frac{\Gamma \vdash \hat{A}}{\Gamma; \Sigma \vdash \mathtt{blame}\ \ell : \hat{A}}\ (\vdash\textit{blame})$$

Fig. 16. Typing rules for cast calculus $\mathrm{CC}^{\alpha,\star}_{\mathsf{Ref}}$

$$\Gamma \vdash \Sigma \quad \triangleq \quad \forall a, \hat{A}.\ \text{if } \Sigma(a) = \hat{A} \text{ then } \Gamma \vdash \hat{A}$$

$$\Gamma; \Sigma \vdash^{H} \mu \quad \triangleq \quad \mathrm{dom}(\mu) = \mathrm{dom}(\Sigma) \ \wedge\ (\forall a, V, \hat{A}.\ \text{if } \mu(a) = V \text{ and } \Sigma(a) = \hat{A} \text{ then } \Gamma; \Sigma \vdash V : \hat{A})$$

$$\boxed{\Sigma \vdash^{C} C : \hat{A}} \qquad \boxed{\Sigma; e \vdash^{C} C^{\circ} : T}$$

$$\frac{\vdash \Gamma \qquad \forall x, \hat{B}.\ x : \hat{B} \notin \Gamma \qquad \Gamma \vdash \Sigma \qquad \Gamma; \Sigma \vdash^{H} \mu \qquad \Gamma; \Sigma \vdash M : \hat{A}}{\Sigma \vdash^{C} \langle \Gamma; M; \mu \rangle : \hat{A}}\;(\vdash\textit{meta-config})$$

$$\frac{\vdash \Gamma \qquad \forall x, \hat{B}.\ x : \hat{B} \notin \Gamma \qquad \Gamma \vdash \Sigma \qquad \Gamma; \Sigma \vdash^{H} \mu \qquad \Gamma; \Sigma; e \vdash M^{\circ} : T}{\Sigma; e \vdash^{C} \langle \Gamma; M^{\circ}; \mu \rangle : T}\;(\vdash\textit{code-config})$$

Fig. 17. Well-typed runtime configurations for $\mathrm{CC}^{\alpha,\star}_{\mathrm{Ref}}$

$$\boxed{M^{\circ} \longrightarrow^{\circ} N^{\circ}}$$

$$\frac{}{\sim \text{``} V^{\circ} \text{''} e_1 \longrightarrow^{\circ} V^{\circ}}\;(\textit{splice-}\beta) \qquad \frac{}{\sim((\text{``} V^{\circ} \text{''} e_1)\langle \text{``}[\,]\text{''}\ [e_1{\uparrow}e_2]\rangle) \longrightarrow^{\circ} V^{\circ}}\;(\textit{splice-sub})$$

$$\frac{}{F^{\circ}[\mathtt{blame}\ \ell] \longrightarrow^{\circ} \mathtt{blame}\ \ell}\;(\xi\textit{-code-blame}) \qquad \frac{}{\overline{\lambda}(x{:}T)^{\alpha}.\,\mathtt{blame}\ \ell \longrightarrow^{\circ} \mathtt{blame}\ \ell}\;(\xi\textit{-code-}\overline{\lambda}\textit{-blame})$$

$$\frac{}{\sim(\mathtt{blame}\ \ell) \longrightarrow^{\circ} \mathtt{blame}\ \ell}\;(\textit{splice-blame})$$

$$\boxed{e \vdash C^{\circ} \longrightarrow^{\circ} \mathcal{D}^{\circ}}$$

$$\frac{M^{\circ} \longrightarrow^{\circ} N^{\circ}}{e \vdash \langle \Gamma; M^{\circ}; \mu \rangle \longrightarrow^{\circ} \langle \Gamma; N^{\circ}; \mu \rangle}\;(\textit{pure-code})$$

$$\frac{e \vdash \langle \Gamma; M^{\circ}; \mu \rangle \longrightarrow^{\circ} \langle \Gamma'; N^{\circ}; \mu' \rangle}{e \vdash \langle \Gamma; F^{\circ}[M^{\circ}]; \mu \rangle \longrightarrow^{\circ} \langle \Gamma'; F^{\circ}[N^{\circ}]; \mu' \rangle}\;(\xi\textit{-code}) \qquad \frac{\langle \Gamma; M; \mu \rangle \longrightarrow^{m} \langle \Gamma'; N; \mu' \rangle}{e \vdash \langle \Gamma; {\sim}M; \mu \rangle \longrightarrow^{\circ} \langle \Gamma'; {\sim}N; \mu' \rangle}\;(\xi\textit{-splice})$$

$$\frac{\beta, y \notin \Gamma}{e \vdash \langle \Gamma; \lambda(x{:}T)^{\alpha}.\,N^{\circ}; \mu \rangle \longrightarrow^{\circ} \langle (\Gamma,\ \beta_{\mathrm{glob}}, (e{<:}\beta)_{\mathrm{glob}}, (y{:}T)^{\beta}_{\mathrm{glob}}); \overline{\lambda}(y{:}T)^{\beta}.\,N^{\circ}[\alpha := \beta][x := y]; \mu \rangle}\;(\textit{code-}\lambda)$$

$$\frac{\alpha \vdash \langle \Gamma; N^{\circ}; \mu \rangle \longrightarrow^{\circ} \langle \Gamma'; N^{\circ\prime}; \mu' \rangle}{e \vdash \langle \Gamma; \overline{\lambda}(x{:}T)^{\alpha}.\,N^{\circ}; \mu \rangle \longrightarrow^{\circ} \langle \Gamma'; \overline{\lambda}(x{:}T)^{\alpha}.\,N^{\circ\prime}; \mu' \rangle}\;(\xi\textit{-code-}\overline{\lambda})$$

Fig. 18. Code reduction rules for $\mathrm{CC}^{\alpha,\star}_{\mathrm{Ref}}$

$$\boxed{\Theta \vdash M \longrightarrow^m N}$$

$$\frac{}{\Theta \vdash F[\texttt{blame}\ \ell] \longrightarrow^m \texttt{blame}\ \ell}\ (\xi\text{-meta-blame}) \qquad \frac{}{\Theta \vdash \text{``}\texttt{blame}\ \ell\text{''}e \longrightarrow^m \texttt{blame}\ \ell}\ (\text{quote-blame})$$

$$\frac{}{\Theta \vdash (\lambda x.\, N)\, V \longrightarrow^m N[x := V]}\ (\beta) \qquad \frac{\mathsf{Inert}([\bar{c} \to \bar{d}])}{\Theta \vdash (U\langle[\bar{c} \to \bar{d}]\rangle)\, W \longrightarrow^m (U\ (W\langle\bar{c}\rangle))\langle\bar{d}\rangle}\ (\beta\text{-cast})$$

$$\frac{}{\Theta \vdash (\Lambda\alpha.\, M)[e] \longrightarrow^m M[\alpha := e]}\ (\beta\text{-ec}) \qquad \frac{}{\Theta \vdash (U\langle[\forall\alpha.\, \bar{c}]\rangle)[e] \longrightarrow^m U[e]\langle\bar{c}[\alpha := e]\rangle}\ (\beta\text{-ec-cast})$$

$$\frac{}{\Theta \vdash (e_1 {<:} e_2 {\Rightarrow} M)\bullet \longrightarrow^m M}\ (\beta\text{-sub}) \qquad \frac{}{\Theta \vdash (U\langle[e_1 {<:} e_2 {\Rightarrow} \bar{c}]\rangle)\bullet \longrightarrow^m (U\bullet)\langle\bar{c}\rangle}\ (\beta\text{-sub-cast})$$

$$\frac{\Theta \vdash \bar{c} \longrightarrow \bar{d}}{\Theta \vdash U\langle\bar{c}\rangle \longrightarrow^m U\langle\bar{d}\rangle}\ (\text{cast-reduce}) \qquad \frac{}{\Theta \vdash U\langle\bar{c}\rangle\langle\bar{d}\rangle \longrightarrow^m U\langle\bar{c} \mathbin{+\!\!+} \bar{d}\rangle}\ (\text{cast-compose})$$

$$\frac{}{\Theta \vdash U\langle[\,]\rangle \longrightarrow^m U}\ (\text{cast-id}) \qquad \frac{}{\Theta \vdash U\langle[\bot^{\ell}]\rangle \longrightarrow^m \texttt{blame}\ \ell}\ (\text{cast-blame})$$

$$\frac{\Theta \vdash e_1 {<:} e_2 \qquad \mathsf{Inert}(\bar{c} \mathbin{+\!\!+} [(e_1 {<:} e_2 {\Rightarrow} \star)!])}{\Theta \vdash U\langle\bar{c} \mathbin{+\!\!+} [(e_1 {<:} e_2 {\Rightarrow} \star)!]\rangle\bullet\star^{\ell} \longrightarrow^m (U\langle\bar{c}\rangle)\bullet}\ (\text{sub-cast-success})$$

$$\frac{\Theta \nvdash e_1 {<:} e_2 \qquad \mathsf{Inert}(\bar{c} \mathbin{+\!\!+} [(e_1 {<:} e_2 {\Rightarrow} \star)!])}{\Theta \vdash U\langle\bar{c} \mathbin{+\!\!+} [(e_1 {<:} e_2 {\Rightarrow} \star)!]\rangle\bullet\star^{\ell} \longrightarrow^m \texttt{blame}\ \ell}\ (\text{sub-cast-sub-fail})$$

$$\frac{\mathsf{Inert}(\bar{c} \mathbin{+\!\!+} [G!]) \qquad \forall e_1\, e_2.\ G \neq (e_1 {<:} e_2 {\Rightarrow} \star)}{\Theta \vdash U\langle\bar{c} \mathbin{+\!\!+} [G!]\rangle\bullet\star^{\ell} \longrightarrow^m \texttt{blame}\ \ell}\ (\text{sub-cast-fail})$$

$$\frac{\mathsf{Inert}([\texttt{Ref}\ \bar{c}\ \bar{d}])}{\Theta \vdash !(a\langle[\texttt{Ref}\ \bar{c}\ \bar{d}]\rangle) \longrightarrow^m (!a)\langle\bar{d}\rangle}\ (\text{deref-cast}) \qquad \frac{\mathsf{Inert}([\texttt{Ref}\ \bar{c}\ \bar{d}])}{\Theta \vdash (a\langle[\texttt{Ref}\ \bar{c}\ \bar{d}]\rangle) := V \longrightarrow^m a := (V\langle\bar{c}\rangle)}\ (\text{assign-cast})$$

$$\boxed{C \longrightarrow^m \mathcal{D}}$$

$$\frac{\mathsf{Sub}(\Gamma) \vdash M \longrightarrow^m N}{\langle\Gamma; M; \mu\rangle \longrightarrow^m \langle\Gamma; N; \mu\rangle}\ (\text{pure-meta}) \qquad \frac{\langle\Gamma; M; \mu\rangle \longrightarrow^m \langle\Gamma'; N; \mu'\rangle}{\langle\Gamma; F[M]; \mu\rangle \longrightarrow^m \langle\Gamma'; F[N]; \mu'\rangle}\ (\xi\text{-meta})$$

$$\frac{e \vdash \langle\Gamma; M^\circ; \mu\rangle \longrightarrow^\circ \langle\Gamma'; N^\circ; \mu'\rangle}{\langle\Gamma; \text{``}M^\circ\text{''}e; \mu\rangle \longrightarrow^m \langle\Gamma'; \text{``}N^\circ\text{''}e; \mu'\rangle}\ (\xi\text{-quote})$$

$$\frac{a \notin \mathrm{dom}(\mu)}{\langle\Gamma; \texttt{ref}\ V; \mu\rangle \longrightarrow^m \langle\Gamma; a; \mu[a \mapsto V]\rangle}\ (\text{ref}) \qquad \frac{\mu(a) = V}{\langle\Gamma; !a; \mu\rangle \longrightarrow^m \langle\Gamma; V; \mu\rangle}\ (\text{deref})$$

$$\frac{}{\langle\Gamma; a := V; \mu\rangle \longrightarrow^m \langle\Gamma; (); \mu[a \mapsto V]\rangle}\ (\text{assign})$$

Fig. 19. Meta reduction rules for $\mathrm{CC}^{\alpha,\star}_{\mathsf{Ref}}$

$$\boxed{\text{coerce}^{e}(\hat{e}_1, \hat{e}_2, \ell) = \bar{c}^{e}}$$

$$\begin{aligned}
\text{coerce}^{e}(\star, \star, \ell) &= [\star{\uparrow}\star] \\
\text{coerce}^{e}(\star, e, \ell) &= [e?^{\ell}] \\
\text{coerce}^{e}(e, \star, \ell) &= [e!] \\
\text{coerce}^{e}(e_1, e_2, \ell) &= [e_1{\uparrow}e_2]
\end{aligned}$$

$$\boxed{\text{coerce}^{\circ}(\hat{T}_1, \hat{T}_2, \ell) = \bar{c}^{\circ}}$$

$$\begin{aligned}
\text{coerce}^{\circ}(\star, \star, \ell) &= [] \\
\text{coerce}^{\circ}(\star, \iota, \ell) &= [\iota?^{\ell}] \\
\text{coerce}^{\circ}(\iota, \star, \ell) &= [\iota!] \\
\text{coerce}^{\circ}(\iota, \iota, \ell) &= [] \\
\text{coerce}^{\circ}(\star, \hat{T}_1{\rightarrow}\hat{T}_2, \ell) &= [\star{\rightarrow}\star?^{\ell}] \mathbin{+\!\!+} [\text{coerce}^{\circ}(\hat{T}_1, \star, \ell){\rightarrow}\text{coerce}^{\circ}(\star, \hat{T}_2, \ell)] \\
\text{coerce}^{\circ}(\hat{T}_1{\rightarrow}\hat{T}_2, \star, \ell) &= [\text{coerce}^{\circ}(\star, \hat{T}_1, \ell){\rightarrow}\text{coerce}^{\circ}(\hat{T}_2, \star, \ell)] \mathbin{+\!\!+} [\star{\rightarrow}\star!] \\
\text{coerce}^{\circ}(\hat{T}_1{\rightarrow}\hat{T}_2, \hat{T}_1'{\rightarrow}\hat{T}_2', \ell) &= [\text{coerce}^{\circ}(\hat{T}_1', \hat{T}_1, \ell){\rightarrow}\text{coerce}^{\circ}(\hat{T}_2, \hat{T}_2', \ell)]
\end{aligned}$$

$$\boxed{\text{coerce}(\hat{A}, \hat{B}, \ell) = \bar{c}}$$

$$\begin{aligned}
&\vdots \qquad \text{(base type and function cases analogous to coerce}^{\circ}\text{)} \\
\text{coerce}(\star, \texttt{Ref}\ \hat{A}, \ell) &= [\texttt{Ref}\ \star?^{\ell}] \mathbin{+\!\!+} [\texttt{Ref}\ \text{coerce}(\hat{A}, \star, \ell)\ \text{coerce}(\star, \hat{A}, \ell)] \\
\text{coerce}(\texttt{Ref}\ \hat{A}, \star, \ell) &= [\texttt{Ref}\ \text{coerce}(\star, \hat{A}, \ell)\ \text{coerce}(\hat{A}, \star, \ell)] \mathbin{+\!\!+} [\texttt{Ref}\ \star!] \\
\text{coerce}(\texttt{Ref}\ \hat{A}, \texttt{Ref}\ \hat{B}, \ell) &= [\texttt{Ref}\ \text{coerce}(\hat{B}, \hat{A}, \ell)\ \text{coerce}(\hat{A}, \hat{B}, \ell)] \\
\text{coerce}(\star, \text{``}\hat{T}\text{''}\hat{e}, \ell) &= [\text{``}\star\text{''}\star?^{\ell}] \mathbin{+\!\!+} [\text{``}\text{coerce}^{\circ}(\star, \hat{T}, \ell)\text{''}\text{coerce}^{e}(\star, \hat{e}, \ell)] \\
\text{coerce}(\text{``}\hat{T}\text{''}\hat{e}, \star, \ell) &= [\text{``}\text{coerce}^{\circ}(\hat{T}, \star, \ell)\text{''}\text{coerce}^{e}(\hat{e}, \star, \ell)] \mathbin{+\!\!+} [\text{``}\star\text{''}\star!] \\
\text{coerce}(\text{``}\hat{T}_1\text{''}\hat{e}_1, \text{``}\hat{T}_2\text{''}\hat{e}_2, \ell) &= [\text{``}\text{coerce}^{\circ}(\hat{T}_1, \hat{T}_2, \ell)\text{''}\text{coerce}^{e}(\hat{e}_1, \hat{e}_2, \ell)] \\
\text{coerce}(\star, \forall\alpha.\hat{A}, \ell) &= [\forall\alpha.\star?^{\ell}] \mathbin{+\!\!+} [\forall\alpha.\text{coerce}(\star, \hat{A}, \ell)] \\
\text{coerce}(\forall\alpha.\hat{A}, \star, \ell) &= [\forall\alpha.\text{coerce}(\hat{A}, \star, \ell)] \mathbin{+\!\!+} [\forall\alpha.\star!] \\
\text{coerce}(\forall\alpha.\hat{A}, \forall\alpha.\hat{B}, \ell) &= [\forall\alpha.\text{coerce}(\hat{A}, \hat{B}, \ell)] \\
\text{coerce}(\star, e_1{<:}e_2{\Rightarrow}\hat{A}, \ell) &= [e_1{<:}e_2{\Rightarrow}\star?^{\ell}] \mathbin{+\!\!+} [e_1{<:}e_2{\Rightarrow}\text{coerce}(\star, \hat{A}, \ell)] \\
\text{coerce}(e_1{<:}e_2{\Rightarrow}\hat{A}, \star, \ell) &= [e_1{<:}e_2{\Rightarrow}\text{coerce}(\hat{A}, \star, \ell)] \mathbin{+\!\!+} [e_1{<:}e_2{\Rightarrow}\star!] \\
\text{coerce}(e_1{<:}e_2{\Rightarrow}\hat{A}, e_1{<:}e_2{\Rightarrow}\hat{B}, \ell) &= [e_1{<:}e_2{\Rightarrow}\text{coerce}(\hat{A}, \hat{B}, \ell)]
\end{aligned}$$

Fig. 20. Generating coercions from source and target

$$\boxed{\Gamma \vdash M : \hat{A} \leadsto M}$$

$$\frac{k : \iota}{\Gamma \vdash k : \iota \leadsto k} \qquad \frac{x : \hat{A} \in \Gamma}{\Gamma \vdash x : \hat{A} \leadsto x} \qquad \frac{\Gamma \vdash \hat{A} \qquad (\Gamma, x : \hat{A}) \vdash N : \hat{B} \leadsto N'}{\Gamma \vdash \lambda(x : \hat{A}) N : \hat{A} \to \hat{B} \leadsto \lambda(x : \hat{A}) N'}$$

$$\frac{\Gamma \vdash L : \hat{A}_1 \to \hat{A}_2 \leadsto L' \qquad \Gamma \vdash M : \hat{B} \leadsto M'}{\Gamma \vdash (L\ M)^{\ell} : \hat{A}_2 \leadsto L'(M'\langle \mathsf{coerce}(\hat{B}, \hat{A}_1, \ell)\rangle)}$$

$$\frac{\Gamma \vdash L : \star \leadsto L' \qquad \Gamma \vdash M : \hat{A} \leadsto M'}{\Gamma \vdash (L\ M)^{\ell} : \star \leadsto (L'\langle \mathsf{coerce}(\star, (\star \to \star), \ell)\rangle)(M'\langle \mathsf{coerce}(\hat{A}, \star, \ell)\rangle)} \qquad \frac{\Gamma, \alpha_{\mathsf{lex}} \vdash N : \hat{A} \leadsto N'}{\Gamma \vdash \Lambda\alpha.N : \forall\alpha.\hat{A} \leadsto \Lambda\alpha.N'}$$

$$\frac{\Gamma \vdash M : \forall\alpha.\hat{A} \leadsto M' \qquad \Gamma \vdash e}{\Gamma \vdash M[e]^{\ell} : \hat{A}[\alpha := e] \leadsto M'[e]} \qquad \frac{\Gamma \vdash M : \star \leadsto M' \qquad \Gamma \vdash e}{\Gamma \vdash M[e]^{\ell} : \star \leadsto (M'\langle \mathsf{coerce}(\star, (\forall\beta.\star), \ell)\rangle)[e]}$$

$$\frac{\Gamma \vdash e_1 \qquad \Gamma \vdash e_2 \qquad \Gamma, (e_1 {<:} e_2)_{\mathsf{lex}} \vdash M : \hat{A} \leadsto M'}{\Gamma \vdash (e_1 <: e_2 \Rightarrow M) : (e_1 <: e_2 \Rightarrow \hat{A}) \leadsto (e_1 <: e_2 \Rightarrow M')}$$

$$\frac{\Gamma \vdash M : (e_1 <: e_2 \Rightarrow \hat{A}) \leadsto M' \qquad \mathsf{Sub}(\Gamma) \vdash e_1 <: e_2}{\Gamma \vdash M \bullet^{\ell} : \hat{A} \leadsto M' \bullet} \qquad \frac{\Gamma \vdash M : \star \leadsto M'}{\Gamma \vdash M \bullet^{\ell} : \star \leadsto (M' \bullet^{\ell}_{\star})}$$

$$\frac{\Gamma; e \vdash M^{\circ} \Rightarrow T \leadsto M^{\circ\prime} \qquad \Gamma \vdash e}{\Gamma \vdash \text{``}M^{\circ}\text{''}e : \text{``}T\text{''}e \leadsto \text{``}M^{\circ\prime}\text{''}e} \qquad \frac{\Gamma \vdash M : \hat{A} \leadsto M'}{\Gamma \vdash \mathsf{ref}\ M : \mathsf{Ref}\ \hat{A} \leadsto \mathsf{ref}\ M'}$$

$$\frac{\Gamma \vdash L : \mathsf{Ref}\ \hat{A} \leadsto L' \qquad \Gamma \vdash M : \hat{B} \leadsto M'}{\Gamma \vdash L :=^{\ell} M : \mathsf{Unit} \leadsto L' := (M'\langle \mathsf{coerce}(\hat{B}, \hat{A}, \ell)\rangle)}$$

$$\frac{\Gamma \vdash L : \star \leadsto L' \qquad \Gamma \vdash M : \hat{A} \leadsto M'}{\Gamma \vdash L :=^{\ell} M : \mathsf{Unit} \leadsto (L'\langle \mathsf{coerce}(\star, (\mathsf{Ref}\ \star), \ell)\rangle) := (M'\langle \mathsf{coerce}(\hat{A}, \star, \ell)\rangle)}$$

$$\frac{\Gamma \vdash M : \mathsf{Ref}\ \hat{A} \leadsto M'}{\Gamma \vdash {!}M : \hat{A} \leadsto {!}M'} \qquad \frac{\Gamma \vdash M : \star \leadsto M'}{\Gamma \vdash {!}^{\ell} M : \star \leadsto {!}(M'\langle \mathsf{coerce}(\star, \mathsf{Ref}\ \star, \ell)\rangle)}$$

$$\boxed{\Gamma; e \vdash M^{\circ} \Rightarrow T \leadsto M^{\circ}}$$

$$\frac{\begin{array}{c}(x{:}T)^{\alpha}_{\rho} \in \Gamma \\ \mathsf{Sub}(\Gamma) \vdash \alpha <: e\end{array}}{\Gamma; e \vdash x \Rightarrow T \leadsto x} \qquad \frac{\begin{array}{c}\Gamma; e \vdash L^{\circ} \Rightarrow T_1 \to T_2 \leadsto L^{\circ\prime} \\ \Gamma; e \vdash M^{\circ} \Leftarrow T_1 \leadsto M^{\circ\prime}\end{array}}{\Gamma; e \vdash L^{\circ}\ M^{\circ} \Rightarrow T_2 \leadsto L^{\circ\prime}\ M^{\circ\prime}}$$

$$\frac{\Gamma, \alpha_{\mathsf{lex}}, (e <: \alpha)_{\mathsf{lex}}, (x{:}T_1)^{\alpha}_{\mathsf{lex}}; \alpha \vdash N^{\circ} \Rightarrow T_2 \leadsto N^{\circ\prime} \qquad \alpha \notin \Gamma}{\Gamma; e \vdash \lambda(x{:}T_1)^{\alpha}.N^{\circ} \Rightarrow T_1 \to T_2 \leadsto \lambda(x{:}T_1)^{\alpha}.N^{\circ\prime}}$$

$$\boxed{\Gamma; e \vdash M^{\circ} \Leftarrow T \leadsto M^{\circ}}$$

$$\frac{\Gamma \vdash M : \hat{A} \leadsto M'}{\Gamma; e \vdash {\sim}^{\ell} M \Leftarrow T \leadsto {\sim}M'\langle \mathsf{coerce}(\hat{A}, \text{``}T\text{''}e, \ell)\rangle} \qquad \frac{\Gamma; e \vdash M^{\circ} \Rightarrow T \leadsto M'}{\Gamma; e \vdash M^{\circ} \Leftarrow T \leadsto M'}$$

Fig. 21. Elaboration from $\lambda^{\alpha,\star}_{\mathsf{Ref}}$ to $\mathsf{CC}^{\alpha,\star}_{\mathsf{Ref}}$

$$
\begin{array}{rcl}
\mathsf{EC}(\Gamma) & \triangleq & \text{the ordered subsequence of entries } \alpha_\rho \text{ in } \Gamma \\
\mathsf{Code}(\Gamma) & \triangleq & \text{the ordered subsequence of entries } (x{:}T)^\alpha_\rho \text{ in } \Gamma \\
\mathsf{Sub}(\Gamma) & \triangleq & \text{the ordered subsequence of entries } (e_1{<:}e_2)_\rho \text{ in } \Gamma \\
\mathsf{Ctx}(\langle \Gamma; M; \mu \rangle) & \triangleq & \Gamma \\
\mathsf{Ctx}(\langle \Gamma; M^\circ; \mu \rangle) & \triangleq & \Gamma
\end{array}
$$

$$
\begin{array}{rcl}
\mathsf{EC}(\Gamma_1) \subseteq \mathsf{EC}(\Gamma_2) & \triangleq & \forall \alpha, \rho.\ \text{if } \alpha_\rho \in \mathsf{EC}(\Gamma_1) \text{ then } \alpha_\rho \in \mathsf{EC}(\Gamma_2) \\
\mathsf{Code}(\Gamma_1) \subseteq \mathsf{Code}(\Gamma_2) & \triangleq & \forall x, T, \alpha, \rho.\ \text{if } (x{:}T)^\alpha_\rho \in \mathsf{Code}(\Gamma_1) \text{ then } (x{:}T)^\alpha_\rho \in \mathsf{Code}(\Gamma_2) \\
\mathsf{Sub}(\Gamma_1) \subseteq \mathsf{Sub}(\Gamma_2) & \triangleq & \quad \forall e_1, e_2.\ \text{if } \mathsf{Sub}(\Gamma_1) \vdash e_1{<:}e_2 \text{ then } \mathsf{Sub}(\Gamma_2) \vdash e_1{<:}e_2 \\
 & & \wedge\, \forall e, \alpha.\ \text{if } (e{<:}\alpha)_{\mathsf{glob}} \in \mathsf{Sub}(\Gamma_1) \text{ then } (e{<:}\alpha)_{\mathsf{glob}} \in \mathsf{Sub}(\Gamma_2) \\
\Sigma_1 \subseteq \Sigma_2 & \triangleq & \forall a, \hat{A}.\ \text{if } a : \hat{A} \in \Sigma_1 \text{ then } a : \hat{A} \in \Sigma_2 \\
\Gamma_1 \subseteq \Gamma_2 & \triangleq & \mathsf{EC}(\Gamma_1) \subseteq \mathsf{EC}(\Gamma_2) \wedge \mathsf{Code}(\Gamma_1) \subseteq \mathsf{Code}(\Gamma_2) \wedge \mathsf{Sub}(\Gamma_1) \subseteq \mathsf{Sub}(\Gamma_2).
\end{array}
$$

Fig. 22. Projection functions and inclusion relations used by preservation for $\mathsf{CC}^{\alpha,\star}_{\mathsf{Ref}}$

```
1  let rec sort : Array Int → Int → Int → Unit =
2    λ (a : Array ★) (lo : Int) (hi : Int).
3      if lo < hi then
4        let pivot : Int = partition a lo hi in
5        sort a lo (pivot - 1);
6        sort a (pivot + 1) hi
7
8  let swap : Array Int → Int → Int → Unit =
9    λ (a : Array Int) (i : Int) (j : Int).
10     let temp : Int = a.(i) in
11     a.(i) <- a.(j);
12     a.(j) <- temp
13
14 let partition : Array Int → Int → Int → Int =
15   λ (a : Array Int) (l : Int) (h : Int).
16     let p : Int = a.(h) in
17     let i : Ref Int = ref (h - 1) in
18     for j = l to h do
19       if a.(j) <= p then
20         i := !i + 1;
21         swap a !i j
22     done;
23     swap a (!i + 1) h;
24     !i + 1
```

Fig. 23. Gradual quicksort program with potential catastrophic slowdown

$$
\begin{array}{lrcl}
\text{meta hypercoercions} & c, d & ::= & h; m; t \\
\text{meta head coercions} & h & ::= & H?^{\ell} \mid \texttt{id} \\
\text{meta middle coercions} & m & ::= & c \longrightarrow d \mid \texttt{Ref}\ c\ d \mid \text{``}c^{\circ}\text{''}c^{e} \mid \forall\alpha.c \mid e{<:}e \Rightarrow c \mid \texttt{id}\,\star \mid \texttt{id}\,\iota \mid \bot^{\ell} \\
\text{meta tail coercions} & t & ::= & G! \mid \texttt{id}
\end{array}
$$

$\boxed{\Gamma \vdash c : \hat{A} \Rightarrow \hat{B}}$

$$
\frac{\Gamma \vdash h : \hat{A} \Rightarrow \hat{A}' \qquad \Gamma \vdash m : \hat{A}' \Rightarrow \hat{B}' \qquad \Gamma \vdash t : \hat{B}' \Rightarrow \hat{B}}{\Gamma \vdash h; m; t : \hat{A} \Rightarrow \hat{B}}
$$

$\boxed{\Theta \vdash c_1 \fatsemi c_2 = d}$

$$
\begin{array}{lllcl}
\Theta \vdash & h_1; \bot^{\ell}; t_1 & \fatsemi & h_2; m_2; t_2 & = & h_1; \bot^{\ell}; t_2 \\
\Theta \vdash & h_1; m_1; t_1 & \fatsemi & h_2; \bot^{\ell}; t_2 & = & \text{match } t_1 \fatsemi h_2 \text{ with} \\
 & & & & & \quad \text{case } \bot^{\ell'} \ \Rightarrow\ h_1; \bot^{\ell'}; t_2 \\
 & & & & & \quad \text{case } _ \ \Rightarrow\ h_1; \bot^{\ell}; t_2 \\
\Theta \vdash & h_1; m_1; t_1 & \fatsemi & h_2; m_2; t_2 & = & \text{match } t_1 \fatsemi h_2 \text{ with} \\
 & & & & & \quad \text{case } \texttt{id} \ \Rightarrow\ h_1; (\Theta \vdash m_1 \fatsemi m_2); t_2 \\
 & & & & & \quad \text{case } G! \ \Rightarrow\ h_1; m_1; G! \\
 & & & & & \quad \text{case } H?^{\ell} \ \Rightarrow\ H?^{\ell}; m_2; t_2 \\
 & & & & & \quad \text{case } \bot^{\ell} \ \Rightarrow\ h_1; \bot^{\ell}; t_2
\end{array}
$$

$\boxed{t \fatsemi h = t \cup h \cup \bot^{\ell}}$

$$
\begin{array}{rcll}
t & \fatsemi & \texttt{id} & = t \\
\texttt{id} & \fatsemi & h & = h \\
G! & \fatsemi & G?^{\ell} & = \texttt{id} \\
G! & \fatsemi & H?^{\ell} & = \bot^{\ell} \quad , \text{if } G \neq H
\end{array}
$$

$\boxed{\Theta \vdash m_1 \fatsemi m_2 = m}$

$$
\begin{array}{llclcl}
\Theta \vdash & m_1 & \fatsemi & \texttt{id} & = & m_1 \\
\Theta \vdash & \texttt{id} & \fatsemi & m_2 & = & m_2 \\
\Theta \vdash & c_1 \longrightarrow d_1 & \fatsemi & c_2 \longrightarrow d_2 & = & (\Theta \vdash c_2 \fatsemi c_1) \longrightarrow (\Theta \vdash d_1 \fatsemi d_2) \\
\Theta \vdash & \text{``}c_1^{\circ}\text{''}c_1^{e} & \fatsemi & \text{``}c_2^{\circ}\text{''}c_2^{e} & = & \text{match } (c_1^{\circ} \fatsemi c_2^{\circ})\ (\Theta \vdash c_1^{e} \fatsemi c_2^{e}) \text{ with} \\
 & & & & & \quad \text{case } (h^{\circ}; \bot^{\ell}; t^{\circ}) \quad _ \quad \Rightarrow\ \bot^{\ell} \\
 & & & & & \quad \text{case } _ \quad (h^{e}; \bot^{\ell}; t^{e}) \quad \Rightarrow\ \bot^{\ell} \\
 & & & & & \quad \text{case } c^{\circ} \quad c^{e} \quad \Rightarrow\ \text{``}c^{\circ}\text{''}c^{e} \\
\Theta \vdash & \texttt{Ref}\ c_1\ d_1 & \fatsemi & \texttt{Ref}\ c_2\ d_2 & = & \texttt{Ref}\ (\Theta \vdash c_2 \fatsemi c_1)\ (\Theta \vdash d_1 \fatsemi d_2) \\
\Theta \vdash & \forall\alpha.c & \fatsemi & \forall\alpha.d & = & \forall\alpha.(\Theta \vdash c \fatsemi d) \\
\Theta \vdash & e_1{<:}e_2 \Rightarrow c & \fatsemi & e_1{<:}e_2 \Rightarrow d & = & e_1{<:}e_2 \Rightarrow (\Theta, e_1{<:}e_2 \vdash c \fatsemi d) \\
\Theta \vdash & m_1 & \fatsemi & \bot^{\ell} & = & \bot^{\ell} \\
\Theta \vdash & \bot^{\ell} & \fatsemi & m_2 & = & \bot^{\ell}
\end{array}
$$

Fig. 24. Meta hypercoercions: syntax, typing, and composition

$$
\begin{array}{lrcl}
\text{EC hypercoercions} & c^e, d^e & ::= & h^e; m^e; t^e \\
\text{EC head coercions} & h^e & ::= & e?^{\ell} \mid \mathtt{id} \\
\text{EC middle coercions} & m^e & ::= & e \uparrow e \mid \bot^{\ell} \\
\text{EC tail coercions} & t & ::= & e! \mid \mathtt{id}
\end{array}
$$

$$\boxed{\Gamma \vdash c : \hat{e_1} \Rightarrow \hat{e_2}}$$

$\vdots$ (analogous to meta hypercoercion typing)

$$\boxed{\Theta \vdash c_1^e \fatsemi c_2^e = d^e}$$

$$
\begin{array}{llcll}
\Theta \vdash & h_1^e; \bot^{\ell}; t_1^e & \fatsemi & h_2^e; m_2^e; t_2^e & = h_1^e; \bot^{\ell}; t_2^e \\
\Theta \vdash & h_1^e; m_1^e; t_1^e & \fatsemi & h_2^e; \bot^{\ell}; t_2^e & = \text{match } t_1^e \fatsemi h_2^e \text{ with} \\
& & & & \quad \text{case } \bot^{\ell'} \;\Rightarrow\; h_1^e; \bot^{\ell'}; t_2^e \\
& & & & \quad \text{case } _ \;\Rightarrow\; h_1^e; \bot^{\ell}; t_2^e \\
\Theta \vdash & h_1^e; m_1^e; t_1^e & \fatsemi & h_2^e; m_2^e; t_2^e & = \text{match } (\Theta \vdash t_1^e \fatsemi h_2^e) \text{ with} \\
& & & & \quad \text{case } e_1 \uparrow e_2 \;\Rightarrow\; h_1^e; (m_1^e \fatsemi (e_1 \uparrow e_2 \fatsemi m_2^e)); t_2^e \\
& & & & \quad \text{case } \mathtt{id} \;\Rightarrow\; h_1^e; (m_1^e \fatsemi m_2^e); t_2^e \\
& & & & \quad \text{case } e! \;\Rightarrow\; h_1^e; m_1^e; e! \\
& & & & \quad \text{case } e?^{\ell} \;\Rightarrow\; e?^{\ell}; m_2^e; t_2^e \\
& & & & \quad \text{case } \bot^{\ell} \;\Rightarrow\; h_1^e; \bot^{\ell}; t_2^e
\end{array}
$$

$$\boxed{\Theta \vdash t^e \fatsemi h^e = t^e \cup h^e \cup m^e}$$

$$
\begin{array}{lcll}
t^e & \fatsemi & \mathtt{id} & = t^e \\
\mathtt{id} & \fatsemi & h^e & = h^e \\
e_1! & \fatsemi & e_2?^{\ell} & = e_1 \uparrow e_2 \quad , \text{if } \Theta \vdash e_1 <: e_2 \\
e_1! & \fatsemi & e_2?^{\ell} & = \bot^{\ell} \quad , \text{if } \Theta \vdash e_1 \not<: e_2
\end{array}
$$

$$\boxed{m_1^e \fatsemi m_2^e = m^e}$$

$$
\begin{array}{lcccl}
e_1 \uparrow e_2 & \fatsemi & e_2 \uparrow e_3 & = & e_1 \uparrow e_3 \\
m_1^e & \fatsemi & \bot^{\ell} & = & \bot^{\ell} \\
\bot^{\ell} & \fatsemi & m_2^e & = & \bot^{\ell}
\end{array}
$$

Fig. 25. EC hypercoercions: syntax, typing and composition

$$
\begin{array}{llll}
\text{code hypercoercions} & c^\circ, d^\circ & ::= & h^\circ; m^\circ; t^\circ \\
\text{code head coercions} & h^\circ & ::= & H^\circ?^\ell \mid \mathtt{id} \\
\text{code middle coercions} & m^\circ & ::= & c^\circ \longrightarrow d^\circ \mid \mathtt{id}\,\star \mid \mathtt{id}\,\iota \mid \bot^\ell \\
\text{code tail coercions} & t^\circ & ::= & G^\circ! \mid \mathtt{id}
\end{array}
$$

$\boxed{\vdash c : \hat{T}_1 \Rightarrow \hat{T}_2}$

$\vdots$ (analogous to meta hypercoercion typing)

$\boxed{c_1^\circ \fatsemi c_2^\circ = d^\circ}$

$\vdots$ (analogous to meta hypercoercion composition)

$\boxed{t^\circ \fatsemi h^\circ = t^\circ \cup h^\circ \cup \bot^\ell}$

$\vdots$ (analogous to meta head-tail coercion composition)

$\boxed{m_1^\circ \fatsemi m_2^\circ = m^\circ}$

$$
\begin{array}{lllllll}
m_1^\circ & \fatsemi & \mathtt{id} & = & m_1^\circ & & \\
\mathtt{id} & \fatsemi & m_2^\circ & = & m_2^\circ & & \\
c_1^\circ \longrightarrow d_1^\circ & \fatsemi & c_2^\circ \longrightarrow d_2^\circ & = & \multicolumn{3}{l}{\text{match}\,(c_2^\circ \fatsemi c_1^\circ)\,(d_1^\circ \fatsemi d_2^\circ)\,\text{with}} \\
 & & & & \text{case}\quad (h^\circ; \bot^\ell; t^\circ) & _ & \Rightarrow \bot^\ell \\
 & & & & \text{case}\quad _ & (h^\circ; \bot^\ell; t^\circ) & \Rightarrow \bot^\ell \\
 & & & & \text{case}\quad c^\circ & d^\circ & \Rightarrow c^\circ \longrightarrow d^\circ \\
m_1^\circ & \fatsemi & \bot^\ell & = & \bot^\ell & & \\
\bot^\ell & \fatsemi & m_2^\circ & = & \bot^\ell & &
\end{array}
$$

Fig. 26. Code hypercoercions: syntax, typing and composition

$$\boxed{\mathsf{coerce}^e(\hat{e}_1, \hat{e}_2, \ell) = c^e}$$

$$
\begin{aligned}
\mathsf{coerce}^e(\star, \star, \ell) &= \mathsf{id}; \star{\uparrow}\star; \mathsf{id} \\
\mathsf{coerce}^e(\star, e, \ell) &= e?^{\ell}; e{\uparrow}e; \mathsf{id} \\
\mathsf{coerce}^e(e, \star, \ell) &= \mathsf{id}; e{\uparrow}e; e! \\
\mathsf{coerce}^e(e_1, e_2, \ell) &= \mathsf{id}; e_1{\uparrow}e_2; \mathsf{id}
\end{aligned}
$$

$$\boxed{\mathsf{coerce}^\circ(\hat{T}_1, \hat{T}_2, \ell) = c^\circ}$$

$$
\begin{aligned}
\mathsf{coerce}^\circ(\star, \star, \ell) &= \mathsf{id}; \mathsf{id}; \mathsf{id} \\
\mathsf{coerce}^\circ(\star, \iota, \ell) &= \iota?^{\ell}; \mathsf{id}; \mathsf{id} \\
\mathsf{coerce}^\circ(\iota, \star, \ell) &= \mathsf{id}; \mathsf{id}; \iota! \\
\mathsf{coerce}^\circ(\iota, \iota, \ell) &= \mathsf{id}; \mathsf{id}; \mathsf{id} \\
\mathsf{coerce}^\circ(\star, \hat{T}_1 {\to} \hat{T}_2, \ell) &= \star{\to}\star?^{\ell}; \mathsf{coerce}^\circ(\hat{T}_1, \star, \ell) {\to} \mathsf{coerce}^\circ(\star, \hat{T}_2, \ell); \mathsf{id} \\
\mathsf{coerce}^\circ(\hat{T}_1 {\to} \hat{T}_2, \star, \ell) &= \mathsf{id}; \mathsf{coerce}^\circ(\star, \hat{T}_1, \ell) {\to} \mathsf{coerce}^\circ(\hat{T}_2, \star, \ell); \star{\to}\star! \\
\mathsf{coerce}^\circ(\hat{T}_1 {\to} \hat{T}_2, \hat{T}_1' {\to} \hat{T}_2', \ell) &= \mathsf{id}; \mathsf{coerce}^\circ(\hat{T}_1', \hat{T}_1, \ell) {\to} \mathsf{coerce}^\circ(\hat{T}_2, \hat{T}_2', \ell); \mathsf{id}
\end{aligned}
$$

$$\boxed{\mathsf{coerce}(\hat{A}, \hat{B}, \ell) = c}$$

$$
\begin{aligned}
&\vdots \qquad \text{(base type and function cases analogous to } \mathsf{coerce}^\circ\text{)} \\
\mathsf{coerce}(\star, \mathsf{Ref}\ \hat{A}, \ell) &= \mathsf{Ref}\ \star?^{\ell}; \mathsf{Ref}\ \mathsf{coerce}(\hat{A}, \star, \ell)\ \mathsf{coerce}(\star, \hat{A}, \ell); \mathsf{id} \\
\mathsf{coerce}(\mathsf{Ref}\ \hat{A}, \star, \ell) &= \mathsf{id}; \mathsf{Ref}\ \mathsf{coerce}(\star, \hat{A}, \ell)\ \mathsf{coerce}(\hat{A}, \star, \ell); \mathsf{Ref}\ \star! \\
\mathsf{coerce}(\mathsf{Ref}\ \hat{A}, \mathsf{Ref}\ \hat{B}, \ell) &= \mathsf{id}; \mathsf{Ref}\ \mathsf{coerce}(\hat{B}, \hat{A}, \ell)\ \mathsf{coerce}(\hat{A}, \hat{B}, \ell); \mathsf{id} \\
\mathsf{coerce}(\star, \text{``}\hat{T}\text{''}\hat{e}, \ell) &= \text{``}\star\text{''}\star?^{\ell}; \text{``}\mathsf{coerce}^\circ(\star, \hat{T}, \ell)\text{''}\mathsf{coerce}^e(\star, \hat{e}, \ell); \mathsf{id} \\
\mathsf{coerce}(\text{``}\hat{T}\text{''}\hat{e}, \star, \ell) &= \mathsf{id}; \text{``}\mathsf{coerce}^\circ(\hat{T}, \star, \ell)\text{''}\mathsf{coerce}^e(\hat{e}, \star, \ell); \text{``}\star\text{''}\star! \\
\mathsf{coerce}(\text{``}\hat{T}_1\text{''}\hat{e}_1, \text{``}\hat{T}_2\text{''}\hat{e}_2, \ell) &= \mathsf{id}; \text{``}\mathsf{coerce}^\circ(\hat{T}_1, \hat{T}_2, \ell)\text{''}\mathsf{coerce}^e(\hat{e}_1, \hat{e}_2, \ell); \mathsf{id} \\
\mathsf{coerce}(\star, \forall\alpha.\hat{A}, \ell) &= \forall\alpha.\star?^{\ell}; \forall\alpha.\mathsf{coerce}(\star, \hat{A}, \ell); \mathsf{id} \\
\mathsf{coerce}(\forall\alpha.\hat{A}, \star, \ell) &= \mathsf{id}; \forall\alpha.\mathsf{coerce}(\hat{A}, \star, \ell); \forall\alpha.\star! \\
\mathsf{coerce}(\forall\alpha.\hat{A}, \forall\alpha.\hat{B}, \ell) &= \mathsf{id}; \forall\alpha.\mathsf{coerce}(\hat{A}, \hat{B}, \ell); \mathsf{id} \\
\mathsf{coerce}(\star, e_1{<:}e_2{\Rightarrow}\hat{A}, \ell) &= e_1{<:}e_2{\Rightarrow}\star?^{\ell}; e_1{<:}e_2{\Rightarrow}\mathsf{coerce}(\star, \hat{A}, \ell); \mathsf{id} \\
\mathsf{coerce}(e_1{<:}e_2{\Rightarrow}\hat{A}, \star, \ell) &= \mathsf{id}; e_1{<:}e_2{\Rightarrow}\mathsf{coerce}(\hat{A}, \star, \ell); e_1{<:}e_2{\Rightarrow}\star! \\
\mathsf{coerce}(e_1{<:}e_2{\Rightarrow}\hat{A}, e_1{<:}e_2{\Rightarrow}\hat{B}, \ell) &= \mathsf{id}; e_1{<:}e_2{\Rightarrow}\mathsf{coerce}(\hat{A}, \hat{B}, \ell); \mathsf{id}
\end{aligned}
$$

Fig. 27. Generating hypercoercions from source and target

$\boxed{\text{Updated syntax for } V, F}$

$$
\begin{array}{lrcl}
\text{meta values} & V, W & ::= & U \mid U\langle c_i \rangle \\
\text{meta frame} & F & ::= & \square\, M \mid V\, \square \mid \mathtt{ref}\, \square \mid \square := M \mid V := \square \mid !\square \mid \square[e] \mid \square\, \bullet \mid \square\, \bullet\star^{\ell} \\
\text{inert coercions} & c_i & ::= & (\mathtt{id}; m_1; G!)^{\dagger_1} \mid (\mathtt{id}; m_2; \mathtt{id})^{\dagger_2} \\
& & & {}^{\dagger_1}\neg\mathsf{Blame}(m_1) \quad {}^{\dagger_2}\neg(\mathsf{Id}(m_2) \wedge \mathsf{Blame}(m_2))
\end{array}
$$

$$
\mathsf{Id}(m) \triangleq (m = \mathtt{id}\,\star) \vee (m = \mathtt{id}\,\iota) \qquad \mathsf{Blame}(m) \triangleq \exists \ell.\ m = \bot^{\ell}
$$

$$
\text{ctx} ::= \text{Any} \mid \text{NonCast}
$$

$\boxed{M \longrightarrow^{\mathrm{A}} N}$

$$
\begin{array}{rclrcl}
(\lambda(x{:}A)\, N)V & \longrightarrow^{\mathrm{A}} & N[x := V] & (U\langle \mathtt{id}; c \rightarrow d; \mathtt{id}\rangle)\, W & \longrightarrow^{\mathrm{A}} & (U\ (W\langle c \rangle))\langle d \rangle \\
(\Lambda\alpha.N)[e] & \longrightarrow^{\mathrm{A}} & N[\alpha := e] & U\langle \mathtt{id}; \mathtt{Ref}\ c\ d; \mathtt{id}\rangle := V & \longrightarrow^{\mathrm{A}} & U := (V\langle c \rangle) \\
(e_1{<:}e_2 \Rightarrow N)\, \bullet & \longrightarrow^{\mathrm{A}} & N & !(U\langle \mathtt{id}; \mathtt{Ref}\ c\ d; \mathtt{id}\rangle) & \longrightarrow^{\mathrm{A}} & (!U)\langle d \rangle \\
F[\mathtt{blame}\ \ell] & \longrightarrow^{\mathrm{A}} & \mathtt{blame}\ \ell & U\langle \mathtt{id}; \forall\alpha.c; \mathtt{id}\rangle[e] & \longrightarrow^{\mathrm{A}} & U[e]\langle c[\alpha := e] \rangle \\
\text{``}{\sim}(\mathtt{blame}\ \ell)\text{''}e & \longrightarrow^{\mathrm{A}} & \mathtt{blame}\ \ell & U\langle \mathtt{id}; e_1{<:}e_2 \Rightarrow c; \mathtt{id}\rangle\, \bullet & \longrightarrow^{\mathrm{A}} & (U\ \bullet)\langle c \rangle
\end{array}
$$

$\boxed{M \longrightarrow^{\mathrm{NC}} N}$

$$
U\langle \mathtt{id}; \mathtt{id}; \mathtt{id}\rangle \longrightarrow^{\mathrm{NC}} U \qquad\qquad U\langle \mathtt{id}; \bot^{\ell}; t\rangle \longrightarrow^{\mathrm{NC}} \mathtt{blame}\ \ell
$$

$\boxed{\text{ctx} \vdash \mathcal{C} \longrightarrow^{m} \mathcal{D}}$

$$
\begin{array}{rcll}
\text{NonCast} \vdash \langle \Gamma; M\langle c \rangle\langle d \rangle; \mu \rangle & \longrightarrow^{m} & \langle \Gamma; M\langle \mathrm{Sub}(\Gamma) \vdash c \,\mathbin{;}\, d \rangle; \mu \rangle & \\
\text{Any} \vdash \langle \Gamma_1; \text{``}M^{\circ}\text{''}e; \mu_1 \rangle & \longrightarrow^{m} & \langle \Gamma_2; \text{``}N^{\circ}\text{''}e; \mu_2 \rangle & \text{if } \langle \Gamma_1; M^{\circ}; \mu_1 \rangle \mid e \longrightarrow^{\circ} \langle \Gamma_2; N^{\circ}; \mu_2 \rangle \\
\text{Any} \vdash \langle \Gamma; \text{``}V^{\circ}\text{''}e; \mu \rangle & \longrightarrow^{m} & \langle \Gamma; \text{`}V^{\circ}\text{'}e; \mu \rangle & \\
\text{NonCast} \vdash \langle \Gamma; M; \mu \rangle & \longrightarrow^{m} & \langle \Gamma; N; \mu \rangle & \text{if } M \longrightarrow^{\mathrm{NC}} N \\
\text{Any} \vdash \langle \Gamma; M; \mu \rangle & \longrightarrow^{m} & \langle \Gamma; N; \mu \rangle & \text{if } M \longrightarrow^{\mathrm{A}} N \\
\text{Any} \vdash \langle \Gamma; \mathtt{ref}\ V; \mu \rangle & \longrightarrow^{m} & \langle \Gamma; a; \mu[a \mapsto V] \rangle & \text{if } a \notin \mathrm{dom}(\mu) \\
\text{Any} \vdash \langle \Gamma; a := V; \mu \rangle & \longrightarrow^{m} & \langle \Gamma; (); \mu[a \mapsto V] \rangle & \\
\text{Any} \vdash \langle \Gamma; !a; \mu \rangle & \longrightarrow^{m} & \langle \Gamma; \mu(a); \mu \rangle & \\
\text{Any} \vdash \langle \Gamma; F[M]; \mu \rangle & \longrightarrow^{m} & \langle \Gamma'; F[N]; \mu' \rangle & \text{if ctx} \vdash \langle \Gamma; M; \mu \rangle \longrightarrow^{m} \langle \Gamma'; N; \mu' \rangle \\
\text{NonCast} \vdash \langle \Gamma; M\langle c \rangle; \mu \rangle & \longrightarrow^{m} & \langle \Gamma'; N\langle c \rangle; \mu' \rangle & \text{if Any} \vdash \langle \Gamma; M; \mu \rangle \longrightarrow^{m} \langle \Gamma'; N; \mu' \rangle \\
\text{Any} \vdash \langle \Gamma; U\langle \mathtt{id}; m; (e_1{<:}e_2 \Rightarrow \star)! \rangle \bullet\star^{\ell}; \mu \rangle & \longrightarrow^{m} & \langle \Gamma; U\langle \mathtt{id}; m; \mathtt{id} \rangle \bullet; \mu \rangle & \text{if } \Gamma \vdash e_1{<:}e_2 \\
\text{Any} \vdash \langle \Gamma; U\langle \mathtt{id}; m; (e_1{<:}e_2 \Rightarrow \star)! \rangle \bullet\star^{\ell}; \mu \rangle & \longrightarrow^{m} & \langle \Gamma; \mathtt{blame}\ \ell; \mu \rangle & \text{if } \Gamma \vdash e_1 \not<: e_2 \\
\text{Any} \vdash \langle \Gamma; U\langle \mathtt{id}; m; G! \rangle \bullet \star\ell; \mu \rangle & \longrightarrow^{m} & \langle \Gamma; \mathtt{blame}\ \ell; \mu \rangle & \text{if } \forall e_1 e_2.\ G \neq (e_1{<:}e_2 \Rightarrow \star)
\end{array}
$$

$\boxed{M^{\circ} \longrightarrow^{\circ} N^{\circ}}$

$$
\frac{}{{\sim}((\text{`}V^{\circ}\text{'}e_1)\langle \mathtt{id}; \text{``}\mathtt{id}; m^{\circ}; \mathtt{id}\text{''}(\mathtt{id}; e_1{\uparrow}e_2; \mathtt{id}); \mathtt{id} \rangle) \longrightarrow^{\circ} V^{\circ}}\ \textit{(splice-sub)}
$$

⋮ (remaining rules same as Figure 18)

$\boxed{e \vdash \mathcal{C}^{\circ} \longrightarrow^{\circ} \mathcal{D}^{\circ}}$

$$
\frac{\text{ctx} \vdash \langle \Gamma; M; \mu \rangle \longrightarrow^{m} \langle \Gamma'; N; \mu' \rangle}{e \vdash \langle \Gamma; {\sim}M; \mu \rangle \longrightarrow^{\circ} \langle \Gamma'; {\sim}N; \mu' \rangle}\ \textit{($\xi$-splice)}
$$

⋮ (remaining rules same as Figure 18)

Fig. 28. Space-efficient reduction rules for $\mathrm{CC}^{\alpha,\star}_{\mathrm{Ref}}$ using hypercoercions

$$d \in \mathbb{B} \qquad n \in \mathbb{N}$$

$\boxed{d \vdash^{\circ} M^{\circ}\ \mathsf{ok}}$

$$\frac{n \mid d \vdash^{m} M\ \mathsf{ok}}{d \vdash^{\circ} {\sim}M\ \mathsf{ok}}\ (\textit{ok-splice}) \qquad \frac{}{d \vdash^{\circ} k\ \mathsf{ok}}\ (\textit{ok-code-const}) \qquad \frac{}{d \vdash^{\circ} x\ \mathsf{ok}}\ (\textit{ok-code-var})$$

$$\frac{d \vdash^{\circ} N^{\circ}\ \mathsf{ok}}{d \vdash^{\circ} \lambda(x{:}T)^{\alpha}.\,N^{\circ}\ \mathsf{ok}}\ (\textit{ok-code-abs}) \qquad \frac{d \vdash^{\circ} N^{\circ}\ \mathsf{ok}}{d \vdash^{\circ} \overline{\lambda}(x{:}T)^{\alpha}.\,N^{\circ}\ \mathsf{ok}}\ (\textit{ok-code-abs-val})$$

$$\frac{d \vdash^{\circ} L^{\circ}\ \mathsf{ok} \qquad d \vdash^{\circ} M^{\circ}\ \mathsf{ok}}{d \vdash^{\circ} L^{\circ}\, M^{\circ}\ \mathsf{ok}}\ (\textit{ok-code-app}) \qquad \frac{}{d \vdash^{\circ} \mathtt{blame}\ \ell\ \mathsf{ok}}\ (\textit{ok-code-blame})$$

$\boxed{n \mid d \vdash^{m} M\ \mathsf{ok}}$

$$\frac{n \mid \mathsf{true} \vdash^{m} M\ \mathsf{ok} \qquad n \leq 1}{n+1 \mid \mathsf{true} \vdash^{m} M\langle c\rangle\ \mathsf{ok}}\ (\textit{ok-cast-delayed}) \qquad \frac{n \mid \mathsf{false} \vdash^{m} M\ \mathsf{ok} \qquad n \leq 2}{n+1 \mid \mathsf{false} \vdash^{m} M\langle c\rangle\ \mathsf{ok}}\ (\textit{ok-cast})$$

$$\frac{}{0 \mid d \vdash^{m} k\ \mathsf{ok}}\ (\textit{ok-meta-const}) \qquad \frac{}{1 \mid d \vdash^{m} x\ \mathsf{ok}}\ (\textit{ok-meta-var}) \qquad \frac{}{0 \mid d \vdash^{m} a\ \mathsf{ok}}\ (\textit{ok-addr})$$

$$\frac{n \mid \mathsf{true} \vdash^{m} N\ \mathsf{ok}}{0 \mid d \vdash^{m} \lambda x.\,N\ \mathsf{ok}}\ (\textit{ok-meta-abs}) \qquad \frac{n \mid d \vdash^{m} L\ \mathsf{ok} \qquad n' \mid d \vdash^{m} M\ \mathsf{ok}}{0 \mid d \vdash^{m} L\, M\ \mathsf{ok}}\ (\textit{ok-meta-app})$$

$$\frac{n \mid \mathsf{true} \vdash^{m} N\ \mathsf{ok}}{0 \mid d \vdash^{m} \Lambda\alpha.\,N\ \mathsf{ok}}\ (\textit{ok-ec-abs}) \qquad \frac{n \mid d \vdash^{m} M\ \mathsf{ok}}{0 \mid d \vdash^{m} M[e]\ \mathsf{ok}}\ (\textit{ok-ec-app})$$

$$\frac{n \mid \mathsf{true} \vdash^{m} N\ \mathsf{ok}}{0 \mid d \vdash^{m} e_1{<:}e_2{\Rightarrow}N\ \mathsf{ok}}\ (\textit{ok-sub-intro}) \qquad \frac{n \mid d \vdash^{m} M\ \mathsf{ok}}{0 \mid d \vdash^{m} M\bullet\ \mathsf{ok}}\ (\textit{ok-sub-elim})$$

$$\frac{n \mid d \vdash^{m} M\ \mathsf{ok}}{0 \mid d \vdash^{m} M\bullet\star^{\ell}\ \mathsf{ok}}\ (\textit{ok-sub-elim-}\star) \qquad \frac{d \vdash^{\circ} M^{\circ}\ \mathsf{ok}}{0 \mid d \vdash^{m} \text{``}M^{\circ}\text{''}e\ \mathsf{ok}}\ (\textit{ok-quote})$$

$$\frac{n \mid d \vdash^{m} M\ \mathsf{ok}}{0 \mid d \vdash^{m} \mathtt{ref}\ M\ \mathsf{ok}}\ (\textit{ok-ref}) \qquad \frac{n \mid d \vdash^{m} L\ \mathsf{ok} \qquad n' \mid d \vdash^{m} M\ \mathsf{ok}}{0 \mid d \vdash^{m} L := M\ \mathsf{ok}}\ (\textit{ok-assign})$$

$$\frac{n \mid d \vdash^{m} M\ \mathsf{ok}}{0 \mid d \vdash^{m}\ !M\ \mathsf{ok}}\ (\textit{ok-deref}) \qquad \frac{}{0 \mid d \vdash^{m} \mathtt{blame}\ \ell\ \mathsf{ok}}\ (\textit{ok-meta-blame})$$

Fig. 29. Adjacent cast judgments for code and meta terms

$\boxed{\mu\ \mathsf{ok}}$

$$\frac{}{\emptyset\ \mathsf{ok}}\ (\mu\textit{ok-empty}) \qquad \frac{n \mid \mathsf{false} \vdash^{m} V\ \mathsf{ok} \qquad \mu\ \mathsf{ok}}{(\mu[a \mapsto V])\ \mathsf{ok}}\ (\mu\textit{ok-cons})$$

Fig. 30. Adjacent cast judgment for heaps

$$\boxed{\mathsf{height} : c \cup c^{\circ} \longrightarrow \mathbb{N}}$$

$$
\begin{aligned}
\mathsf{height}(h^{o}; m^{o}; t^{o}) &= \mathsf{height}(m^{o}) \\
\mathsf{height}(h; m; t) &= \mathsf{height}(m)
\end{aligned}
$$

$$\boxed{\mathsf{height} : c^{e} \longrightarrow \mathbb{N}}$$

$$\mathsf{height}(c^{e}) = 1$$

$$\boxed{\mathsf{height} : m \cup m^{\circ} \longrightarrow \mathbb{N}}$$

$$
\begin{array}{lll}
\mathsf{height}(\mathtt{id}) & = \mathsf{height}(\bot^{\ell}) & = 1 \\
\mathsf{height}(c_1 \longrightarrow c_2) & = \mathsf{height}(\mathtt{Ref}\ c_1\ c_2) & = 1 + \max(\mathsf{height}(c_1), \mathsf{height}(c_2)) \\
\mathsf{height}(c_1^{\circ} \longrightarrow c_2^{\circ}) & & = 1 + \max(\mathsf{height}(c_1^{\circ}), \mathsf{height}(c_2^{\circ})) \\
\mathsf{height}(\text{``}\, c^{\circ}\, \text{''}\, c^{e}) & & = 1 + \max(\mathsf{height}(c^{\circ}), \mathsf{height}(c^{e})) \\
\mathsf{height}(\forall \alpha.c) & = \mathsf{height}(e_1 <: e_2 \Rightarrow c) & = 1 + \mathsf{height}(c)
\end{array}
$$

Fig. 31. Hypercoercion height

$$\boxed{\mathsf{size} : c \cup c^{\circ} \longrightarrow \mathbb{N}}$$

$$
\begin{aligned}
\mathsf{size}(h^{o}; m^{o}; t^{o}) &= 4 + \mathsf{size}(m^{o}) \\
\mathsf{size}(h; m; t) &= 4 + \mathsf{size}(m)
\end{aligned}
$$

$$\boxed{\mathsf{size} : c^{e} \longrightarrow \mathbb{N}}$$

$$\mathsf{size}(c^{e}) = 5$$

$$\boxed{\mathsf{size} : m \cup m^{\circ} \longrightarrow \mathbb{N}}$$

$$
\begin{array}{lll}
\mathsf{size}(\mathtt{id}) & = \mathsf{size}(\bot^{\ell}) & = 1 \\
\mathsf{size}(c_1 \longrightarrow c_2) & = \mathsf{size}(\mathtt{Ref}\ c_1\ c_2) & = 1 + \mathsf{size}(c_1) + \mathsf{size}(c_2) \\
\mathsf{size}(c_1^{\circ} \longrightarrow c_2^{\circ}) & & = 1 + \mathsf{size}(c_1^{\circ}) + \mathsf{size}(c_2^{\circ}) \\
\mathsf{size}(\text{``}\, c^{\circ}\, \text{''}\, c^{e}) & & = 1 + \mathsf{size}(c^{\circ}) + \mathsf{size}(c^{e}) \\
\mathsf{size}(\forall \alpha.c) & = \mathsf{size}(e_1 <: e_2 \Rightarrow c) & = 1 + \mathsf{size}(c)
\end{array}
$$

Fig. 32. Hypercoercion size

$$\boxed{\text{c_height} : M \cup M^\circ \longrightarrow \mathbb{N}}$$

$$
\begin{array}{rcl}
\text{c_height}(k) & = & 0 \\
\text{c_height}(x) & = & 0 \\
\text{c_height}(a) & = & 0 \\
\text{c_height}(\texttt{blame}\ \ell) & = & 0 \\
\text{c_height}(\lambda(x{:}T)^\alpha.\, N^\circ) & = & \text{c_height}(N^\circ) \\
\text{c_height}(\overline{\lambda}(x{:}T)^\alpha.\, N^\circ) & = & \text{c_height}(N^\circ) \\
\text{c_height}(L^\circ\, M^\circ) & = & \max(\text{c_height}(L^\circ), \text{c_height}(M^\circ)) \\
\text{c_height}({\sim} M) & = & \text{c_height}(M) \\
\text{c_height}(\lambda(x : \hat{A}).\, N) & = & \text{c_height}(N) \\
\text{c_height}(\Lambda\alpha.\, N) & = & \text{c_height}(N) \\
\text{c_height}(e_1{<:}e_2{\Rightarrow}N) & = & \text{c_height}(N) \\
\text{c_height}(``M^\circ\,''e) & = & \text{c_height}(M^\circ) \\
\text{c_height}(L\, M) & = & \max(\text{c_height}(L), \text{c_height}(M)) \\
\text{c_height}(L := M) & = & \max(\text{c_height}(L), \text{c_height}(M)) \\
\text{c_height}(M[e]) & = & \text{c_height}(M) \\
\text{c_height}(M\bullet) & = & \text{c_height}(M) \\
\text{c_height}(M{\bullet}{\star}^\ell) & = & \text{c_height}(M) \\
\text{c_height}(\texttt{ref}\ M) & = & \text{c_height}(M) \\
\text{c_height}(!\, M) & = & \text{c_height}(M) \\
\text{c_height}(M\langle c\rangle) & = & \max(\text{c_height}(M), \text{height}(c))
\end{array}
$$

Fig. 33. Cast height for code and meta terms

$$\boxed{\text{size} : M \cup M^\circ \longrightarrow \mathbb{N}}$$

$$
\begin{array}{rl}
\text{size}(k) = \text{size}(x) = \text{size}(\texttt{blame}\ \ell) = \text{size}(a) & = 1 \\
\text{size}(\lambda(x{:}T)^\alpha.\, N^\circ) = \text{size}(\overline{\lambda}(x{:}T)^\alpha.\, N^\circ) & = 1 + \text{size}^\circ(N^\circ) \\
\text{size}^\circ(L^\circ\, M^\circ) & = 1 + \text{size}(L^\circ) + \text{size}(M^\circ) \\
\text{size}({\sim} M) & = 1 + \text{size}(M) \\
\text{size}(\lambda(x : \hat{A}).\, N) = \text{size}(\Lambda\alpha.\, N) = \text{size}(e_1{<:}e_2{\Rightarrow}N) & = 1 + \text{size}(N) \\
\text{size}(L\, M) = \text{size}(L := M) & = 1 + \text{size}(L) + \text{size}(M) \\
\text{size}(M[e]) = \text{size}(M\bullet) = \text{size}(M{\bullet}{\star}^\ell) = \text{size}(\texttt{ref}\ M) = \text{size}(!\, M) & = 1 + \text{size}(M) \\
\text{size}(``M^\circ\,''e) & = 1 + \text{size}(M^\circ) \\
\text{size}(M\langle c\rangle) & = \text{size}(M) + \text{size}(c)
\end{array}
$$

$$\boxed{\text{num} : M \cup M^\circ \longrightarrow \mathbb{N}}$$

$$\vdots \qquad \text{(remaining cases are analogous to size)}$$

$$\text{num}(M\langle c\rangle) = \text{num}(M) + 1$$

$$\boxed{\text{ideal_size} : M \cup M^\circ \longrightarrow \mathbb{N}}$$

$$\vdots \qquad \text{(remaining cases are analogous to size)}$$

$$\text{ideal_size}(M\langle c\rangle) = \text{ideal_size}(M)$$

Fig. 34. size, num, and ideal_size for code and meta terms

$$\boxed{\text{c_height}, \text{size}, \text{ideal_size} : \mu \longrightarrow \mathbb{N}}$$

$$\begin{array}{rclrcl}
\text{c_height}(\emptyset) & = & 0 & \text{c_height}(\mu[a \mapsto V]) & = & \max(\text{c_height}(V), \text{c_height}(\mu)) \\
\text{size}(\emptyset) & = & 0 & \text{size}(\mu[a \mapsto V]) & = & \text{size}(V) + \text{size}(\mu) \\
\text{ideal_size}(\emptyset) & = & 0 & \text{ideal_size}(\mu[a \mapsto V]) & = & \text{ideal_size}(V) + \text{ideal_size}(\mu)
\end{array}$$

Fig. 35. Sizes and height functions for heaps

$$\boxed{\text{c_height}, \text{size}, \text{ideal_size} : C \cup C^{\circ} \longrightarrow \mathbb{N}}$$

$$\begin{array}{rcl}
\text{c_height}(\langle \Gamma; M; \mu \rangle) & = & \max(\text{c_height}(\mu), \text{c_height}(M)) \\
\text{size}(\langle \Gamma; M; \mu \rangle) & = & \text{size}(\mu) + \text{size}(M) \\
\text{ideal_size}(\langle \Gamma; M; \mu \rangle) & = & \text{ideal_size}(\mu) + \text{ideal_size}(M)
\end{array}$$

Fig. 36. Sizes and height functions for configurations